\documentclass[preprint,nofootinbib,longbibliography,
amsmath,amssymb,aps,prd]{revtex4-2}

\usepackage[utf8]{inputenc}
\usepackage{amsmath}
\usepackage{amsfonts,dsfont}
\usepackage{mathrsfs}
\usepackage{cancel} 
\usepackage{amssymb,ulem}
\usepackage{amsmath}
\usepackage{romannum}
\usepackage{subfigure}
\usepackage{adjustbox}
\usepackage{dcolumn}
\usepackage{graphicx}
\usepackage{bm}
\usepackage{enumerate}
\usepackage[colorlinks=true,linkcolor=blue,urlcolor=blue,filecolor=black,
citecolor=red,pdfstartview=FitV,pdftitle={},pdfsubject={},pdfkeywords={},
pdfpagemode=None,bookmarksopen=true]{hyperref}

\usepackage{xcolor}
\usepackage{float}
\usepackage{enumitem}  
\usepackage{bbold}
\usepackage{tikz}
\usepackage{multirow}
\usepackage{longtable}
\usepackage{supertabular,rotating,adjustbox}
\newcommand{\dd}{{\text{d}}}

\begin{document}
	\preprint{APS/PRD}
	\title{\boldmath
		The  Global Fits of New Physics  in  \texorpdfstring{$b \to s $}{b to s} after 
		$R_{K^{(*)}}$ 2022 Release}
	
	\author{Qiaoyi Wen}
	\author{Fanrong Xu}
	\email{fanrongxu@jnu.edu.cn}
	
	\affiliation{Department of Physics and Siyuan Laboratory, Jinan University,
		Guangzhou 510632, P.R. China}
	
	\begin{abstract} 
	
The measurement of lepton universality parameters $R_{K^{(*)}}$ was updated by LHCb in December 2022, which indicated that the well-known anomalies in flavor-changing neutral current (FCNC) processes of B meson decays have faded away. However, does this mean that all new physics possibilities related to $b\to s\ell^+\ell^-$ have been excluded? We aim to answer this question in this work.
The state-of-the-art effective Hamiltonian is adopted to describe $b \to s$ transition, while BSM (beyond the Standard Model) new physics effects are encoded in Wilson coefficients (WCs). Using around 200 observables in leptonic and semileptonic decays of B mesons and bottom baryons, measured by LHCb, CMS, ATLAS, Belle, and BaBar, we perform global fits of these Wilson coefficients in four different scenarios. In particular, lepton flavors in WCs are specified in some of the working scenarios. To see the change of new physics parameters, we use both the data before and after the 2022 release of $R_{K^{(*)}}$ in two separate sets of fits.
We find that in three of the four scenarios, $\Delta C_9^\mu$ still has a deviation around or
 more than $4\sigma$ from the Standard Model. 
 The lepton flavor in WCs is distinguishable for $\Delta C_{9}$ 
 at the $1\sigma$ level, 
  but at the $2\sigma$ level all the operators are flavor identical. 
 We demonstrate numerically that there is no chirality for muon type of scalar operator and it is kept at the $1\sigma$ level for their electron type dual ones, while chiral difference exists for $\mathcal{O}_{9}^\mu$ at least at the $2\sigma$ level. 
 Moreover, it can be deduced that the scalar operators $\mathcal{O}_{S,P}^{(')\mu}$ become null
 if new physics emerges in terms of SMEFT (Standard Model Effective Field Theory) up to dimension-6.

	\end{abstract}
	\keywords{FCNC, B anomalies, LFU}
	\maketitle
	\clearpage
	\clearpage
	\newpage
	\pagenumbering{arabic}

	\section{Introduction}\label{sec: Intro}
	
	The quest for new physics in FCNC process $b\to s \ell^+ \ell^-$ has lasted for more than one decade.
	It was expected that in  $B\to K^* \mu^+\mu^-$ new physics effect would emerge by measuring the forward-backward asymmetry  (AFB), and the early measurements carried out by Belle \cite{Belle:2009zue}, BaBar \cite{BaBar:2008fao} and CDF \cite{CDF:2011grz} indeed  seemed to prefer the new physics with a flipped $C_7$.  In 2011, however, a SM-like behavior of AFB was confirmed by LHCb even with its $309 \, \textrm{pb}^{-1}$ data \cite{LHCb:2011tcp}. Later on, more  observables, including branching fractions, angular distributions and
	lepton universality parameters,  were  measured more and more precisely and deviations from SM predictions 
	in particular bins were found by  Belle \cite{Belle:2019oag,BELLE:2019xld,Belle:2016fev}, 
	LHCb \cite{LHCb:2014vgu,Aaij:2019wad,LHCb:2021trn,LHCb:2021lvy,LHCb:2017avl,LHCb:2014cxe,LHCb:2016ykl,LHCb:2020lmf,LHCb:2020gog}, ATLAS \cite{ATLAS:2018gqc} and CMS \cite{CMS:2015bcy,CMS:2017rzx}.

	The non-universality of lepton flavor, characterized by  $ R_{K^{(*)}} \equiv\mathcal{B}({B}\to K^{(*)}\mu^+\mu^-)/\mathcal{B}({B} \to K^{(*)}e^+e^-)$,
	is among one of the anomalies appearing in semileptonic B meson decays.
	The $2.6 \sigma$ deviation  of $R_K$ in charged B meson decay was firstly reported  in 2014 \cite{LHCb:2014vgu} and updated in 2019 \cite{Aaij:2019wad} 
	by LHCb. It was found by Belle in 2019  that a similar $2.6 \sigma$ standard deviation also occurred at $q^2 \in (1.0, 6.0) {\rm{GeV}}^2/c^4$ \cite{BELLE:2019xld}.  Later in 2021, with Run I and Run II  dataset LHCb reported 
	$ R_{K} =0.846^{+0.042}_{-0.039}(\text{stat})^{+0.013}_{-0.012}(\text{sys})$ as well as $R_{K_{S}}=0.66^{+0.20}_{-0.14}(\text{stat})^{+0.02}_{-0.04}(\text{sys})$ at the level of $ 3.1\sigma$ and $ 1.4\sigma $, respectively. 
	As for $B\to V\ell^+ \ell^-$ sector, in 2017 LHCb found $ 2.1$ - $2.3\sigma $ and $ 2.4$ - $2.5\sigma $ deviations  for low-$q^2$ bins and central-$q^2$ bins in neutral B decay
	$B^0\to K^{*0} \ell^+ \ell^-$ \cite{LHCb:2017avl}. 
	Later in 2021, in charged B decay $B^+\to K^{*+}\ell^+ \ell^-$ LHCb reported its measurement
	as 
	$R_{K^{\ast+}}=0.70^{+0.18}_{-0.13}(\text{stat})^{+0.03}_{-0.04}(\text{sys})$,  which only deviated by $1.5 \sigma$ \cite{LHCb:2021lvy}. And a SM consistent results was also
	obtained by Belle in 2019  \cite{Belle:2019oag}.
	At the end of 2022,  LHCb reported its updated measurement of $ R_{K^{(\ast)}} $ at both low-$q^2$ and central-$q^2$ region 
	by correcting previous underestimations on electron mode contribution \cite{LHCb:2022zom}, giving
	\begin{equation}
		\text{low-$q^2$}\left\{
		\begin{aligned}
			&R_K = 0.994^{+0.090}_{-0.082}(\text{stat.})^{+0.029}_{-0.027}(\text{syst.})\\
			&R_{K^\ast} = 0.997^{+0.093}_{-0.087}(\text{stat.})^{+0.036}_{-0.035}(\text{syst.}),
		\end{aligned}
		\right.\nonumber
	\end{equation}
	\vspace{-0.6pt}
	\begin{equation}
		\text{central-$q^2$}\left\{
		\begin{aligned}
			&R_K = 0.949^{+0.042}_{-0.041}(\text{stat.})^{+0.022}_{-0.022}(\text{syst.})\\
			&R_{K^\ast} = 1.027^{+0.072}_{-0.068}(\text{stat.})^{+0.027}_{-0.026}(\text{syst.}).
		\end{aligned}
		\right.
	\end{equation}
	The overall 0.2$\sigma$ deviation, compatible with SM, indicates that the expected new physics 
	in form of lepton flavor non-universality has faded away.
	Regarding to the existence of several deviations in branching fractions and 
	angular distribution, the remaining new physics opportunities  in the $b\to s\ell^+\ell^-$ window are naturally of great interest.
	Therefore, it is timely to carry out updated global fits in combination with all the related data to help to understand current status.
	
	So far there have been rich data on leptonic decay $B_{s,d}\to \ell^+\ell^-$ and semileptonic decay 
	$B\to K \ell^+\ell^-, B\to K^* \ell^+ \ell^-, B\to X_s\ell^+ \ell^-,  \Lambda_b\to \Lambda \ell^+ \ell^-$. We 
	collect all available observables of them as parts of inputs in fitting analysis. As for theoretical description,
	we adopt the state-of-the-art effective Hamiltonian, in which high energy particles are integrated out and
	absorbed in Wilson coefficients (WCs). In SM, there are only 4 effective operators contributed in the $b\to s\ell^+\ell^-$
	effective Hamiltonian. New physics effects are brought in either by extra operators  (the dual operators as well as
	scalar operators, see Sec. \ref{sec:frame}) or modifications of WCs. 
	At hadron energy scale, we rely on QCDF approach to deal with B meson semileptonic decays \cite{Beneke:2004dp,Beneke:2001at}. 
	To extend our exploration in model-independent analysis, 
	the global fits of  four different cases with particular operator combinations
	are performed based on Bayesian statistics,
	taking into account both theoretical and experimental errors.  
	Especially, the fit based on 20-D WCs by specifying lepton flavors
	as one of the working scenarios is  provided.  
	We make the fits both before and after the 2022 $R_{K^{(*)}}$ release, and make comparisons with some of the similar model-independent analyses done by other groups \cite{Alguero:2021anc,Altmannshofer:2021qrr,Hurth:2022lnw,Geng:2021nhg, MunirBhutta:2020ber, Gubernari:2022hxn, SinghChundawat:2022zdf}. 
	Although  WCs slightly differ from each other in our four scenarios, we find commonly for $\Delta C_9^\mu$ 
	the around $4 \sigma$ deviation from SM
	still exists after the 2022 $R_{K^{(*)}}$ release, which indicates that the new physics opportunity cannot be excluded.  
	Meanwhile, we find that, at $1\sigma$ level,  $\Delta C_{10}^\mu$ is indistinguishable from $\Delta C_{10}^e$ 
	but $\Delta C_{9,S,P}^\mu$ differ from $\Delta C_{9,S,P}^e$.  The WCs $\Delta C_{S,P}^\mu$ are strictly chirality independent,
	while the equivalence between $\Delta C_{S,P}^e$ and $\Delta C_{S,P}^{'e}$ is obeyed at $1\sigma$ level. 
	It can be demonstrated numerically that $\Delta C_{9,10}$ violate this chirality identity at least at $2\sigma$ level.
	By combining the data, we also find that scalar operators become null
        if the new physics is described within the framework of SMEFT.	
	
	The remaining parts of the paper are organized as follows.  
	In Sec. \ref{sec:frame}, we provide the whole working frame of the analysis, including
	theoretical framework of effective Hamiltonian approach and the adopted four different working scenarios (denoted as the muon-specific scenario, lepton-universal scenario, lepton-specific scenario and the full scenario), the related observables in all the involved decay processes as well as the fitting schemes.  Then numerical analysis is given in Sec. \ref{sec: num}, with
	inputs summarized in \ref{sub: num_input}, results presented in \ref{sub: num_res} and discussions
	carried on in \ref{sub:disc}.
	We conclude the paper in Sec.\ref{sec:con}. More details on theoretical formulas and
	experimental inputs can be referred to
	Appendix \ref{app:Theo}  and \ref{app:exp_input}, respectively.

	\section{The Working Frame}
	\label{sec:frame}
	
	\subsection{Theoretical Framework}
	\label{sub: theo}
	
	The state-of-the-art effective Hamiltonian is adopted to describe $ b \to s $ transition,
	in which high energy information is contained in Wilson coefficients while remaining low energy part 
	resorted to effective operators and their corresponding matrix elements.
	The Wilson coefficients can be obtained at electroweak scale by integrating out heavy particles and 
	running into B meson scale with RGE,  
	leading to the effective Hamiltonian 
	\begin{equation}
		\begin{aligned}
			\mathscr{H}=-\frac{4G_F}{\sqrt{2}}V_{tb}V^*_{ts}\frac{e^2}{16\pi^2}\sum_i\left(C_i \mathcal{O}_i+C'_i\mathcal{O}'_i\right)+\text{h.c.},
		\end{aligned}
	\end{equation}
	in which the effective operators are defined as
	\begin{equation}
		\begin{aligned}
			&\mathcal{O}_7=\frac{m_b}{e}(\bar{s}\sigma_{\mu\nu}P_R b)F^{\mu\nu},
			&\mathcal{O}'_7=\frac{m_b}{e}(\bar{s}\sigma_{\mu\nu}P_L b)F^{\mu\nu},\\
			&\mathcal{O}_8=\frac{g_sm_b}{e^2}(\bar{s}\sigma_{\mu\nu}T^aP_R b)G^{\mu\nu}_a,
			&\mathcal{O}'_8=\frac{g_sm_b}{e^2}(\bar{s}\sigma_{\mu\nu}T^aP_L b)G^{\mu\nu}_a,\\	
			&\mathcal{O}_9=(\bar{s}\gamma_\mu P_L b)(\bar{\ell}\gamma^\mu \ell),
			&\mathcal{O}'_9=(\bar{s}\gamma_\mu P_R b)(\bar{\ell}\gamma^\mu \ell),\nonumber\\
			&\mathcal{O}_{10}=(\bar{s}\gamma_\mu P_L b)(\bar{\ell}\gamma^\mu\gamma_5 \ell),
			&\mathcal{O}'_{10}=(\bar{s}\gamma_\mu P_R b)(\bar{\ell}\gamma^\mu \gamma_5\ell),\nonumber\\
			&\mathcal{O}_{S}=m_b(\bar{s}P_R b)(\bar{\ell}  \ell),
			&\mathcal{O}'_{S}=m_b(\bar{s}P_L b)(\bar{\ell}  \ell),\;\nonumber\\
			&\mathcal{O}_{P}=m_b(\bar{s}P_R b)(\bar{\ell} \gamma_5 \ell),
			&\mathcal{O}'_{P}=m_b(\bar{s}P_L b)(\bar{\ell} \gamma_5 \ell),\nonumber	
		\end{aligned}
	\end{equation}
	with  strength tensor of electromagnetic field $F_{\mu\nu}$
	and gluon field $G_{\mu\nu}$, respectively.  The new physics effects manifest either in forms of 
	new types of operators and WCs.
	In SM, only operators $\mathcal{O}_{7,8,9,10}$ turn up while the appearance of their chiral-flipped 
	dual operators with a prime as well as scalar operators
	$\mathcal{O}_{S,P}^{(')}$ implicate an existence of new physics (NP).
	As for WCs, they  have been calculated precisely  in SM and can be found in \cite{Altmannshofer:2008dz,Du:2015tda,Hou:2014dza,Blake:2016olu}, 
	of which any deviations from their SM results are indications of NP. Especially, NP being embodied in WCs\footnote{Here 
		we focus on the discussion of CP-conserving NP effects, so these WCs are assumed to be real. The discussion of complex 
		WCs can be referred to \cite{Biswas:2020uaq,Altmannshofer:2021qrr}. $C_7'$ has a little SM contributions proportional to $ m_s/m_b C_7 $.}
	can be denoted as 
	\begin{equation}
		\begin{aligned}
			C_{i;\text{NP}}^{(\prime)\ell }\equiv\Delta C_{i}^{(\prime)\ell }=C_{i}^{(\prime)\ell }-C_{i;\text{SM}}^{(\prime)\ell },
		\end{aligned}
	\end{equation}
	in which lepton flavors ($\ell=e, \mu$) will be specified in part of following numerical analysis. Although we aim to perform a model-independent
	analysis, the true NP model is unknown. Thus in following practical analysis, to explore various NP possibilities
	we discuss different combinations of NP operators as follows.
	\begin{enumerate}[label=\Roman*., start=1]
		\item  The muon-specific scenario.\\
		In this case, we set
		$ \Delta C_{9,10,S,P}^{(\prime)\mu} $  and $ \Delta C_{7,8}^{(\prime)} $ as free parameters by setting
		$\Delta C_{9,10,S,P}^{(\prime)e} = 0$.
		
		\item  The lepton-universal scenario.\\
		With
		$\Delta C_{9,10,S,P}^{(\prime)\mu} =  \Delta C_{9,10,S,P}^{(\prime)e}$, the degree of freedoms are
		$\Delta C_{9,10,S,P}^{(\prime)\mu}$
		and $ \Delta C_{7,8}^{(\prime)} $. 
		
		\item The lepton-specific scenario.\\
		Here the radiative operators and gluon dipole operators vanish  $(\Delta C_7^{(\prime)} = 0, ~ \Delta C_{8}^{(\prime)}=0)$,
	         while the remaining operators  
		$\Delta C_{9,10,S,P}^{(\prime)\mu;e} $ are unconstrained. 
		
		\item The full scenario.\\
		In this case, all the parameters
		$ \Delta C_{7,8}^{(\prime)},~ \Delta C_{9,10,S,P}^{(\prime)\mu;e}$ are left unrestricted.
		
	\end{enumerate}
	
	The following numerical analysis on the WCs will provide the latest model-independent information. And by
	making use of the obtained WCs in $b \to s $ FCNC processes, we will
	discuss some related NP models in a separated paper.

	\subsection{Observables in Decay Processes}\label{sub: fit_obs}
	
	We summarize theoretically and experimentally all available  decay processes  related to $b\to s \ell^+\ell^-$ in this part. 
	As for the choice of  experimental data of observables, 
	the adopted data from
	different collaborations (LHCb, CMS, ATLAS, Belle) will be divided into two datasets. 
	Dataset \textbf{A} contains \textcolor{red}{201} observables  before $R_{K^{(*)}}$ 2022 LHCb release, while Dataset \textbf{B}, containing
	\textcolor{red}{203} observables, is obtained by replacing LHCb earlier $R_{K^{(*)}}$  results by the 2022 updated one \cite{LHCb:2022zom} .  
	
	\begin{enumerate}[label=\Roman*., start=1]
		
		\item $B_{s,d}\to\mu^+\mu^-$ \\
		The branching fraction of leptonic decay $B_{s,d}\to\mu^+\mu^-$ \cite{Altmannshofer:2008dz,Bobeth:2001sq} is given as
		\begin{equation}
			\begin{aligned}
			\resizebox{0.9\textwidth}{!}{
			$\mathcal{B}\left(B_{s,d}\to\mu^+\mu^-\right)=\tau_{B_{s,d}}f_{B_{s,d}}^2m_{B_{s,d}}\frac{\alpha^2_{\text{e}}G_F^2}{16\pi^3}|V_{tb}V_{ts(d)}^\ast|^2\sqrt{1-\frac{4m_\mu^2}{m^2_{B_{s,d}}}}\left[|S|^2\left(1-\frac{4m_\mu^2}{m^2_{B_{s,d}}}\right)+|P|^2\right]$}
			\end{aligned}
		\end{equation}
		with 
		\begin{equation}
			\begin{aligned}
				S=\frac{m_{B_{s,d}}}{2}\left(C_{S}^\mu-C_{S}^{\mu\prime}\right),\qquad P=\frac{m_{B_{s,d}}}{2}\left(C_P^{\mu}-C_P^{\mu\prime}\right)+m_\mu\left(C_{10}^{\mu}-C_{10}^{\mu\prime}\right),
			\end{aligned}
		\end{equation}
		in which the SM situation is contained as an extreme example by setting $S\to 0$ and $C_P^{(')}\to 0$ in $P$. Here the muon flavor
		in WC has been neglected. For  $B_s\to \mu^+\mu^-$, we take into account both the latest LHCb  \cite{LHCb:2021awg} and CMS \cite{CMS:2022mgd} measured value.
		So far the branching fraction of $B_d\to \mu^+\mu^-$ has not been measured, we  adopt fitting results from different group.
		
		\item  $ B\to V\ell^+\ell^- $\\
		Decay modes $B^{+;0}\to K^{\ast+;0}\mu^+\mu^-$ are both classified into $ B\to V\ell\ell $.
		There are several kinds of observables, including  \textit{Lepton-Universality-Ratio} (LUR) $ R_{K^{(\ast)}} $, \textit{Branching Ratios}  (BR)\cite{LHCb:2017avl,LHCb:2021lvy,Belle:2019oag,LHCb:2022zom}, \textit{Angular Distribution Observables} (ADO) $ P_{1,2,3},~P_{4,5,6,8}'$, \textit{Forward-Backward asymmetry} $ A_{FB}$ and \textit{Longitude polarization} $F_{L}$ \cite{LHCb:2021lvy,Belle:2019oag,LHCb:2016ykl,CMS:2015bcy,
			LHCb:2020lmf,CMS:2017rzx,CMS:2015bcy,ATLAS:2018gqc,Belle:2016fev,LHCb:2020gog}. 
		The detailed expressions for the observables are given explicitly as
		\begin{equation}
			\begin{aligned}
				\frac{\dd\Gamma^\ell}{\dd q^2}=\frac{3}{4}\left[2J_{1;s}^\ell\left(C_{7,8,9,10}^{(\prime)\ell}\right)+J_{1;c}^\ell\left(C_{7,8,9,10,S,P}^{(\prime)\ell}\right)\right]-\frac14\left[2J_{2;s}^\ell\left(C_{7,8,9,10}^{(\prime)\ell}\right)+J_{2;c}^\ell\left(C_{7,8,9,10}^{(\prime)\ell}\right)\right],
			\end{aligned}
			\label{eq:BtoVll}
		\end{equation}
		\begin{equation}\label{eq:angular_BVll}
			\begin{aligned}
				&P_1\equiv\frac{J_3}{2J_{2;s}},~P_2\equiv\frac{-J_{6;s}}{8J_{2;s}},~P_3\equiv\frac{J_9}{4J_{2;s}},~P_4^{\prime}\equiv\frac{-J_4}{2\sqrt{-J_{2;s}J_{2:c}}},~P_5^{\prime}\equiv\frac{J_5}{2\sqrt{-J_{2;s}J_{2;c}}},\\
				&P_6^{\prime}\equiv\frac{-J_7}{2\sqrt{-J_{2;s}J_{2;c}}},~P_8^{\prime}\equiv\frac{J_8}{2\sqrt{-J_{2;s}J_{2;c}}},\nonumber\\
				&Q_4\equiv P_4^{\prime\mu}-P_4^{\prime e}, ~Q_5\equiv P_5^{\prime\mu}-P_5^{\prime e},~A_{\text{FB}}\equiv\frac{3}{8}(2S_{6;s}+S_{6;c}),~F_{\text{L}}\equiv-S_{2;c},
			\end{aligned}
		\end{equation}
		among which $ C_{S,P}^{(\prime)} $ only involved  in
		$\Gamma^\ell $, $ P_5' $ and $ A_{\text{FB}} $. 
		The definitions of  $J_i$ and $S_i$ can be referred to Appendix (\ref{app: BVll}). 
		As for $B^0\to K^{\ast0}e^+e^-$ mode, 4 angular observables measured by LHCb\cite{LHCb:2020dof} are also included in this analysis.

		\item $B\to P \ell^+\ell^-$ \\
		Two modes $B^{+;0}\to K^{+;0}\mu^+\mu^-$ have been measured in this class involving  pseudoscalar meson final states, offering LUR\cite{LHCb:2021trn,LHCb:2021lvy,Aaij:2019wad,BELLE:2019xld,LHCb:2022zom} and BR\cite{LHCb:2021lvy,Aaij:2019wad,BELLE:2019xld,LHCb:2014cxe} 
		as observables.
		Branching fractions can be calculated as
		\begin{equation}
			\begin{aligned}
				\frac{\dd\Gamma^\ell}{\dd q^2}&=|V_{tb}V_{ts}^\ast|^2\frac{G_F^2\alpha_e^2M_B^3}{256\pi^5}\lambda\xi_P^2\left[I_a^\ell\left(q^2;C_{7,8,9,10,S,P}^{(\prime)}\right)+\frac{1}{3}I_c^\ell\left(q^2;C_{7,8,9,10}^{(\prime)}\right)\right],
			\end{aligned}
		\end{equation}
		and more details can be found in Appendix (\ref{app: BPll}).

		\item{$B_s\to\phi\mu^+\mu^-$}\\
		Theoretical formula for branching fraction are similar to $B\to V\ell^+\ell^-$ Eq.(\ref{eq:BtoVll}), up to
		form factors (FFs) and spectator effects. LHCb measurement of BR and $ F_L $ \cite{LHCb:2021zwz,LHCb:2021xxq}
		are taken as experimental inputs in the numerical calculations.

		\item{$B\to X_s\ell^+\ell^-$}\\	
		The inclusive process provides complementary information to those exclusive modes. 
		Here we follow the conventions in \cite{Mahmoudi:2008tp,Dai:1996vg,Ghinculov:2003qd} and  the differential branching fraction \footnote{Note that their  $ C_{Q_1;Q_2}$ are related to our definition $ C_{Q_1;Q_2}=m_bC_{S;P}  $, so we re-scale their WCs with MS mass $ \bar{m}_b $.}
		can be written as,
		 \begin{equation}
			\begin{aligned}
				&\frac{d\mathcal{B}(B\to X_s\ell^+\ell^-)}{d\hat{s}}=\frac{\mathcal{B}(B\to X_c\ell\bar{\nu})\alpha_e^2}{4\pi^2f(z)\kappa(z)}\frac{|V_{tb}V_{ts}^\ast|^2}{|V_{cb}|^2}(1-\hat{s})^2\sqrt{1-\frac{4\hat{m}_\ell}{\hat{s}}}\tilde{N}\left(\hat{s};\hat{m}_\ell;C_{7,8,9,S,P}^{(\prime)}\right),
			\end{aligned}
			\label{eq:inclusive}
		\end{equation}
		where  the definitions of $ f(z) $ and $ \kappa(z) $ can be found in \cite{Ghinculov:2003qd}.
	         The latest theoretical formulas incorporating high-order corrections can be found in \cite{Huber:2020vup, Huber:2023qse}.
	         As a part of inputs, the experimental data is taken from BaBar 2014 measurement \cite{BaBar:2013qry}.
		
		\item{Radiative decays: $B\to X_s\gamma, B\to K^0 \gamma, B^+\to K^{\ast+} \gamma$, }		$B\to\phi\gamma$
		\\
		This class of decays, related to  $ b\to s\gamma $,  give stringent constraint on penguin box diagram and hence $ \Delta C_{7}^{(\prime)} $. For the inclusive radiative decay $ B\to X_s\gamma $, we follow \cite{Mahmoudi:2008tp,Misiak:2006zs,Misiak:2006ab} by using matrices $ K_{ij}^{(1,2)} $ from \textit{flavio} \cite{Straub:2018kue} related to $ P(E_0) $  as well as formula $ C $ from \cite{Gambino:2013rza},
		\begin{equation}
			\begin{aligned}
				\mathcal{B}(\bar{B}\to X_s\gamma)\!=\!\mathcal{B}\left(\bar{B}\to X_c e\bar{\nu}\right)_{\text{exp}}\Big|\frac{V_{ts}^\ast V_{tb}}{V_{cb}}\Big|^2\frac{6\alpha_e}{\pi C}\left[P(E_0;C_{7,8}) + N(E_0;C_{7,8}) +\epsilon_{\text{EM}}(C_{7,8})\right],
			\end{aligned}
		\end{equation}
		where the $N(E_0;C_{7,8})$ represents non-perturbative correction as well as $\epsilon_{\text{EM}}(C_{7,8})$ is the electromagnetic correction. Both of their formulae can be found in \cite{Mahmoudi:2008tp}.
		As for $ B\to V \gamma  $ process, the simplified formula \cite{Paul:2016urs} is adopted,
		\begin{equation}
			\begin{aligned}
				\mathcal{B}\left(B_q\to V\gamma\right)=\tau_{B_q}\frac{\alpha_eG_F^2m^3_{B_q}m_b^2}{32\pi^4}\left(1-\frac{m_V^2}{m_B^2}\right)^3|\lambda_t|^2\left(|C_7^{\text{eff}}|^2+|C_7^{\prime\text{eff}}|^2\right)T_1(0),
			\end{aligned}
		\end{equation} 
		where $ T_1(q^2=0) $ can be found in Table \ref{tab:FFs}. 
		Especially, 3 observable $S_{\phi\gamma},~A_{\text{CP}},~A_{\Delta \Gamma}$ in $B\to\phi\gamma$ process are
		also incorporated. Their expressions can be found in \cite{Muheim:2008vu}. 
		The experimental \footnote{ 
		A cut on photon energy $ E_0 $ is extrapolated from 1.9 GeV to 1.6 GeV in practice.} 
		inclusive results are taken from Belle 2014 \cite{Belle:2014nmp} while the exclusive ones are originated from Belle (2014 and 2021) as well as LHCb 2019 \cite{Belle:2014sac,BelleII:2021tzi,LHCb:2019vks} measurements.
				
		\item{$\Lambda_b \to \Lambda \mu^+\mu^-$} \\
		As the $ b\to s\ell\ell $ related bottomed baryon decay, $\Lambda_b \to \Lambda \mu^+\mu^-$,  
		shares some common features  with $ B\to V\ell\ell $. 
		In low-$q^2$ bins, deviations from SM predictions have been found.
		
		Following  \cite{Boer:2014kda,Detmold:2016pkz} , the differential width as well as involved FFs  is given as
		%
		\begin{equation}
			\begin{aligned}
				\frac{d\Gamma^\ell}{dq^2}\equiv2J_{1ss}^\ell\left(C_{7,9,10}^{(\prime)}\right)+J_{1cc}^\ell\left(C_{7,9,10}^{(\prime)}\right).
			\end{aligned}
		\end{equation}
		More details can be referred to Appendix (\ref{app: LBL}). 
		For the experimental observed branching fraction,  we refer to  the LHCb 2015 measurement \cite{LHCb:2015tgy}.
		
	\end{enumerate}
	
	In both cases before and 
	after $R_{K^{(*)}}$ 2022 release, there are more than 
	200 observables related to the above processes, including the binned ones. 
	 A detailed summary of experimental data of these observables are listed in 
	Appendix \ref{app:exp_input}.

	\subsection{Fitting Schemes}
	\label{sub:fit_sch}
	
	In our following fitting work, 
	Bayesian statistics is adopted, based on which  
	some early analysis on B-anomalies  \cite{Ciuchini:2022wbq,Ciuchini:2020gvn,Ciuchini:2019usw,Kowalska:2019ley}
	is also performed.
	The  advantages of carrying out a Bayesian estimation are its robustness and extensibility. For robustness, firstly, Bayesian inference with posterior functions has the advantage of avoiding the danger of insufficient coverage probability compared to the traditional \textit{Profile} method that used to derive confidence intervals. Secondly, more attention can be paid to the distribution of parameters, namely overall effects, rather than to an individual minimum. 
	
	We denote the posterior function according to Bayesian theorem,
	\begin{equation}
		\begin{aligned}
			\mathcal{P}(\vec{\theta}|\mathcal{O}_{\text{expt.}})\propto \mathcal{L}(\mathcal{O}|\vec{\theta})\pi(\vec{\theta}),
		\end{aligned}
	\end{equation}
	where $ \mathcal{L}(\mathcal{O}|\vec{\theta}) $ and $\pi(\vec{\theta})$ stand for Likelihood function and prior we set, respectively. In our model-independent fit, a Negative Log Likelihood (NLL) function is defined as 
	\begin{equation}
		\begin{aligned}
			-2\log\mathcal{L}(\mathcal{O}|\vec{\theta})&=\chi^2(\vec{\theta})\\
			&=(\mathcal{O}_{\text{theo.}}(\vec{\theta})-\mathcal{O}_{\text{expt.}})^\top(V_{\text{expt.}}+V_{\text{theo.}})^{-1}(\mathcal{O}_{\text{theo.}}(\vec{\theta})-\mathcal{O}_{\text{expt.}}),
		\end{aligned}
	\end{equation}
	where $ \mathcal{O}_{\text{theo.}} $ as well as $ \mathcal{O}_{\text{expt.}} $ represent the theoretical predictions of various observables and their corresponding experimental data. 
	The covariance matrices $ V_{\text{theo.}} $ and $ V_{\text{expt.}} $ are consist of 
	theoretical and experimental errors of observables. 
	
	For the experimental correlation matrix, we have taken into account some of the correlations among relevant experiments \cite{LHCb:2022zom,CMS:2022mgd,LHCb:2021awg,LHCb:2019vks,LHCb:2020dof,ATLAS:2018gqc,LHCb:2020gog}, while the error is aligned to the 
	bigger one in the asymmetrical error case.
	Theoretical covariance matrix is formed by assuming a multivariate gaussian distribution of input parameters which would mainly occupy the pie chart of error (form factors error for example). Both matrices are N dimensional, where N is the number of observables up to 203. The parameter matrix $ \vec\theta $ shown above, 
	is  encoded  various WCs,  
	\begin{equation}
		\begin{aligned}
			\vec{\theta} = (\Delta C_{7},\Delta C_{7}',\Delta C_{8},\Delta C_{8}',\Delta C_{9}^\ell,\Delta C_{9}^{\prime\ell},\Delta C_{10}^{\ell},\Delta C_{10}^{\prime\ell},\Delta C_{S}^{\ell},\Delta C_{S}^{\prime\ell},\Delta C_{P}^{\ell},\Delta C_{P}^{\prime\ell}),
		\end{aligned}
	\end{equation}
	where $ \ell = e \text{ or }\mu$, and the dimension can be as large as  20 in some of 
	the working scenarios.
	
	The likelihood function tells us where we are heading, while the prior distribution $\pi(\vec{\theta})$ tells us where to start. The prior function usually implies the extent of our knowledge about the problem we are facing. Namely, in these fits, it represents the more probable starting position (or the coordinate of WCs) as well as their ranges. In our analysis,  the best-fit point obtained from HMMN\cite{Hurth:2022lnw}  is taken as our prior knowledge. 
So a prior of multidimensional Gaussian distribution which is centered at the latest 20-D fit result from HMMN\cite{Hurth:2022lnw}  with
a common standard deviation of $\sqrt{2}$ is utilized.

\section{Numerical Analysis}
\label{sec: num}	
\subsection{Input Parameters}
\label{sub: num_input}
The global fitting analysis, relying on $\chi^2$ function, contains both theoretical and experimental inputs. 
In Appendix \ref{app:Theo}, we present the necessary theoretical formulas for various observables in related decay processes.
For the WCs, we adopted the obtained results from \cite{Blake:2016olu}, which have been calculated at $\mu_b$ scale with two-loop RGE running.
Other basic parameters (masses, lifetimes, Wolfenstein parameters in CKM matrix, decay constants, Weinberg angles, etc.)
and some non-perturbative parameters  (Gegenbaur  expansion coefficients in distribution amplitudes (DA),  FFs, etc.)  have been 
summarized in Table  \ref{tab:Input_para}  and \ref{tab:FFs}, respectively.
As another part of inputs, the experimental data of various observables shown bin by bin, have been presented in Appendix \ref{app:exp_input},
together with their corresponding calculated SM predictions.

\begin{table}[!ht]
\centering
\caption{\label{tab:Input_para}
	Input parameters I: some basic parameters in the numerical analysis. 	}
\renewcommand\arraystretch{1.1}
\begin{tabular}{cccc}
	\hline\hline
	Parameters& Values& Parameters & Values \\
	\hline
	$ m_{b} $& 4.18( $\!^{+3}_{-2} $)~GeV \cite{ParticleDataGroup:2022pth}& $ m_{t} $ & 173.50(30)~GeV\cite{ParticleDataGroup:2022pth} \\
	$ m_{c} $&1.27$ (2) $~GeV\cite{ParticleDataGroup:2022pth}&$ m_{s} $ &93$\left(^{+11}_{-5}\right)$~MeV\cite{ParticleDataGroup:2022pth}\\
	$ m_{d} $&4.67$ (^{+48}_{-17}) $~MeV\cite{ParticleDataGroup:2022pth}&$ m_{u} $&2.16$( ^{+49}_{-26} )$~MeV\cite{ParticleDataGroup:2022pth}\\
	$ m_{e} $&0.5109989461$ (31) $~MeV\cite{ParticleDataGroup:2022pth}&$ m_{\mu} $&105.6583745$ (24) $~MeV\cite{ParticleDataGroup:2022pth}\\
	$ m_{\tau} $&1776.86(12)~MeV\cite{ParticleDataGroup:2022pth}&$ m_{B_s} $&5366.92(10)~MeV\cite{ParticleDataGroup:2022pth}\\
	$ m_{B_d} $&5279.65(12)~MeV\cite{ParticleDataGroup:2022pth}&$ m_\phi $&1019.461(16)~MeV\cite{ParticleDataGroup:2022pth}\\
	$ m_{K^\pm} $&493.677(16)~MeV\cite{ParticleDataGroup:2022pth}&$ m_{K^0} $&497.611(13)~MeV\cite{ParticleDataGroup:2022pth}\\

	$ m_{K^{\ast\pm}} $&891.67(26)~MeV\cite{ParticleDataGroup:2022pth}&$ m_{K^{\ast0}} $&895.55(20)~MeV\cite{ParticleDataGroup:2022pth}\\
	$ m_{B_u} $&5279.34(12)~MeV\cite{ParticleDataGroup:2022pth}&$ m_{\Lambda} $&1115.683(6)~MeV\cite{ParticleDataGroup:2022pth}\\
	$ m_{\Lambda_b} $&5619.60(17)~MeV\cite{ParticleDataGroup:2022pth}&$ \tau_{B_u} $&1.638(4)~ps\cite{ParticleDataGroup:2022pth}\\
	
	$ \tau_{B_s} $&1.520(5)~ps\cite{ParticleDataGroup:2022pth}&$ \tau_{B_{d}} $&1.519$\left(4\right)$~ps\cite{ParticleDataGroup:2022pth}\\
	$ \tau_{\Lambda_b} $&1.471(9)~ps\cite{ParticleDataGroup:2022pth}&&\\
	$ f_{B_s} $&227.7(4.5)~MeV\cite{Bobeth:2013uxa}&$ f_{B_{d}} $&190.5(4.2)~MeV\cite{Bobeth:2013uxa}\\
	$ f_{\Lambda} $&6.0(4)$\times10^{-3}$~GeV$^2$\cite{Aslam:2008hp}&$ f_{\Lambda_b} $&$ 3.9(^{+4}_{-2})\times10^{-3} $~GeV$^2$\cite{Aslam:2008hp}\\
	$ m_t(m_t) $&163.53(83)~GeV\cite{ParticleDataGroup:2022pth}&$ G_F $&1.1663787(6)~GeV$ ^{-2} $\cite{ParticleDataGroup:2022pth}\\
	\hline
	$ \alpha_\Lambda $&0.642(13)\cite{Detmold:2016pkz}&$\alpha_{\text{e}}(m_{Z})$&1/127.944(14)\cite{ParticleDataGroup:2022pth}\\
	$\alpha_s(m_Z)$&0.1179(9)\cite{ParticleDataGroup:2022pth}&$ \sin^2\theta_W $&0.23121(4)\cite{ParticleDataGroup:2022pth}\\
	$y_s$&0.064(4)\cite{ParticleDataGroup:2022pth}&$y_d$&0.0005(50)\cite{ParticleDataGroup:2022pth}\\
	$ \mu^2_{G} $&0.336$\pm0.064 $\cite{Gambino:2013rza}&&\\
	$ \rho_D^3 $&$ 0.153\pm0.45 $\cite{Gambino:2013rza}&$ \rho_{LS}^3 $&$ -0.145\pm0.098 $\cite{Gambino:2013rza}\\
	$ \mathcal{B}(B\to X_ce\bar{\nu})_{\text{exp}} $&$ 0.997(41)$\cite{Belle-II:2021jlu}&$ \mathcal{B}(B\to X_c\ell\bar{\nu})_{\text{exp}} $&$ 0.975(50) $\cite{Belle-II:2021jlu}\\
	\hline
	$ \lambda $&0.22500(67)\cite{ParticleDataGroup:2022pth}&$ A $&0.826$ (^{+18}_{-15}) $\cite{ParticleDataGroup:2022pth}\\
	$\bar{\rho}$&$ 0.159(^{+10}_{-10}) $\cite{ParticleDataGroup:2022pth}&$\bar{\eta}$&$ 0.348(^{+10}_{-10}) $\cite{ParticleDataGroup:2022pth}\\
	\hline\hline
\end{tabular}
\end{table}

\begin{table}[!ht]
	\centering
	\caption{
	Form factors as well as resonance pole mass used in the numerical analysis of $ B\to P\ell^+\ell^- $\cite{Bobeth:2011nj}, $ B\to V\ell^+\ell^-$\cite{Bharucha:2015bzk}, $B\to V\gamma$\cite{Paul:2016urs} and $ \Lambda_b\to \Lambda\ell^+\ell^- $\cite{Detmold:2016pkz}, respectively. }
	\begin{tabular}{cccc}
		\hline\hline
		Parameters& Values& Parameters & Values \\
		\hline
		$ f^i(0) $&$ 0.34^{+0.05}_{-0.02} $&$ b_1^i $&$ -2.1^{+0.9}_{-1.6} $\\
		$ m_{\text{res},+} $&5.412~GeV&&\\
		\hline
		$ a_0^{A_0}(K^\ast) $&$ +0.36\pm0.05 $&$ a_0^{A_0}(\phi) $&$ +0.39\pm0.05 $\\
		$ a_1^{A_0}(K^\ast) $&$ -1.04\pm0.27 $&$ a_1^{A_0}(\phi) $&$ -0.78\pm0.26 $\\
		$ a_2^{A_0}(K^\ast) $&$ +1.12\pm1.35 $&$ a_2^{A_0}(\phi) $&$ +2.41\pm1.48 $\\
		$ a_0^{T_1}(K^\ast) $&$ +0.28\pm0.03 $&$ a_0^{T_1}(\phi) $&$ +0.31\pm0.03 $\\
		$ a_1^{T_1}(K^\ast) $&$ -0.89\pm0.19 $&$ a_1^{T_1}(\phi) $&$ -0.87\pm0.19 $\\
		$ a_2^{T_1}(K^\ast) $&$ +1.95\pm1.10 $&$ a_2^{T_1}(\phi) $&$ +2.75\pm1.19 $\\
		$ m_{\text{res},A_0}(K^\ast) $&5.366 GeV&$ m_{\text{res},A_0}(\phi) $&5.366 GeV\\
		$ m_{\text{res},T_1}(K^\ast) $&5.415 GeV&$ m_{\text{res},T_1}(\phi) $&5.415 GeV\\
		\hline
		$T_1^\phi(0)$&$0.280^{+20}_{-22}$&$T_1^{K^\ast}(0)$&$0.316^{+16}_{-15}$\\
		\hline
		$ a_0^{f_+} $&$ +0.4221\pm0.0188 $&$ a_1^{g_0} $&$ -1.0290\pm0.1614 $\\
		$ a_1^{f_+} $&$ -1.1386\pm0.1683 $&$ a_1^{g_\perp} $&$ -1.1357\pm0.1911 $\\
		$ a_0^{f_0} $&$ +0.3725\pm0.0213 $&$ a_0^{h_+} $&$ +0.4960\pm0.0258 $\\
		$ a_1^{f_0} $&$ -0.9389\pm0.2250 $&$ a_1^{h_+} $&$ -1.1275\pm0.2537 $\\
		$ a_0^{f_\perp} $&$ +0.5182\pm0.0251 $&$ a_0^{h_\perp} $&$ +0.3876\pm0.0172 $\\	
		$ a_1^{f_\perp} $&$ -1.3495\pm0.2413 $&$ a_1^{h_\perp} $&$ -0.9623\pm0.1550 $\\
		$ a_0^{g_\perp,g_+} $&$ 0.3563\pm0.0142 $&$ a_0^{\tilde{h}_\perp,\tilde{h}_+} $&$ +0.3403\pm0.0133 $\\
		$ a_1^{g_+} $&$ -1.0612\pm0.1678 $&$ a_1^{\tilde{h}_+} $&$ -0.7697\pm0.1612 $\\
		$ a_0^{g_0} $&$ 0.4028\pm0.0182 $&$ a_1^{\tilde{h}_\perp} $&$ -0.8008\pm0.1537 $\\
		$ m_{\text{res},(f_+;f_\perp;h_+;h_\perp)} $&5.416 GeV&$ m_{\text{res},(g_+;g_\perp;\tilde{h}_+;\tilde{h}_\perp)}$&5.750 GeV\\
		$ m_{\text{res},(f_0)} $&5.711 GeV&$ m_{\text{res},(g_0)}$&5.367 GeV\\
		\hline\hline
	\end{tabular}
	\label{tab:FFs}
\end{table}
In order to depict the goodness of fit of different scenarios, we adopt a method that is often utilized by frequentists: performing a reduced chi-squared $\chi^2/\text{d.o.f.}$. We first make a histogram of samples from the negative log-likelihood (NLL) distribution. The numerator of the reduced chi-squared is then the $\chi^2$ corresponding to the maximum density of the NLL distribution. This is in contrast to the traditional frequentist approach, which only considers the minimum chi-squared.

To estimate the parameters $\vec{\theta}$, we adopt the median as our estimation of the center value of the parameter for its robustness. We then use the 16th and 84th percentiles to indicate the boundary of our one standard deviation confidence interval. This region has the same coverage probability as the standard normal distribution.

\subsection{Numerical Results}
	\label{sub: num_res}
\begin{figure}[htp]
		\centering
		\includegraphics[width=0.23\linewidth]{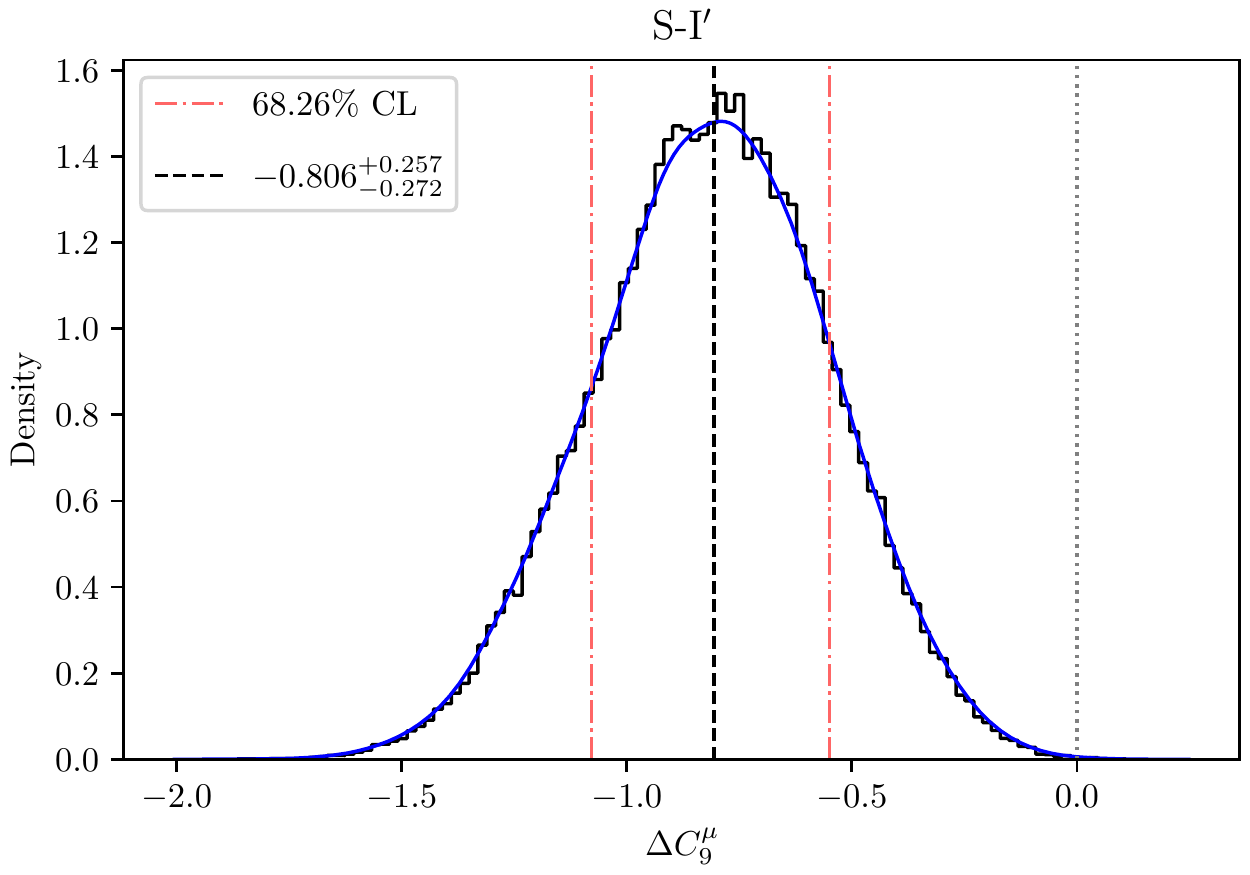}
		\includegraphics[width=0.23\linewidth]{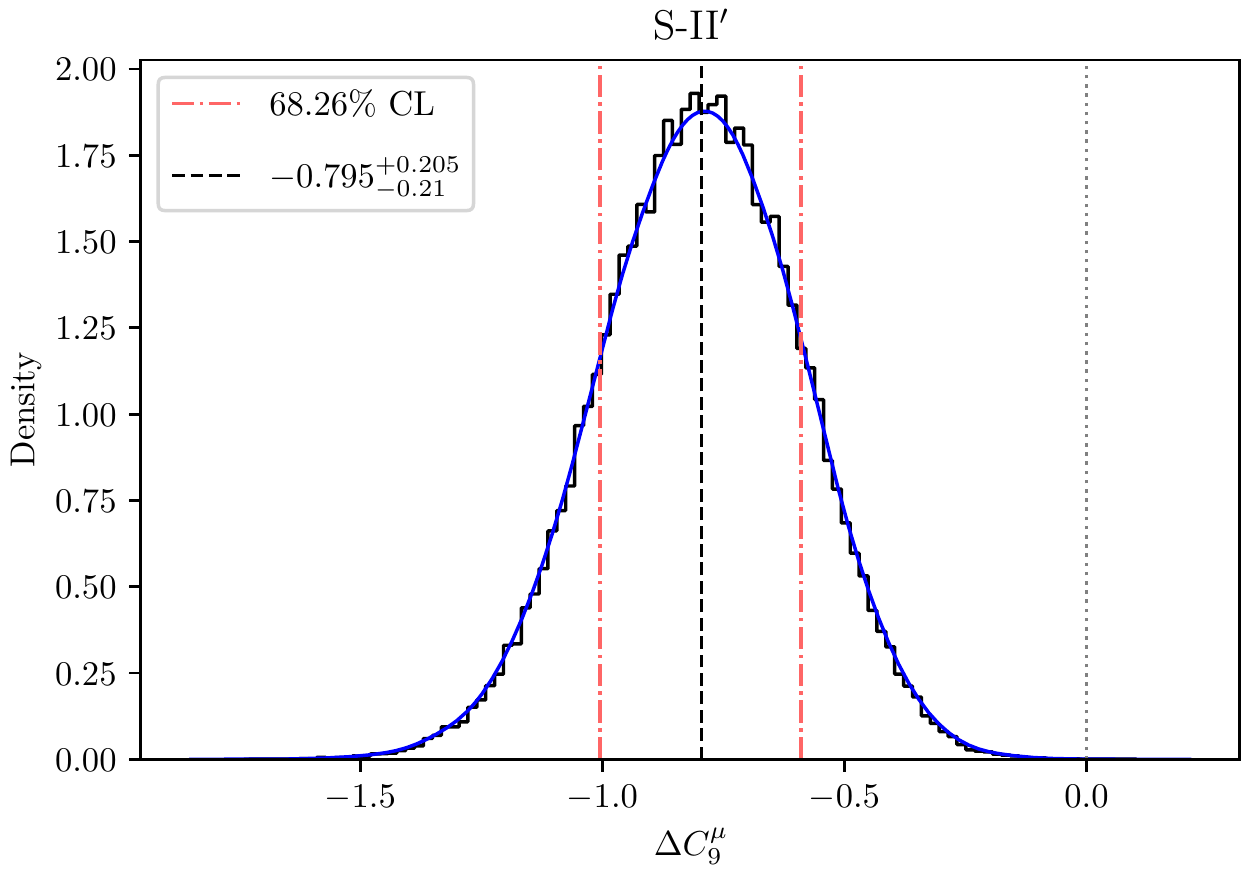}
		\includegraphics[width=0.23\linewidth]{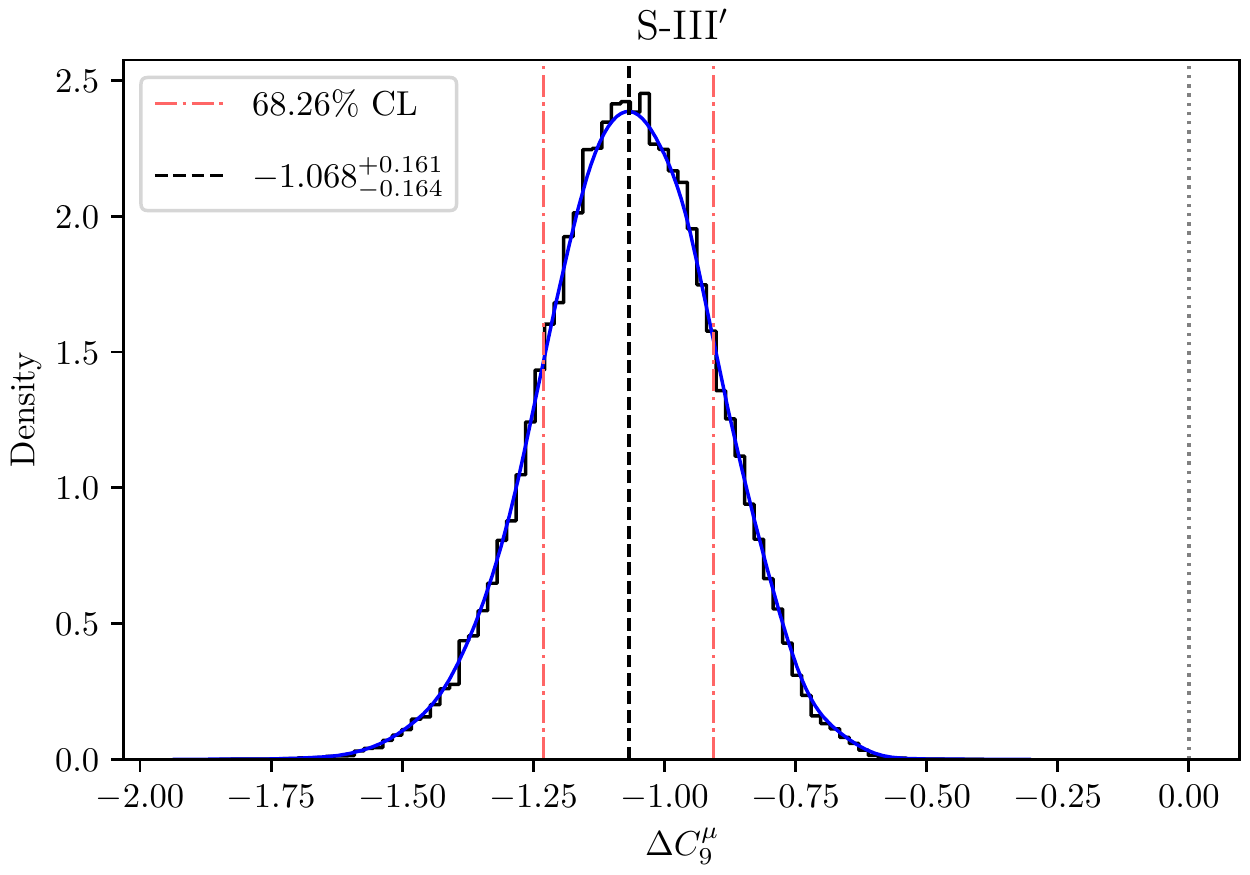}
		\includegraphics[width=0.23\linewidth]{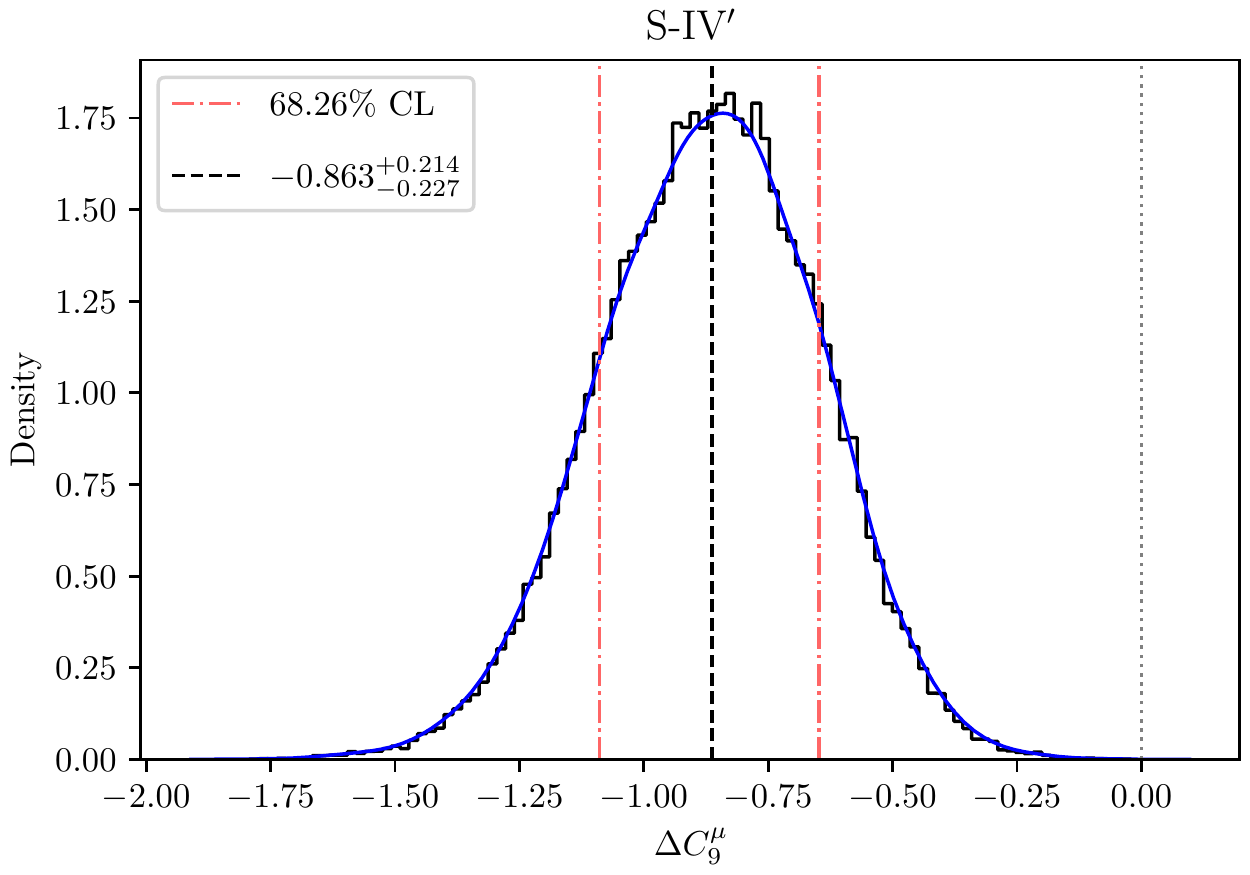}
		\\
		\includegraphics[width=0.23\linewidth]{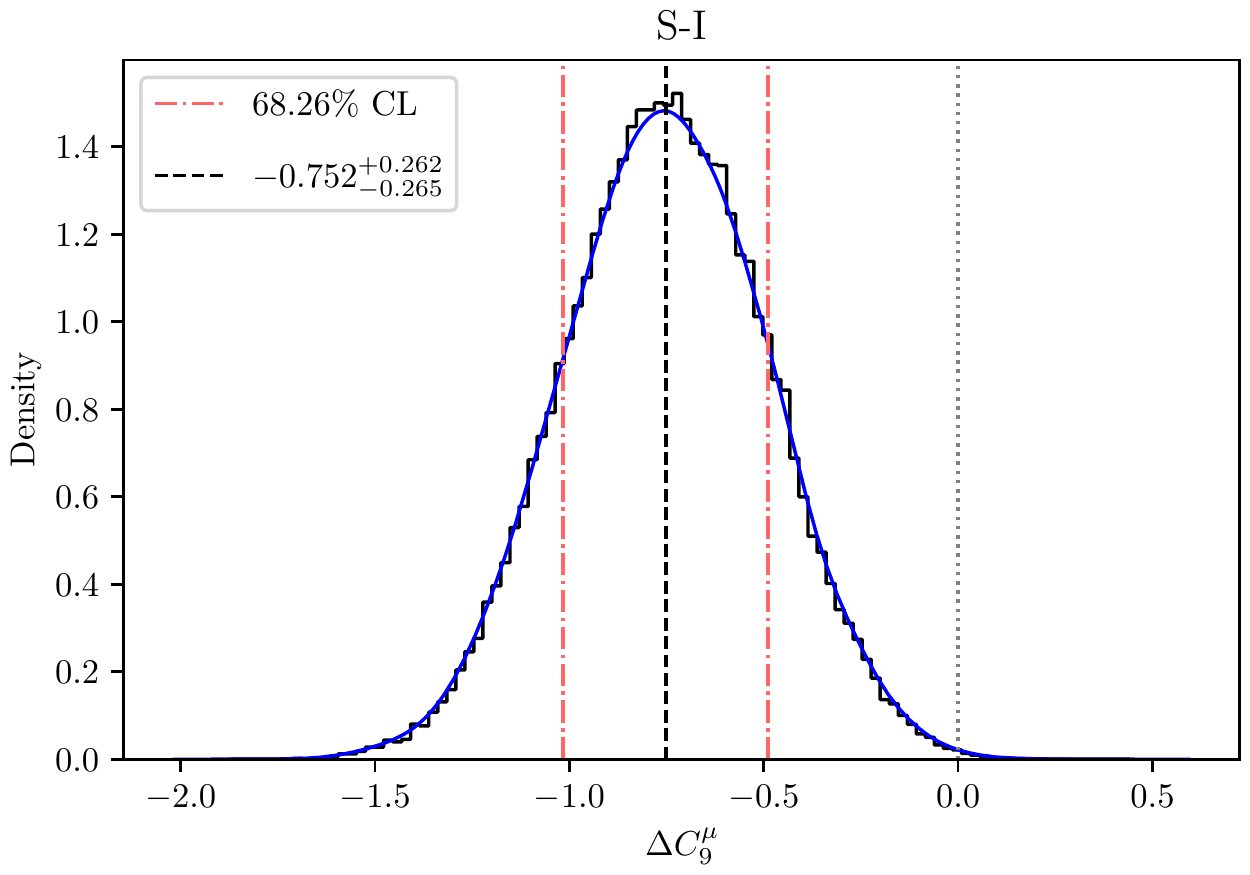}
		\includegraphics[width=0.23\linewidth]{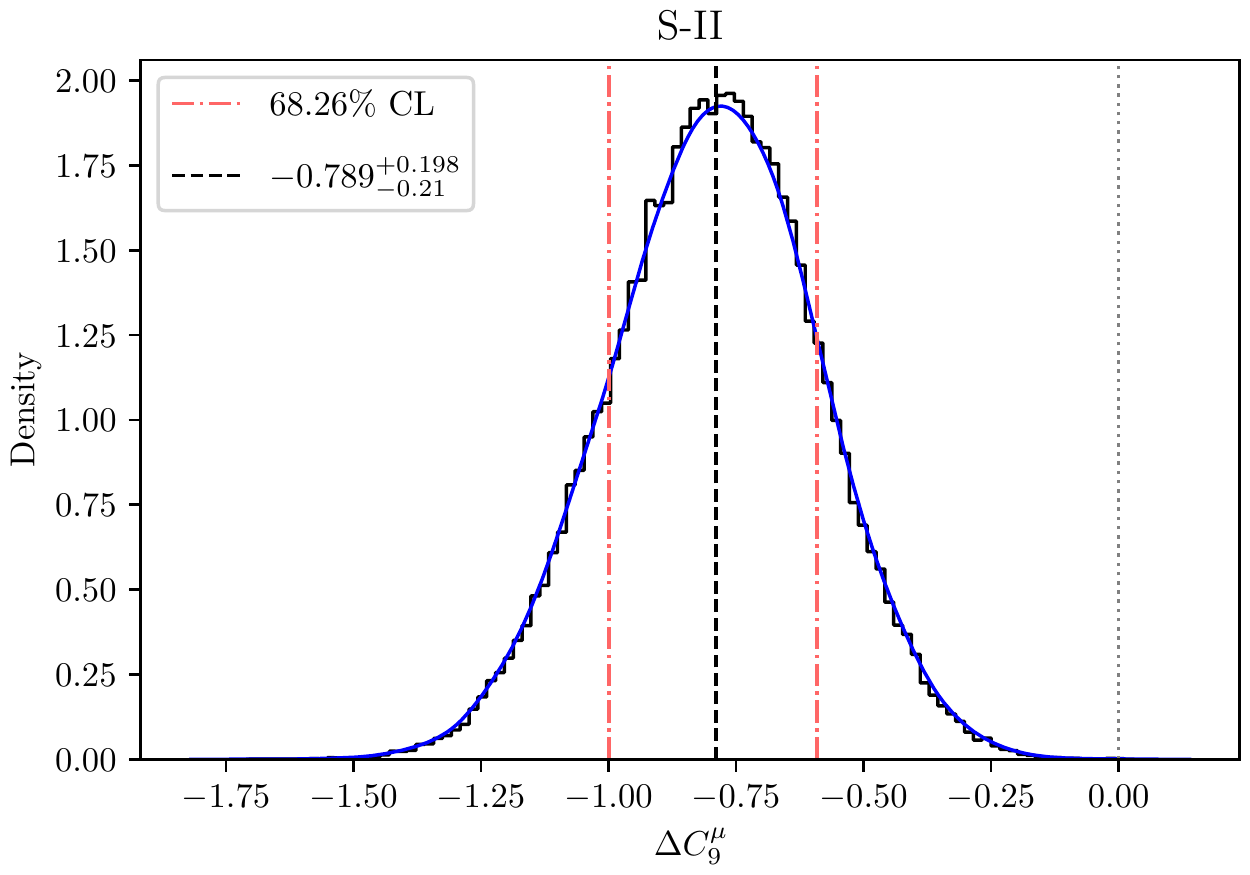}
		\includegraphics[width=0.23\linewidth]{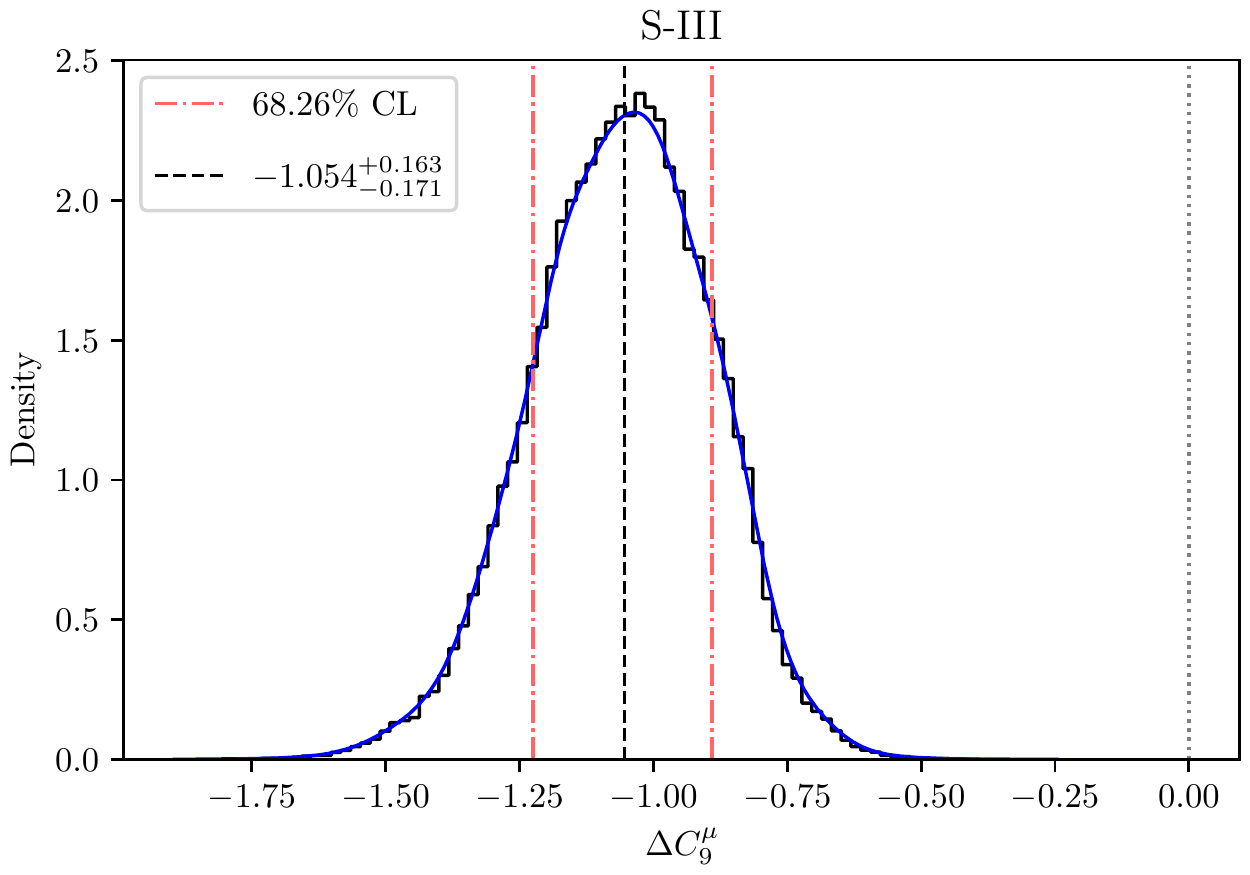}
		\includegraphics[width=0.23\linewidth]{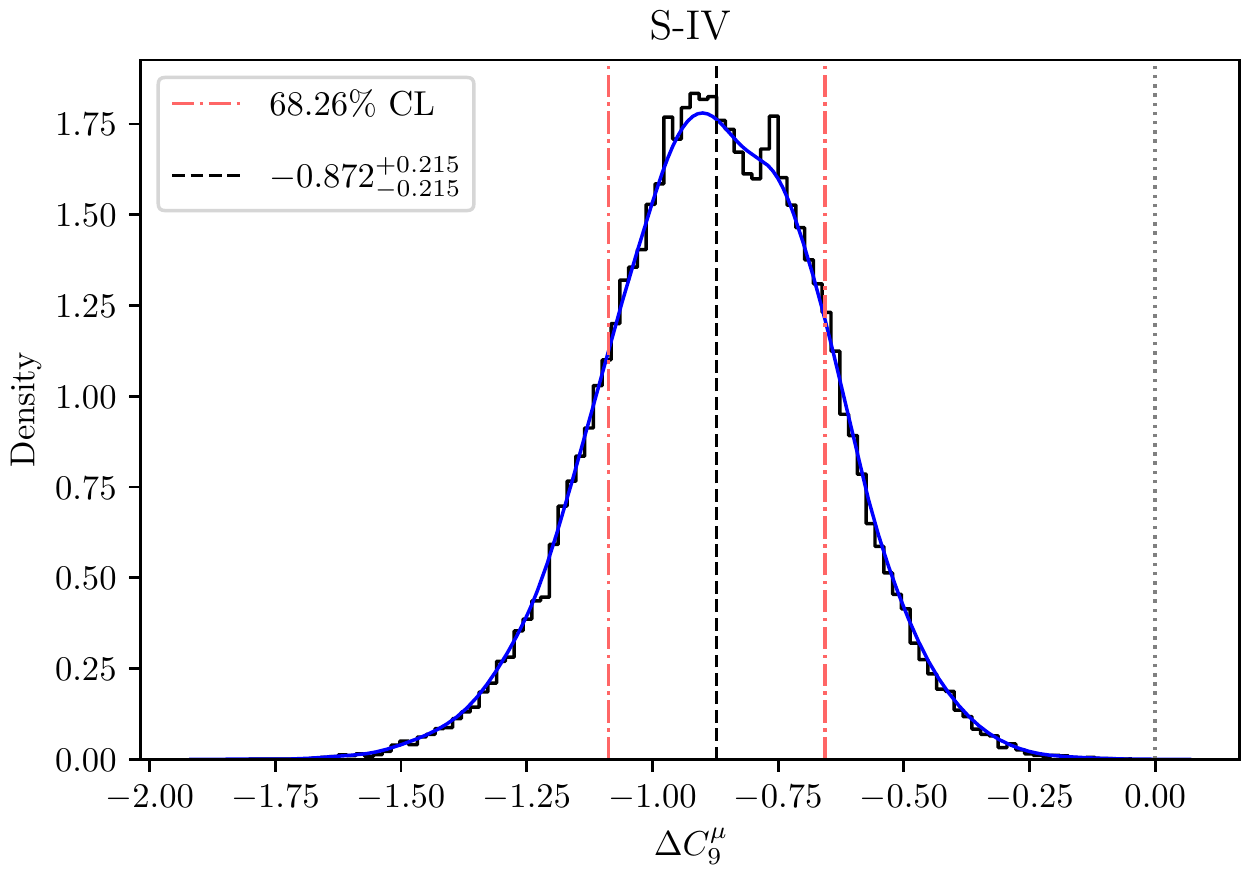}
		\caption{Density(black steps) and Kernel Density Estimation KDE (blue curves) of Wilson coefficients $\Delta C_9^\mu$ in different scenarios varied from old Dataset \textbf{A} to \textbf{B}. Red(dotdashed) lines indicate the Highest Posterior Density (HPD) about 68.26\% which give the error estimations. Black(dashed) lines refer to the Bayesian estimations of $\Delta C_9^\mu$ while gray(dotted) lines refer to SM positions.
		}\label{fig0.1}
	\end{figure}	
\vspace*{-1.08cm}
\begin{figure}[htp]
	\centering
	\includegraphics[width=0.23\linewidth]{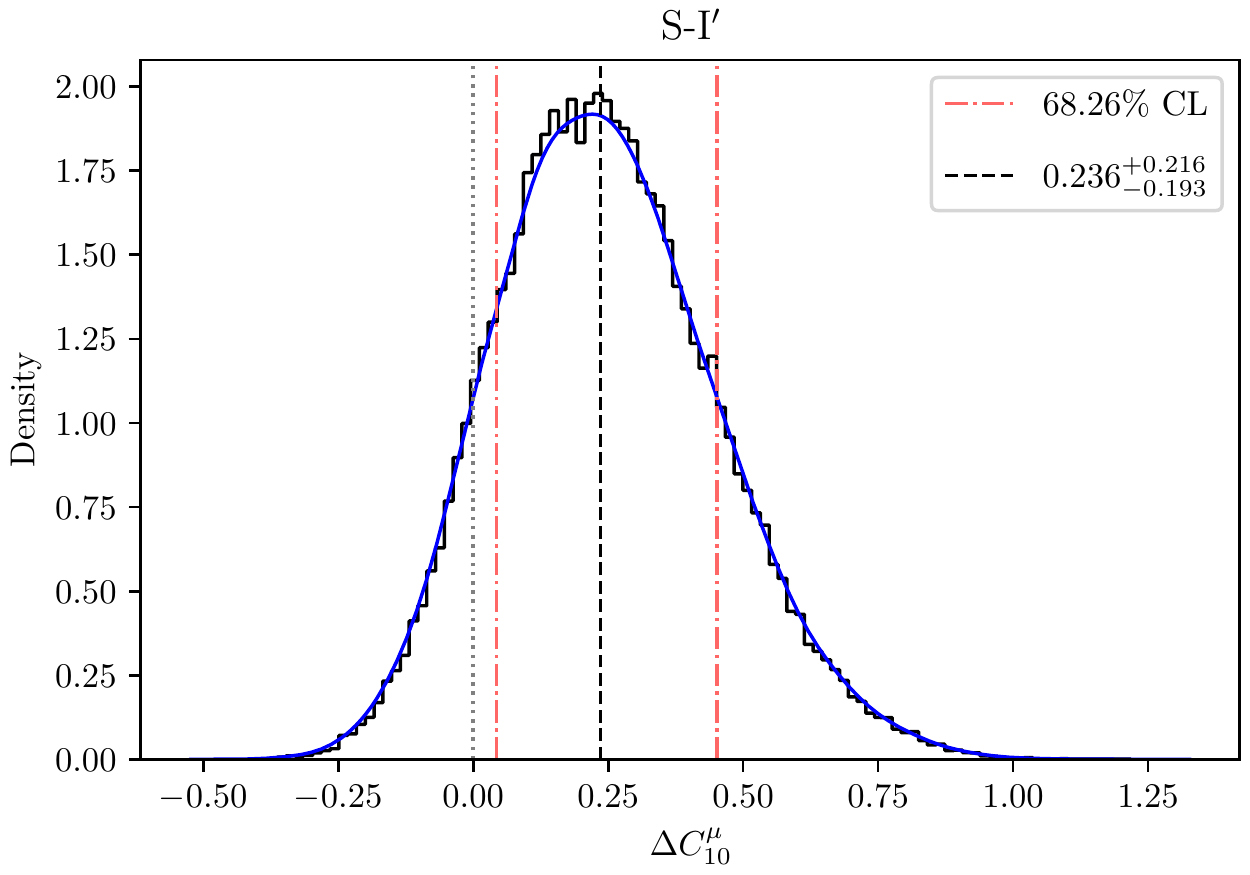}
	\includegraphics[width=0.23\linewidth]{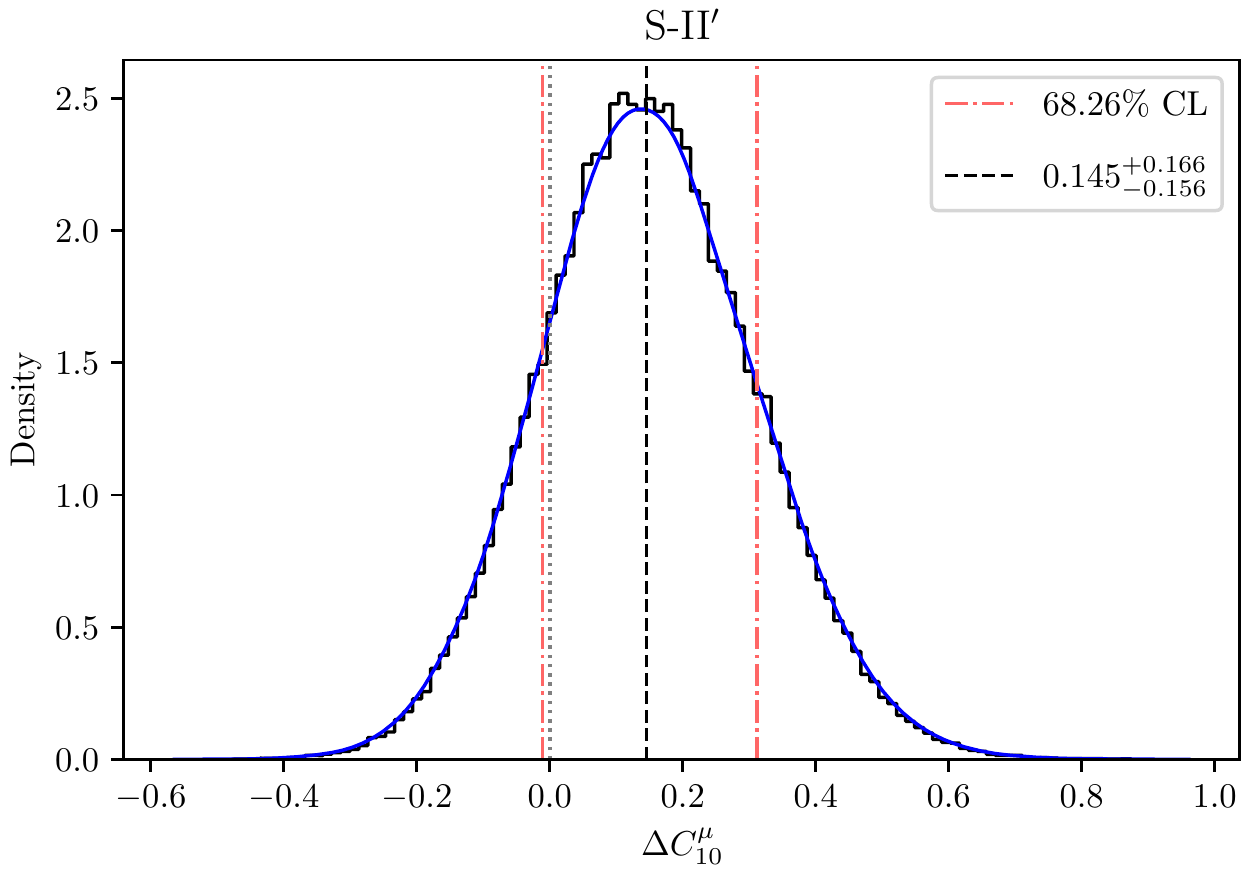}
	\includegraphics[width=0.23\linewidth]{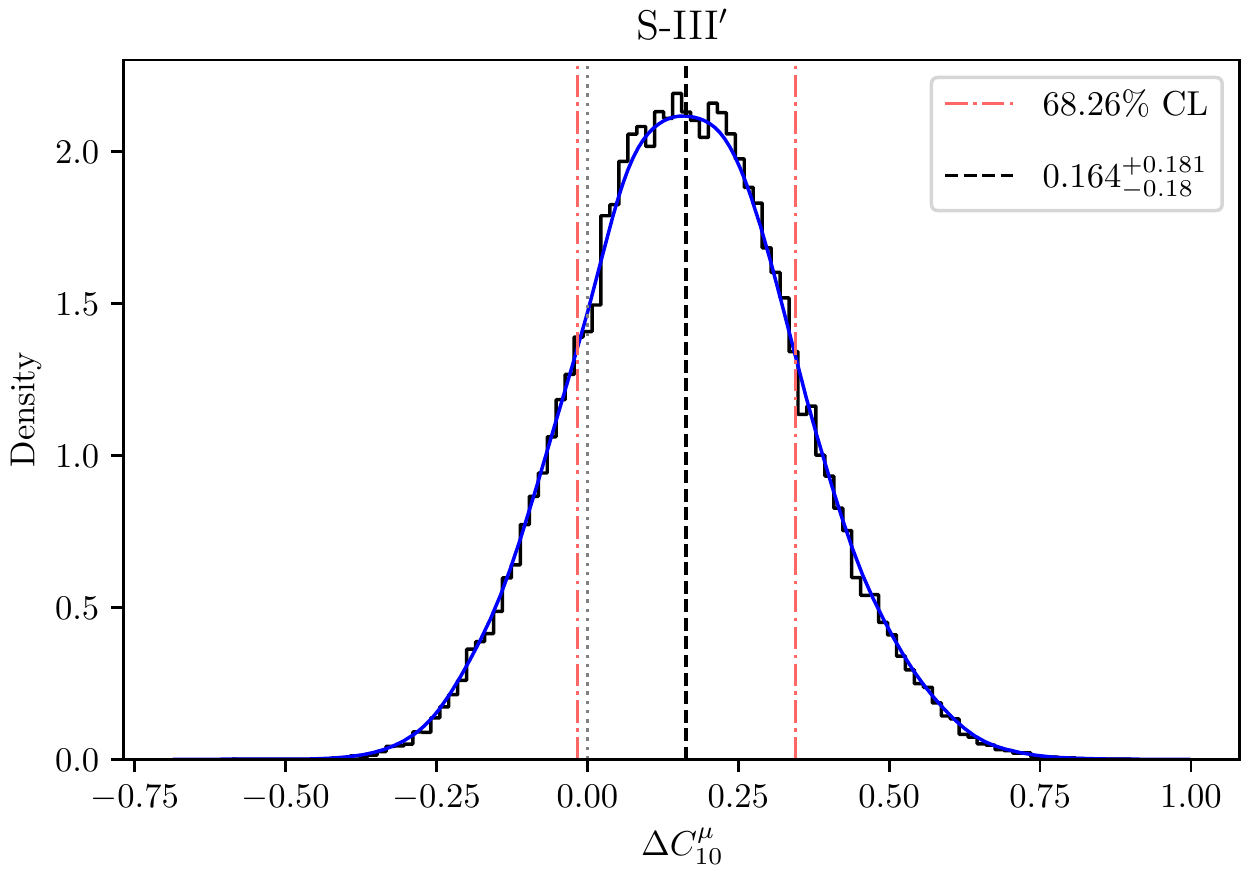}
	\includegraphics[width=0.23\linewidth]{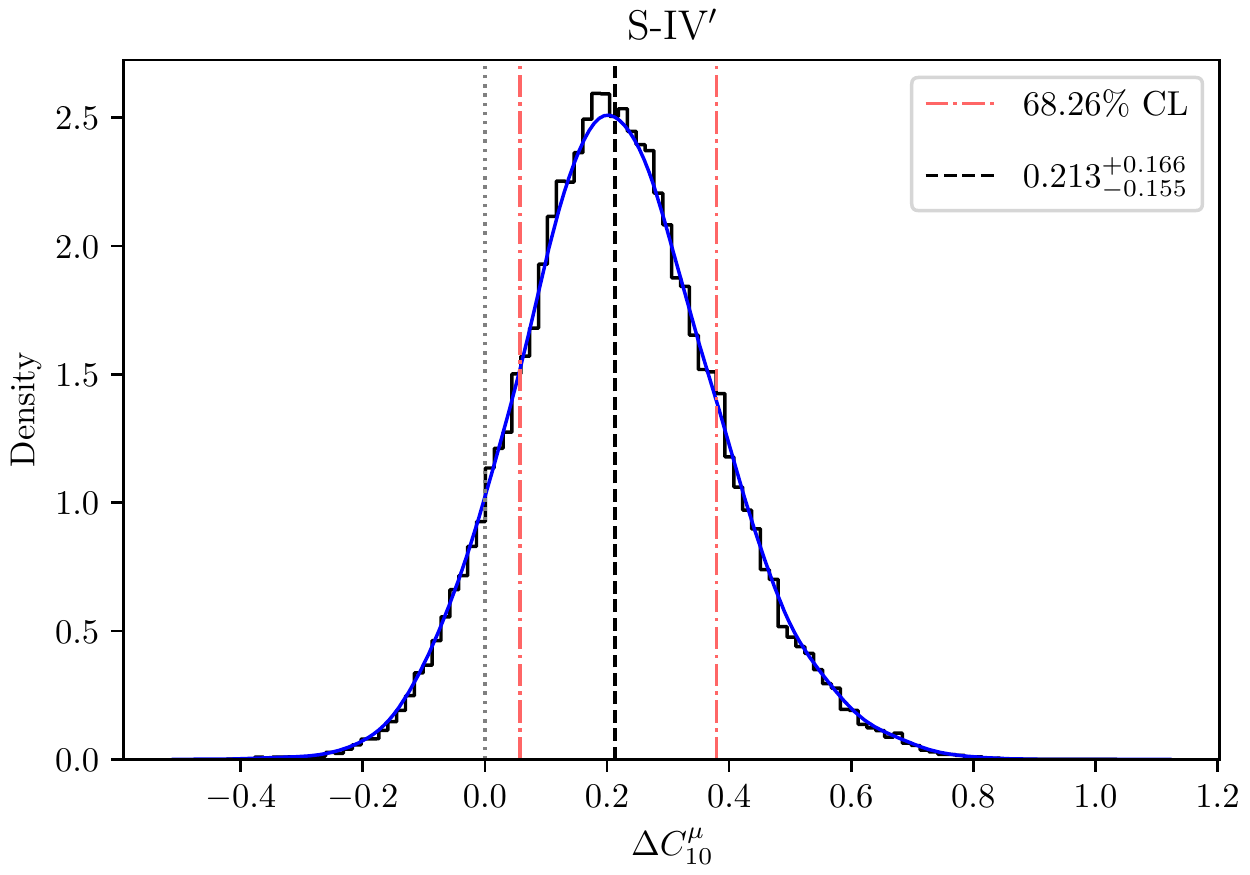}
	\\
	\includegraphics[width=0.23\linewidth]{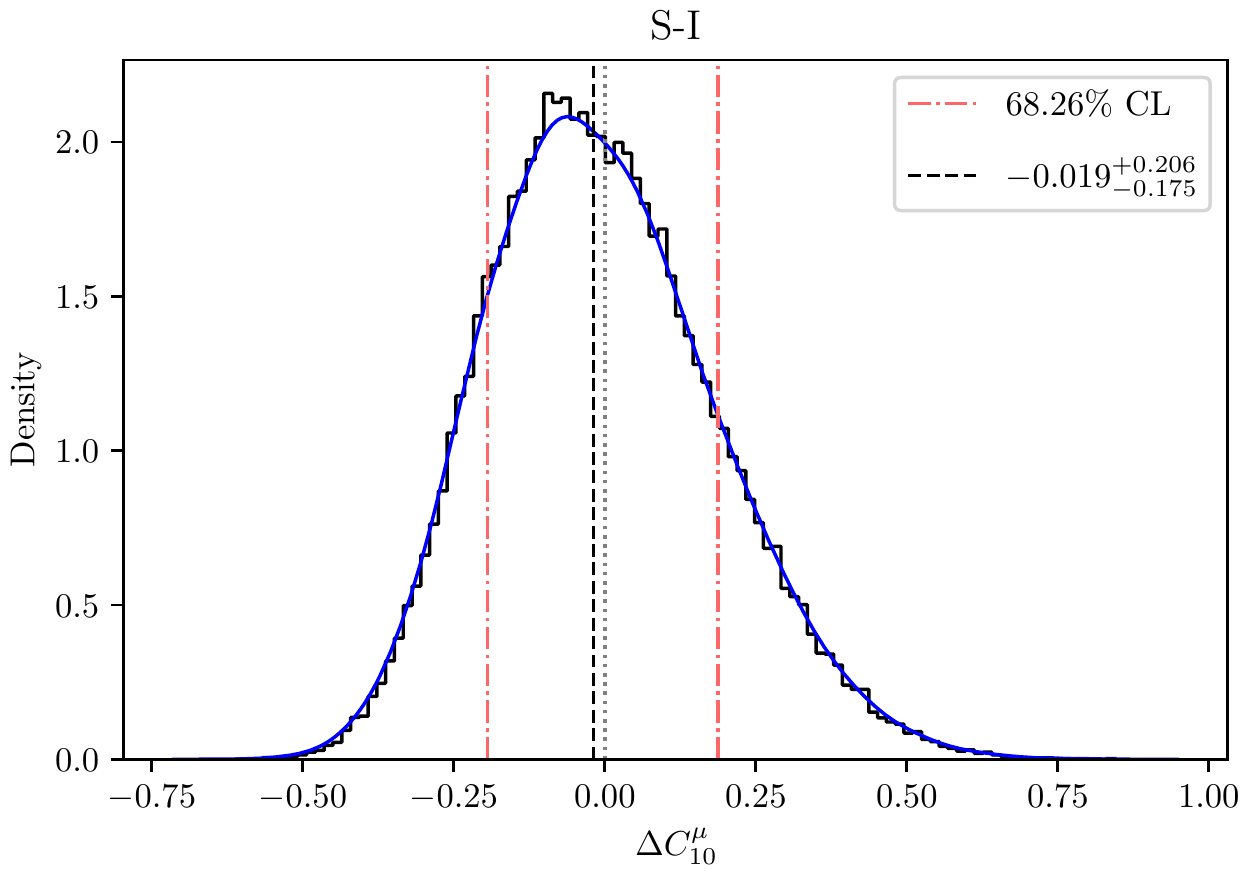}
	\includegraphics[width=0.23\linewidth]{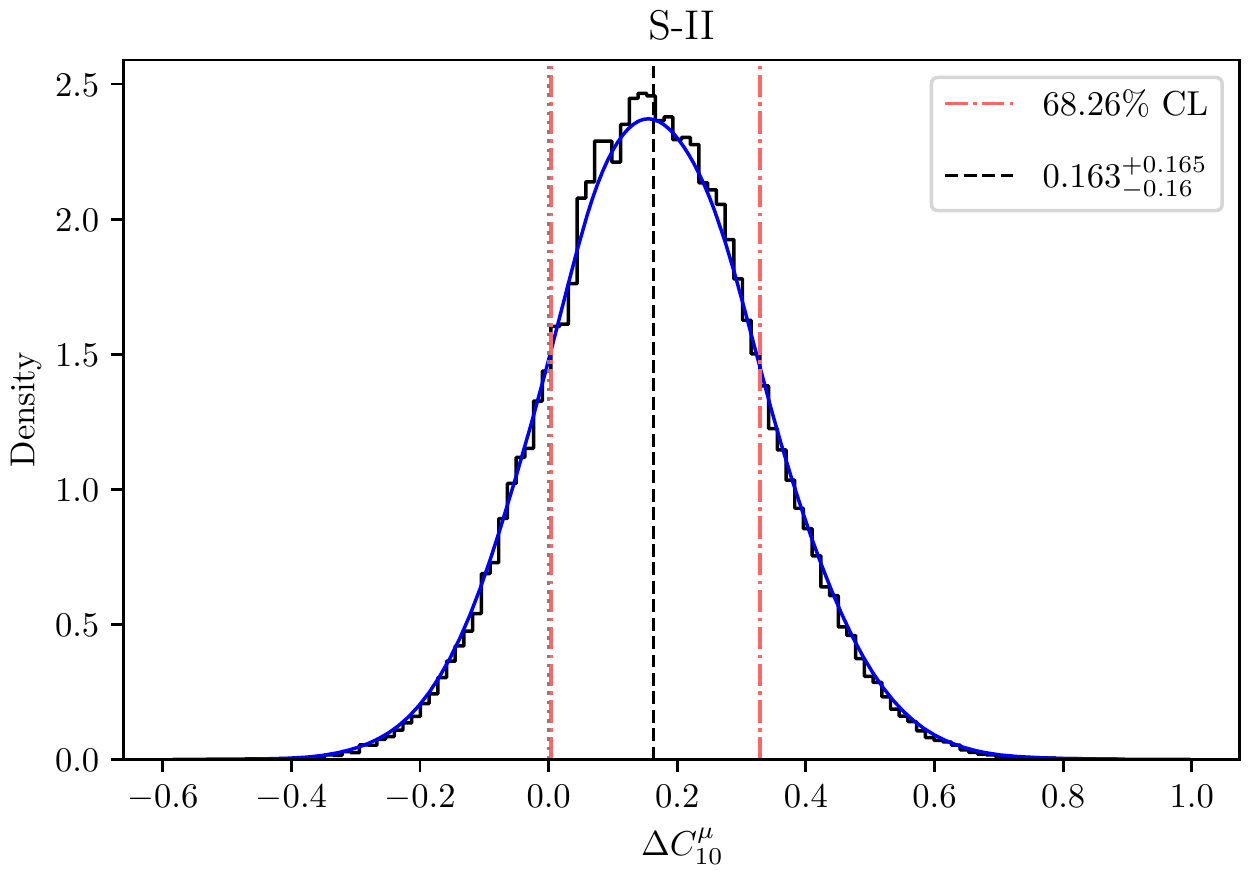}
	\includegraphics[width=0.23\linewidth]{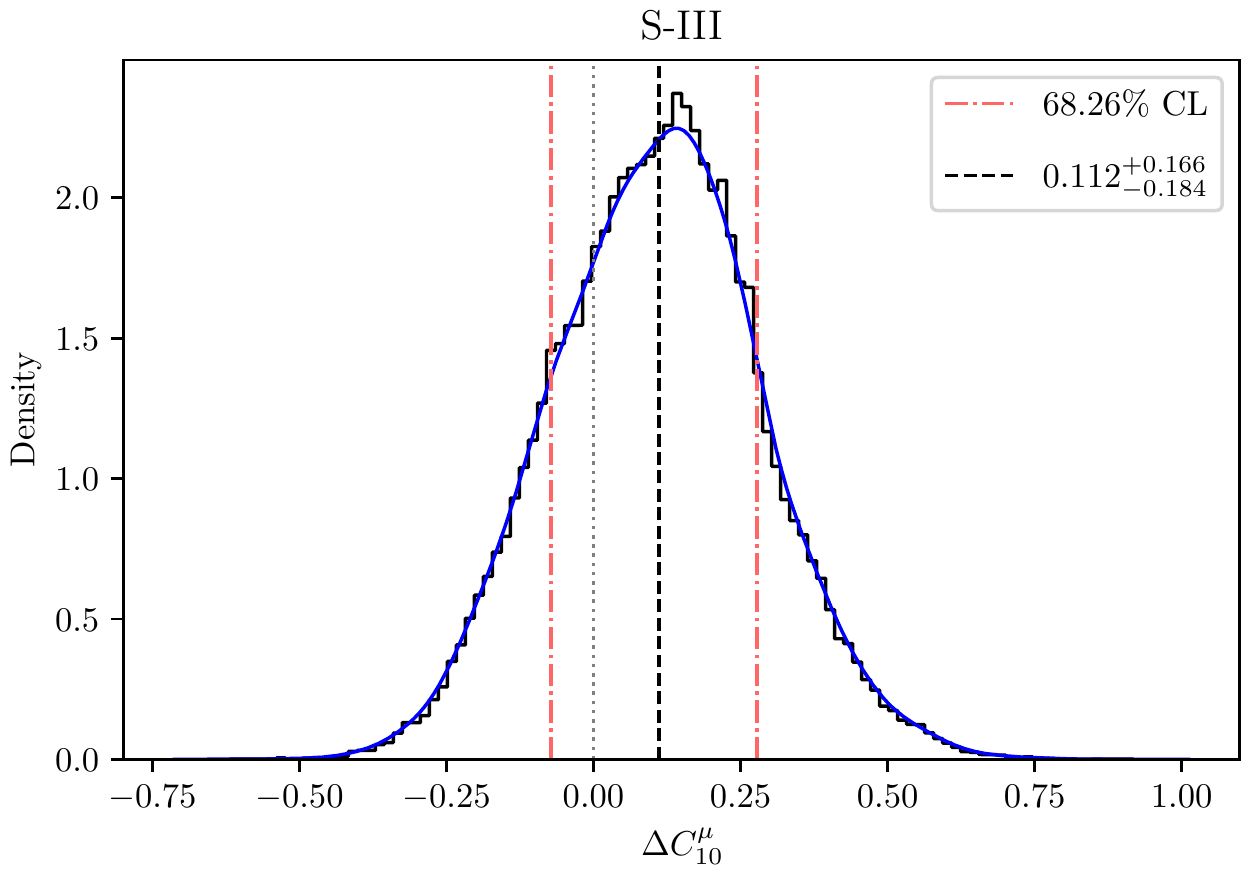}
	\includegraphics[width=0.23\linewidth]{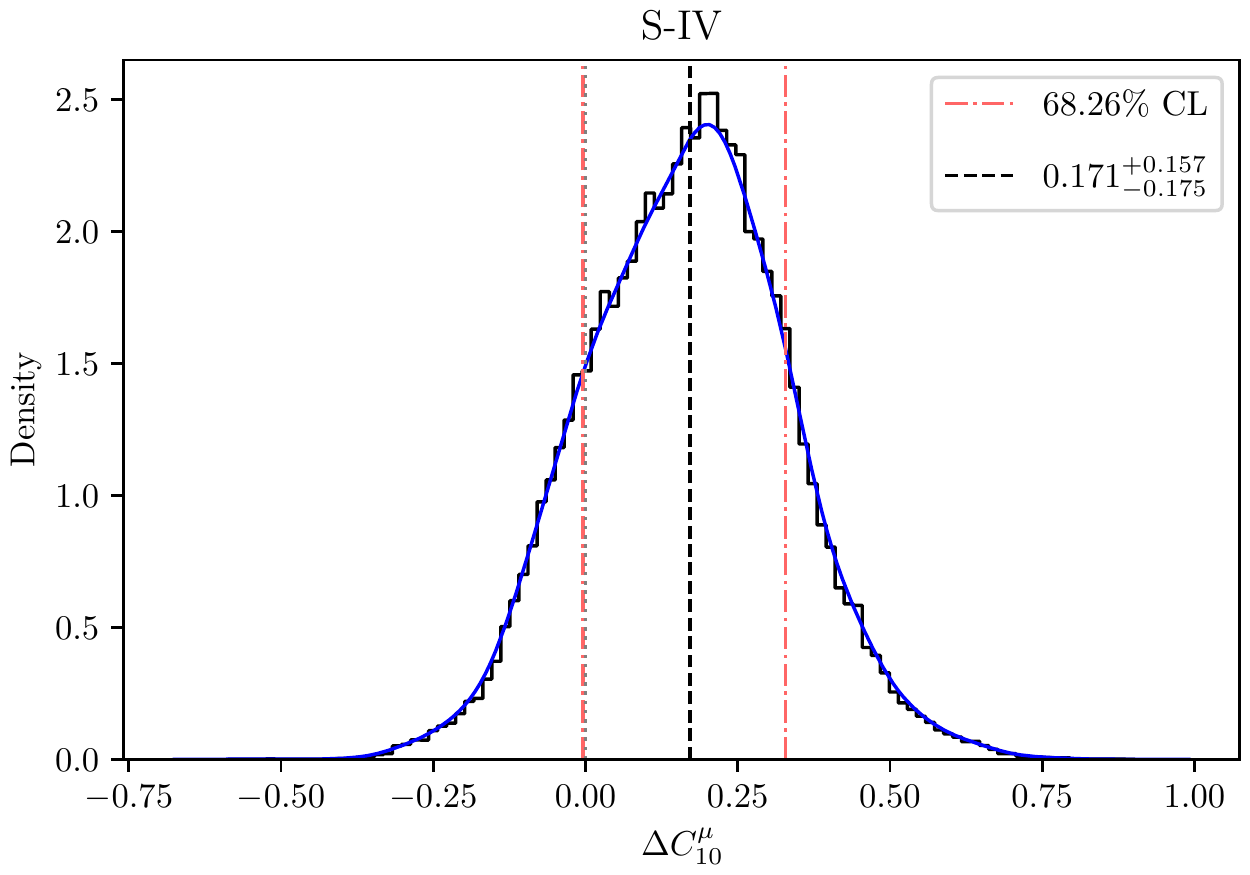}
	\caption{Density(black steps) and KDE (blue curves) of Wilson coefficients $\Delta C_{10}^\mu$ in different scenarios varied from old Dataset \textbf{A} to \textbf{B}. Intervals between the two red(dotdashed) lines cover the HPD about 68.26\% to estimate 1$\sigma$ range. The remained indicators black(dashed) and gray(dotted) lines play the similar role as the description of Fig. \ref{fig0.1}.
	}\label{fig0.2}
\end{figure}
\begin{table}[t]
	
	\caption{ Global fits in various scenarios based on Dataset \textbf{A} with
		$ \chi^2_{\text{SM}} =287.424$ which calculated by setting all WCs to be zero as well as corresponding observables n=201.}
	\label{tab:olddata}
	\vspace{-0.4cm}
	\begin{center}
		\renewcommand\arraystretch{1.2}
		\begin{adjustbox}{scale=0.8,center}
			\begin{tabular}
				{|c | c c c c | c c c c |}
				\hline
				Params & S-I'&S-II'&S-III'&S-IV'
				& ACDMN\cite{Alguero:2021anc} & AS\cite{Altmannshofer:2021qrr} 
				& HMMN\cite{Hurth:2022lnw} & GGJLCS \cite{Geng:2021nhg}
				\\
				\hline
						
				\text{Reduced }$\chi^2$
				&183.404/(n-12)&197.556/(n-12)&182.869/(n-16)&176.807/(n-20)
				&260.66/(254-6)&
				&179.1/(183-20)&96.88/90
				\\
				$\chi^2_{\text{min}}/\text{d.o.f}$
				&= 0.970&= 1.045&= 0.988&= 0.977
				&= 1.05&
				&= 1.1&= 1.08
				\\
			
				\hline				
				
				$\Delta C_7$
				&$-0.003^{+0.020}_{-0.019}$&$-0.001^{+0.015}_{-0.015}$&-&$0.001^{+0.016}_{-0.015}$
				&$0.00_{-0.02}^{+0.01}$&-
				&$0.06_{+0.03}^{-0.03}$&-
				\\
				\hline
				
				$\Delta C_7^{\prime}$ 
				&$0.017^{+0.018}_{-0.019}$&$0.020^{+0.014}_{-0.015}$&-&$0.020^{+0.014}_{-0.014}$
				&$+0.00_{-0.01}^{+0.02}$&-
				&$-0.01_{+0.01}^{-0.01}$&-
				\\
				\hline
				
				$\Delta C_8$
				&$-0.788^{+0.595}_{-0.514}$&$-0.885^{+0.435}_{-0.398}$&-&$-0.773^{+0.451}_{-0.449}$
				&-&-
				&$-0.80_{+0.40}^{-0.40}$&-
				\\
				\hline
				
				$\Delta C_8^{\prime}$ 
				&$-0.073^{+1.089}_{-1.000}$&$-0.093^{+0.921}_{-0.831}$&-&$-0.089^{+0.996}_{-0.922}$
				&-&-
				&$-0.30_{-1.30}^{+1.30}$&-
				\\
				\hline
				
				$\Delta C_9^{\mu}$ 
				&$-0.806^{+0.257}_{-0.272}$&$-0.795^{+0.205}_{-0.210}$&$-1.068^{+0.161}_{-0.164}$&$-0.863^{+0.214}_{-0.227}$
				&$-1.08_{-0.17}^{+0.18}$&$-0.82_{-0.23}^{+0.23}$
				&$-1.14_{+0.19}^{-0.19}$&$-1.07_{-0.29}^{+0.29}$
				\\
				\hline
				
				$\Delta C_9^{\prime\mu}$ 
				&$0.194^{+0.395}_{-0.416}$&$0.056^{+0.338}_{-0.342}$&$0.112^{+0.393}_{-0.397}$&$0.020^{+0.346}_{-0.362}$
				&$0.16_{-0.36}^{+0.37}$&$-0.10_{-0.34}^{+0.34}$
				&$0.05_{-0.32}^{+0.32}$&$0.32_{+0.21}^{-0.21}$
				\\
				\hline
				
				$\Delta C_{10}^{\mu}$ 
				&$0.236^{+0.216}_{-0.193}$&$0.145^{+0.166}_{-0.156}$&$0.164^{+0.181}_{-0.180}$&$0.213^{+0.166}_{-0.155}$
				&$0.15_{-0.13}^{+0.13}$&$+0.14_{-0.23}^{+0.23}$
				&$0.21_{+0.20}^{-0.20}$&$0.21_{+0.14}^{-0.14}$
				\\
				\hline
				
				$\Delta C_{10}^{\prime\mu}$ 
				&$-0.096^{+0.251}_{-0.237}$&$-0.108^{+0.186}_{-0.177}$&$-0.115^{+0.200}_{-0.198}$&$-0.089^{+0.177}_{-0.176}$
				&$-0.18_{-0.18}^{+0.20}$&$-0.33_{-0.23}^{+0.23}$
				&$-0.03_{-0.19}^{+0.19}$&$-0.26_{+0.14}^{-0.14}$
				\\
				\hline
				
				$\Delta C_S^{\mu}$ 
				&$0.066^{+1.091}_{-1.142}$&$-0.004^{+1.102}_{-1.131}$&$-0.008^{+0.883}_{-0.899}$&$-0.043^{+0.842}_{-0.875}$
				&-&-
				&$0.01_{-0.05}^{+0.05}$&-
				\\
				\hline
				
				$\Delta C_S^{\prime\mu}$ 
				&$0.065^{+1.087}_{-1.140}$&$0.003^{+1.103}_{-1.126}$&$-0.002^{+0.873}_{-0.936}$&$-0.059^{+0.844}_{-0.869}$
				&-&-
				&$-0.01_{-0.05}^{+0.05}$&-
				\\
				\hline
				
				$\Delta C_P^{\mu}$ 
				&$0.167^{+1.172}_{-1.225}$&$1.017^{+0.735}_{-0.816}$&$0.092^{+1.076}_{-0.994}$&$0.117^{+0.847}_{-0.894}$
				&-&-
				&$-0.04_{-0.02}^{+0.02}$&-
				\\
				\hline
				
				$\Delta C_P^{\prime\mu}$ 
				&$0.053^{+1.169}_{-1.227}$&$0.891^{+0.729}_{-0.812}$&$0.010^{+1.083}_{-1.002}$&$0.040^{+0.854}_{-0.895}$
				&-&-
				&$-0.04_{-0.02}^{+0.02}$&-
				\\
				\hline

				$\Delta C_9^{e}$ 
				&-&$-0.795^{+0.205}_{-0.210}$&$-1.753^{+0.781}_{-0.772}$&$-1.551^{+0.627}_{-0.599}$
				&-&$-0.24_{-1.17}^{+1.17}$
				&$-6.50_{-1.90}^{+1.90}$&-
				\\
				\hline
				
				$\Delta C_9^{\prime e}$ 
				&-&$0.056^{+0.338}_{-0.342}$&$1.725^{+1.724}_{-2.286}$&$1.710^{+1.466}_{-1.764}$
				&-&-
				&$1.40_{-2.30}^{+2.30}$&-
				\\
				\hline
				
				$\Delta C_{10}^{e}$ 
				&-&$0.145^{+0.166}_{-0.156}$&$0.108^{+1.456}_{-0.661}$&$0.058^{+1.193}_{-0.661}$
				&-&$-0.24_{-0.78}^{+0.78}$
				&$\sim0$&-
				\\
				\hline
				
				$\Delta C_{10}^{\prime e}$ 
				&-&$-0.108^{+0.186}_{-0.177}$&$0.600^{+1.208}_{-1.099}$&$0.655^{+0.958}_{-0.841}$
				&-&-
				&$\sim0$&-
				\\
				\hline
				
				$\Delta C_S^{e}$ 
				&-&$-0.004^{+1.102}_{-1.131}$&$-0.719^{+1.861}_{-1.227}$&$-0.549^{+1.602}_{-1.232}$
				&-&-
				&$-0.38_{-0.41}^{+0.41}$&-
				\\
				\hline
				
				$\Delta C_S^{\prime e}$ 
				&-&$0.003^{+1.103}_{-1.126}$&$-0.699^{+1.837}_{-1.224}$&$-0.550^{+1.618}_{-1.326}$
				&-&-
				&$-0.36_{-0.50}^{+0.50}$&-
				\\
				\hline
				
				$\Delta C_P^{e}$ 
				&-&$1.017^{+0.735}_{-0.816}$&$-1.592^{+1.552}_{-1.079}$&$-1.688^{+1.366}_{-0.978}$
				&-&-
				&$-0.98_{-0.21}^{+0.21}$&-
				
				\\
				\hline
				
				$\Delta C_P^{\prime e}$ 
				&-&$0.891^{+0.729}_{-0.812}$&$-1.360^{+1.318}_{-1.149}$&$-1.431^{+1.212}_{-1.017}$
				&-&-
				&$-0.95_{-0.29}^{+0.29}$&-
				\\
				\hline
			\end{tabular}
		\end{adjustbox}
	\end{center}
\end{table}

To interpret the 2022 release of $R_{K^{(*)}}$ in a global picture of $b\to s\ell^+\ell^-$, we carry out global fits in four aforementioned scenarios based on two different datasets, Dataset \textbf{A} and \textbf{B}. We use Figures \ref{fig0.1} and \ref{fig0.2} to illustrate the central values and errors of typical parameters, $\Delta C_9^\mu$ and $\Delta C_{10}^\mu$, with Bayesian statistics. The first row of figures in Figures \ref{fig0.1} and \ref{fig0.2} are produced based on the early Dataset \textbf{A}, while the second row corresponds to Dataset \textbf{B}. The central values of the parameters, mainly located at around well-known $-1$ and \textcolor{red}{$0.2$}, differ slightly from scenarios in both two sets of fits. On the other hand, to understand how the global change occurs due to the 2022 update of $R_{K^{(*)}}$, a comparison between the results of the two datasets is necessary. Taking $\Delta C_9^\mu$ shown in Figure \ref{fig0.1} as an example, the central values in all the scenarios vary, but not dramatically, while the errors almost keep unchanged.

\begin{table}[t]
	
	\caption{ Global fits in various scenarios based on Dataset \textbf{B} with $ \chi^2_{\text{SM}}=265.888$ originated from $ \Delta C_{i} =0$ and the number of observables n=203.}
	\label{tab:newdata}
	\vspace{-0.4cm}
	\begin{center}
		\renewcommand\arraystretch{1.2}
		\begin{adjustbox}{scale=0.8,center}
			\begin{tabular}
				{|c | c c c c | c c c c |}
				\hline
				Params & S-I &S-II&S-III&S-IV
				& ADCMN\cite{Alguero:2021anc} & AS\cite{Altmannshofer:2021qrr} 
				& HMMN\cite{Hurth:2022lnw} & GGJLCS\cite{Geng:2021nhg}
				\\
				\hline
				
				\text{Reduced }$\chi^2$
				&190.044/(n-12)&177.891/(n-12)&185.386/(n-16)&178.953/(n-20)
				&260.66/(254-6)&
				&179.1/(183-20)&96.88/90
				\\
				$\chi^2_{\text{min}}/\text{d.o.f}$
				&= 0.995&= 0.931&= 0.991&= 0.978
				&= 1.05&
				&= 1.1&= 1.08
				\\
				
				\hline				
				
				$\Delta C_7$
				&$-0.000^{+0.020}_{-0.020}$&$-0.001^{+0.015}_{-0.015}$&-&$-0.000^{+0.016}_{-0.015}$
				&$0.00_{-0.02}^{+0.01}$&-
				&$0.06_{+0.03}^{-0.03}$&-
				\\
				\hline
				
				$\Delta C_7^{\prime}$ 
				&$0.017^{+0.020}_{-0.018}$&$0.020^{+0.015}_{-0.014}$&-&$0.023^{+0.014}_{-0.016}$
				&$+0.00_{-0.01}^{+0.02}$&-
				&$-0.01_{+0.01}^{-0.01}$&-
				\\
				\hline
				
				$\Delta C_8$
				&$-0.995^{+0.540}_{-0.463}$&$-0.921^{+0.443}_{-0.378}$&-&$-0.773^{+0.465}_{-0.424}$
				&-&-
				&$-0.80_{+0.40}^{-0.40}$&-
				\\
				\hline
				
				$\Delta C_8^{\prime}$ 
				&$-0.080^{+1.046}_{-0.942}$&$-0.076^{+0.893}_{-0.833}$&-&$-0.258^{+1.007}_{-0.802}$
				&-&-
				&$-0.30_{-1.30}^{+1.30}$&-
				\\
				\hline
				
				$\Delta C_9^{\mu}$ 
				&$-0.752^{+0.262}_{-0.265}$&$-0.789^{+0.198}_{-0.210}$&$-1.054^{+0.163}_{-0.171}$&$-0.872^{+0.215}_{-0.215}$
				&$-1.08_{-0.17}^{+0.18}$&$-0.82_{-0.23}^{+0.23}$
				&$-1.14_{+0.19}^{-0.19}$&$-1.07_{-0.29}^{+0.29}$
				\\
				\hline
				
				$\Delta C_9^{\prime\mu}$ 
				&$0.174^{+0.434}_{-0.441}$&$0.048^{+0.338}_{-0.348}$&$0.130^{+0.439}_{-0.437}$&$0.088^{+0.342}_{-0.378}$
				&$0.16_{-0.36}^{+0.37}$&$-0.10_{-0.34}^{+0.34}$
				&$0.05_{-0.32}^{+0.32}$&$0.32_{+0.21}^{-0.21}$
				\\
				\hline
				
				$\Delta C_{10}^{\mu}$ 
				&$-0.019^{+0.206}_{-0.175}$&$0.163^{+0.165}_{-0.160}$&$0.112^{+0.166}_{-0.184}$&$0.171^{+0.157}_{-0.175}$
				&$0.15_{-0.13}^{+0.13}$&$+0.14_{-0.23}^{+0.23}$
				&$0.21_{+0.20}^{-0.20}$&$0.21_{+0.14}^{-0.14}$
				\\
				\hline
				
				$\Delta C_{10}^{\prime\mu}$ 
				&$-0.118^{+0.266}_{-0.247}$&$-0.093^{+0.183}_{-0.179}$&$-0.115^{+0.215}_{-0.213}$&$-0.062^{+0.197}_{-0.180}$
				&$-0.18_{-0.18}^{+0.20}$&$-0.33_{-0.23}^{+0.23}$
				&$-0.03_{-0.19}^{+0.19}$&$-0.26_{+0.14}^{-0.14}$
				\\
				\hline
				
				$\Delta C_S^{\mu}$ 
				&$0.023^{+1.064}_{-1.097}$&$0.060^{+1.188}_{-1.230}$&$-0.066^{+0.944}_{-0.929}$&$0.009^{+0.858}_{-0.845}$
				&-&-
				&$0.01_{-0.05}^{+0.05}$&-
				\\
				\hline
				
				$\Delta C_S^{\prime\mu}$ 
				&$0.014^{+1.064}_{-1.086}$&$0.061^{+1.188}_{-1.225}$&$-0.070^{+0.957}_{-0.930}$&$0.012^{+0.858}_{-0.862}$
				&-&-
				&$-0.01_{-0.05}^{+0.05}$&-
				\\
				\hline
				
				$\Delta C_P^{\mu}$ 
				&$0.079^{+1.159}_{-1.146}$&$0.478^{+0.808}_{-0.899}$&$0.189^{+1.018}_{-1.028}$&$0.124^{+0.902}_{-0.910}$
				&-&-
				&$-0.04_{-0.02}^{+0.02}$&-
				\\
				\hline
				
				$\Delta C_P^{\prime\mu}$ 
				&$-0.032^{+1.158}_{-1.145}$&$0.370^{+0.803}_{-0.897}$&$0.098^{+1.009}_{-1.024}$&$0.038^{+0.894}_{-0.913}$
				&-&-
				&$-0.04_{-0.02}^{+0.02}$&-
				\\
				\hline

				$\Delta C_9^{e}$ 
				&-&$-0.789^{+0.198}_{-0.210}$&$-1.623^{+0.662}_{-0.734}$&$-1.511^{+0.561}_{-0.533}$
				&-&$-0.24_{-1.17}^{+1.17}$
				&$-6.50_{-1.90}^{+1.90}$&-
				\\
				\hline
				
				$\Delta C_9^{\prime e}$ 
				&-&$0.048^{+0.338}_{-0.348}$&$1.090^{+1.610}_{-1.793}$&$0.864^{+1.483}_{-1.608}$
				&-&-
				&$1.40_{-2.30}^{+2.30}$&-
				\\
				\hline
				
				$\Delta C_{10}^{e}$ 
				&-&$0.163^{+0.165}_{-0.160}$&$0.555^{+1.042}_{-0.576}$&$0.383^{+0.840}_{-0.424}$
				&-&$-0.24_{-0.78}^{+0.78}$
				&$\sim0$&-
				\\
				\hline
				
				$\Delta C_{10}^{\prime e}$ 
				&-&$-0.093^{+0.183}_{-0.179}$&$0.088^{+0.969}_{-0.956}$&$0.002^{+0.881}_{-0.815}$
				&-&-
				&$\sim0$&-
				\\
				\hline
				
				$\Delta C_S^{e}$ 
				&-&$0.060^{+1.188}_{-1.230}$&$-0.952^{+2.122}_{-1.139}$&$-0.806^{+1.900}_{-1.238}$
				&-&-
				&$-0.38_{-0.41}^{+0.41}$&-
				\\
				\hline
				
				$\Delta C_S^{\prime e}$ 
				&-&$0.061^{+1.188}_{-1.225}$&$-1.051^{+2.251}_{-1.075}$&$-0.803^{+1.861}_{-1.194}$
				&-&-
				&$-0.36_{-0.50}^{+0.50}$&-
				\\
				\hline
				
				$\Delta C_P^{e}$ 
				&-&$0.478^{+0.808}_{-0.899}$&$-1.568^{+1.544}_{-1.149}$&$-1.837^{+1.376}_{-0.930}$
				&-&-
				&$-0.98_{-0.21}^{+0.21}$&-
				
				\\
				\hline
				
				$\Delta C_P^{\prime e}$ 
				&-&$0.370^{+0.803}_{-0.897}$&$-1.477^{+1.409}_{-1.083}$&$-1.652^{+1.200}_{-0.979}$
				&-&-
				&$-0.95_{-0.29}^{+0.29}$&-
				\\
				\hline
			\end{tabular}
		\end{adjustbox}
	\end{center}
\end{table}

Incorporating all the WCs analyzed in various scenarios based on both Dataset \textbf{A} and \textbf{B}. We summarize the results of all the fitted parameters characterizing new physics effects in Tables \ref{tab:olddata} and \ref{tab:newdata}, respectively. The numbers of fitted parameters in our analysis are two sets of 12, one set of 16, and 20, denoted as scenario I, II, III, and IV (S-I, S-II, S-III, and S-IV, or those corresponding ones with a prime). As a comparison, early global fits made by other four independent analyses \cite{Alguero:2021anc,Altmannshofer:2021qrr,Hurth:2022lnw,Geng:2021nhg} (one group of 20-D, one group of 6-D, and two groups of 4-D parameters) are also listed in the two tables.

To confirm the correctness of our numerical calculation, we first perform a calculation based on Dataset \textbf{A}, as shown in Table \ref{tab:olddata}, in different working scenarios (with a prime). Our fitted WCs of muon flavor are consistent not only with each scenario but also with the other four independent groups within the fitted errors.
WCs involving electron flavor have been studied less, and early efforts can be found in the two groups AS\cite{Altmannshofer:2021qrr} and HMMN\cite{Hurth:2022lnw}. 
 With similar errors ($1.2$ and $1.9$) but obvious different central values ($-0.24$ and $-6.50$) for $\Delta C_9^e$, it is not easy to judge how its deviation from the SM prediction. Our calculations in both S-III' ($-1.8\pm0.8$) and S-IV' ($-1.6\pm0.6$) provide self-consistent information and support a negative deviation from the SM about $2.5\sigma$. As for $\Delta C_{10}^e$, it can be concluded that the deviation from the SM is 
within $1\sigma$, 
 combining all our calculations (S-II', III', and IV') as well as the work done by AS\cite{Altmannshofer:2021qrr} and HMMN\cite{Hurth:2022lnw}.
In general, the errors of scalar operator WCs in HMMN\cite{Hurth:2022lnw} are smaller than ours. Though we both make a consistent description (less than $2\sigma$ deviation) for $\Delta C_{S,P}^{(')\mu}$ and $\Delta C_S^{(')e}$, the feature of $\Delta C_{P}^{(')e}$ differs. Our calculation prefers a SM-like behavior while HMMN\cite{Hurth:2022lnw} suggests a deviation of around $4\sigma$, requiring further clarification by incorporating more and more precise data as well as more efforts on fitting.

The impact of the 2022 $R_{K^{(*)}}$ data is detailed in Table  \ref{tab:newdata}. 
The new physics potential in $\Delta C_9^\mu$, which was highly anticipated before, has been widely questioned since the release of $R_{K^{(*)}}^{\text{2022}}$. According to the numerical results in Table \ref{tab:newdata}, the roughly $4 \sigma$ standard deviations still exist in each scenario, with slight shifts in central values and almost unchanged errors. 
The SM-like behavior of $\Delta C_{7,8}^{(')}$, within a $2\sigma$ deviation, remains unchanged from the earlier data. 
The situation for  other muon flavor related fitted parameters depends on the number of fitting parameters.
For example,   the fitted $\Delta C_{9}^\mu$ in S-III exhibits a decreased standard deviation from $6.6\sigma$ in Dataset \textbf{A} to $6.4\sigma$ in 
Dataset \textbf{B}. Therefore, it can be safely said that all the muon type WCs 
except $\Delta C_9^\mu$ 
are all within $2\sigma$ deviations. 
Moreover, the updated $\Delta C_9^e$ values are consistent with the previous values shown in Table \ref{tab:olddata}.
The deviation in Scenario III still keeps around $2.3\sigma$,
%
with
both slight decreased central value and error. 
\begin{figure}[t]
	\centering
	$
	\begin{array}{cc}
		(a) & (b) 
		\\
		\includegraphics[width=0.42\linewidth]{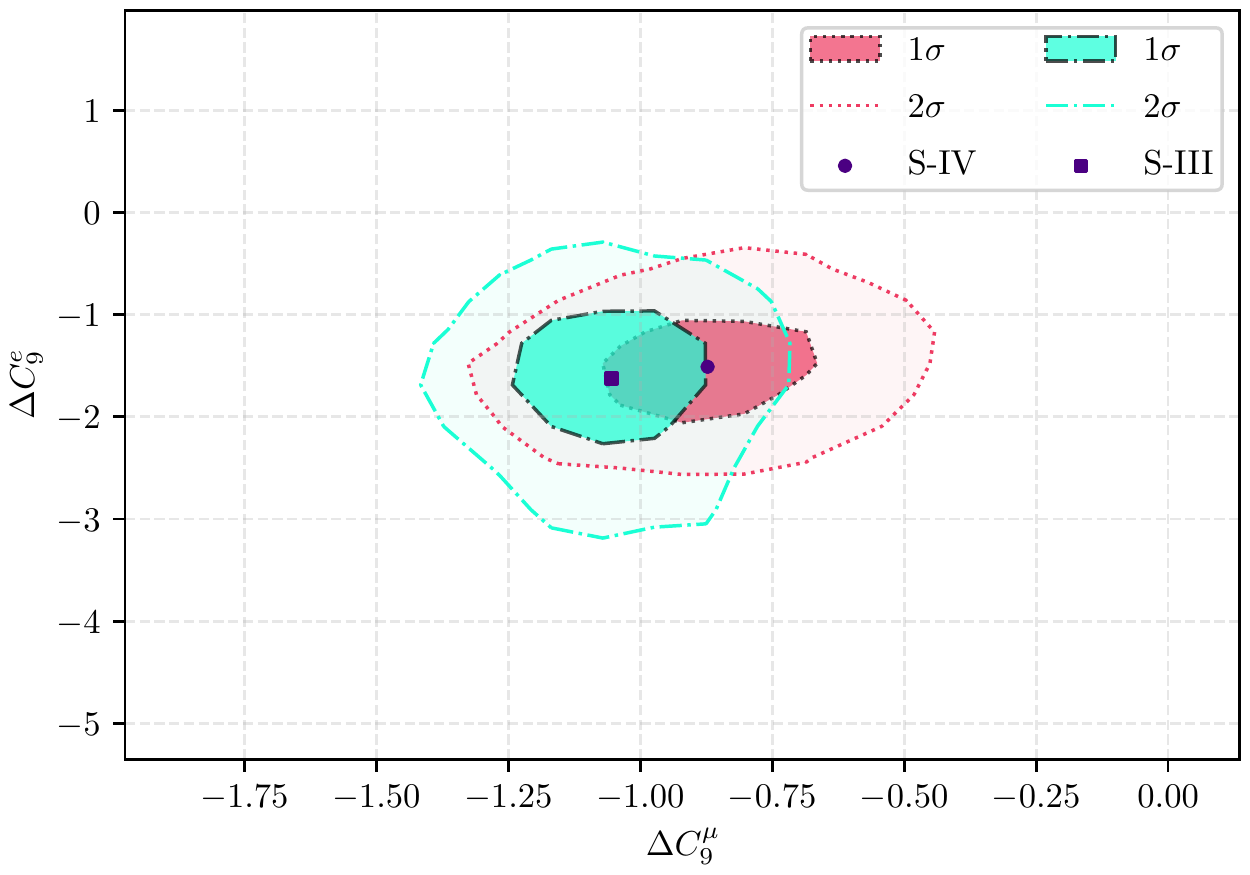}&
		\includegraphics[width=0.42\linewidth]{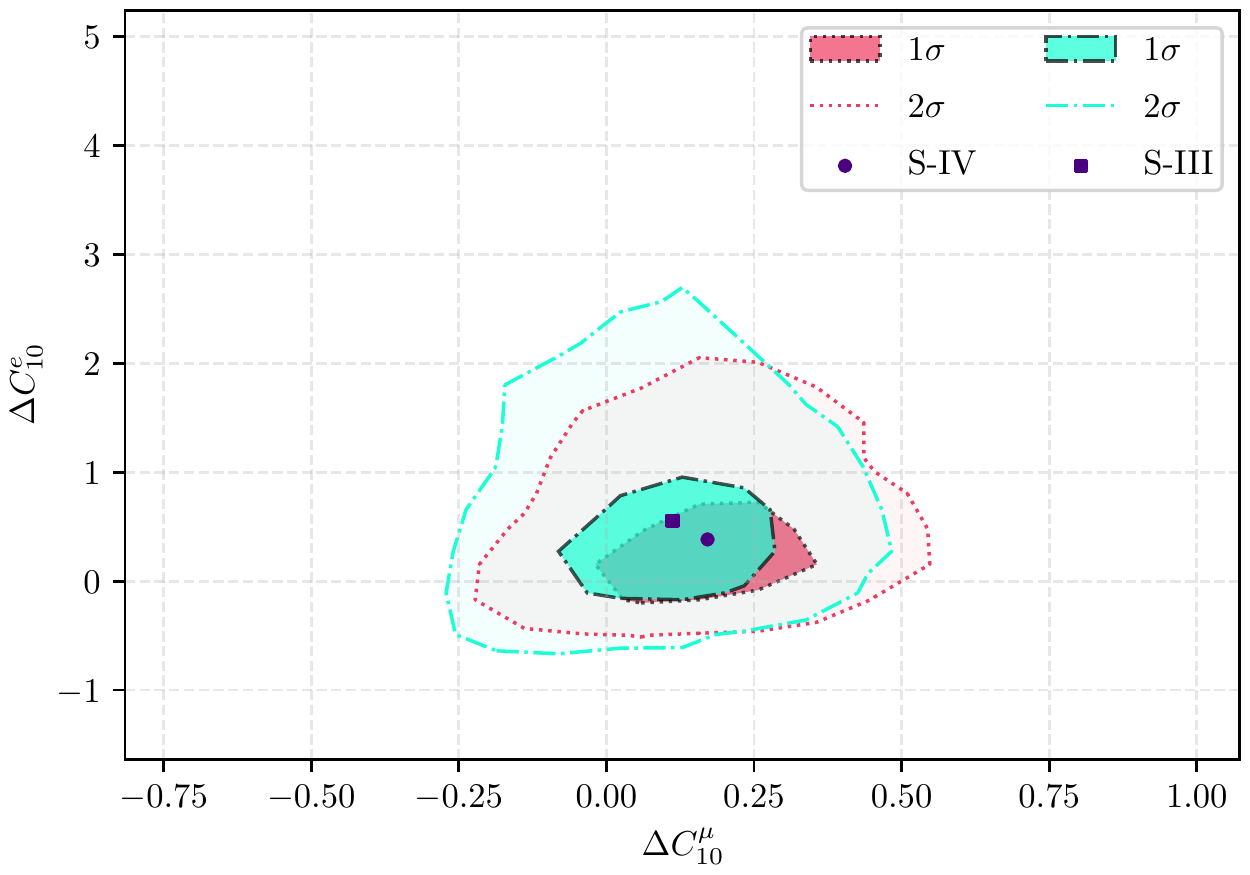}
		\\
		(c) & (d)
		\\
		\includegraphics[width=0.42\linewidth]{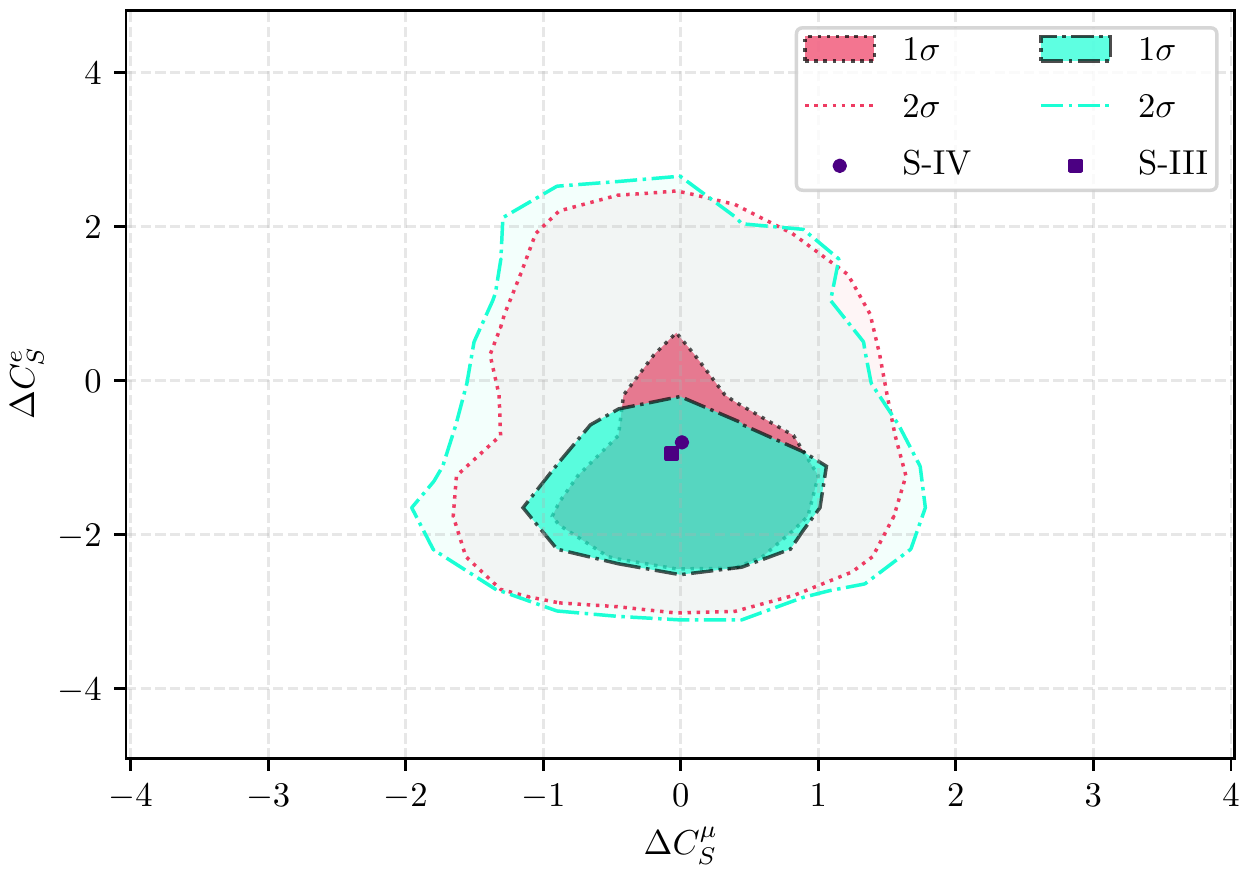}&
		\includegraphics[width=0.42\linewidth]{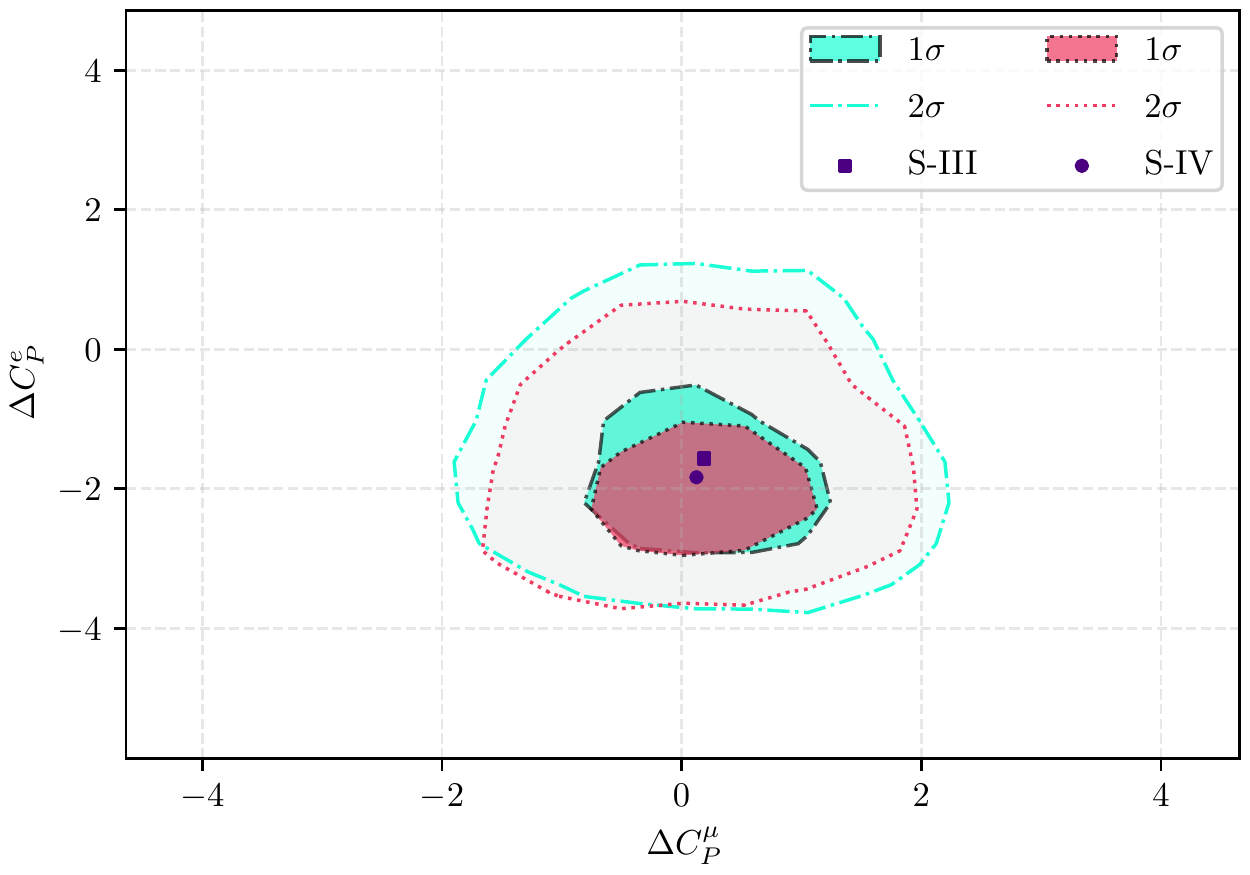}
		
	\end{array}
	$
	\caption{The correlations of corresponding WCs  specified by their lepton flavors with a consideration of $R_{K^{(*)}}^{\text{2022}}$, taking an example of  left-handed operators. The red(dotted) and green(dotdashed) regions represent the S-IV as well as S-III, where darker/lighter part are their corresponding 1/2$\sigma$ regions. Markers circle(square) are estimations of WCs in S-IV(S-III), respectively. The absence of S-II and S-I due to their WCs are not fully independent we set to be.
	}
	\label{Fig:flavor}
\end{figure}
The deviation in Scenario IV 
shifts from $2.6\sigma$ to $2.8\sigma$.
All other electron type WCs, including $\Delta C_{10, S, P}^{(')e}$, are found to be restricted within around $1\sigma$.

In addition to the 1-D parameter projections from the high dimensional full parameter space shown in Tables \ref{tab:olddata} and \ref{tab:newdata}, more information can be drawn from correlations among 2-D parameters. In many previous studies, lepton flavors in $\mathcal{O}_{9,10}^{(')}$ as well as $\mathcal{O}_{S, P}^{(')}$ are usually not discriminated. This assumption is also adopted as one of our working scenarios (S-II or S-II'). However, it is important to keep in mind that relaxing the identical lepton flavor restriction is also possible.
We  present explicitly in Fig. \ref{Fig:flavor}, the correlations between WCs of the same type among by specifying the lepton flavors 
in Scenario III and IV
based on Dataset \textbf{B}. 
The locations of the best fit points and the $1\sigma$ allowed regions are dependent on the working scenario, as shown by the analysis of left-handed operators in Figure \ref{Fig:flavor}. A straight line passing through the origin with a slope of 1 represents lepton flavor independence. In 

\begin{figure}[H]
	\centering
	$
	\begin{array}{cc}
		(a)& (b)\\
		\includegraphics[width=0.42\linewidth]{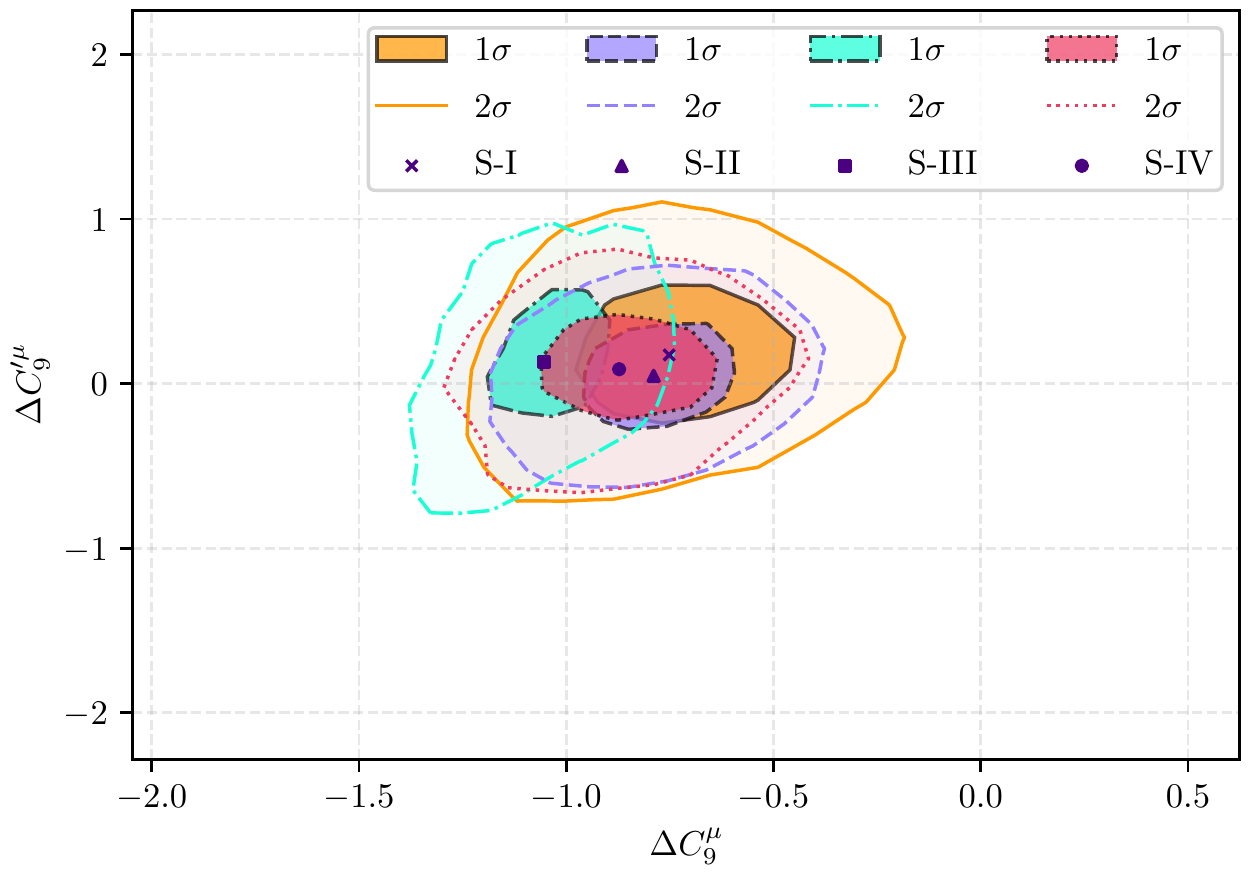}
		&	\includegraphics[width=0.42\linewidth]{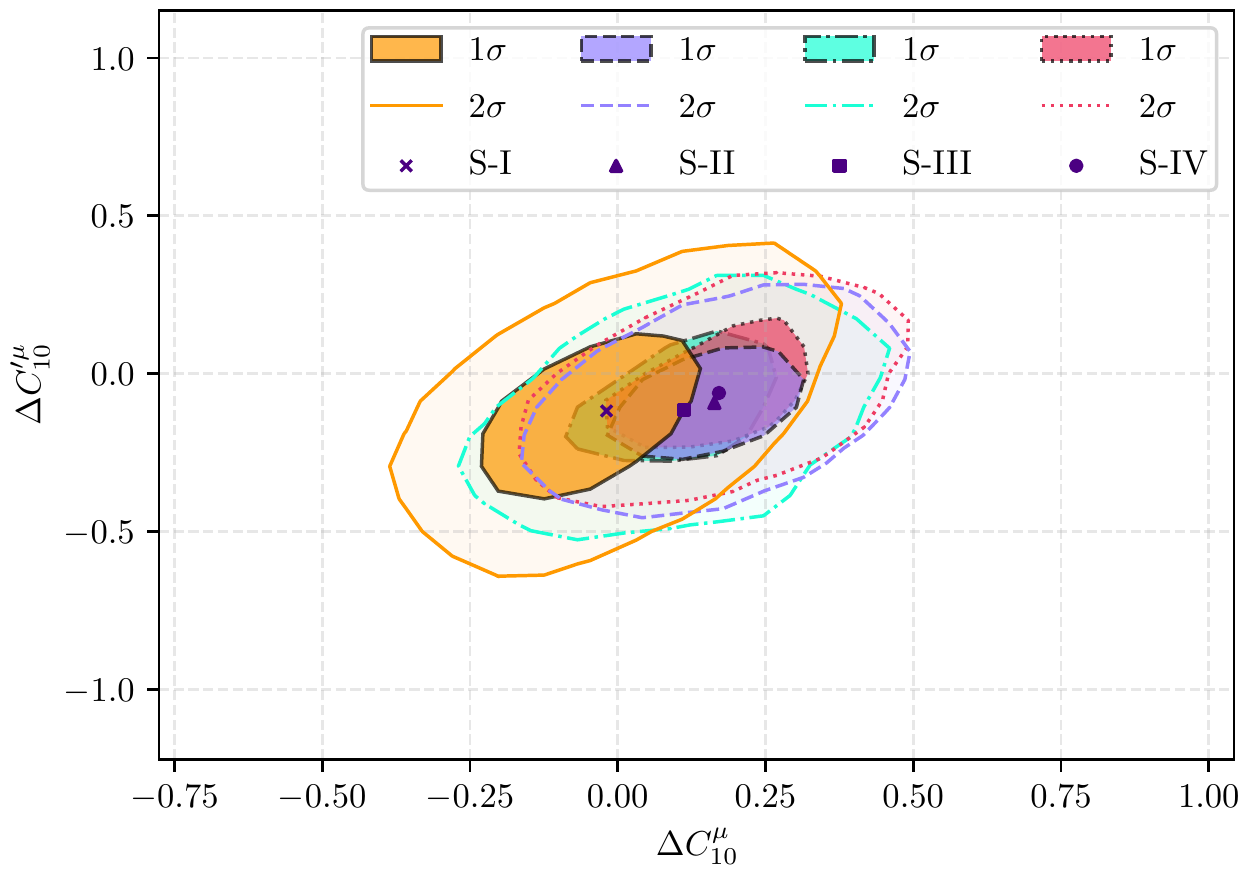}
		\\
		(c) & (d)
		\\
		\includegraphics[width=0.42\linewidth]{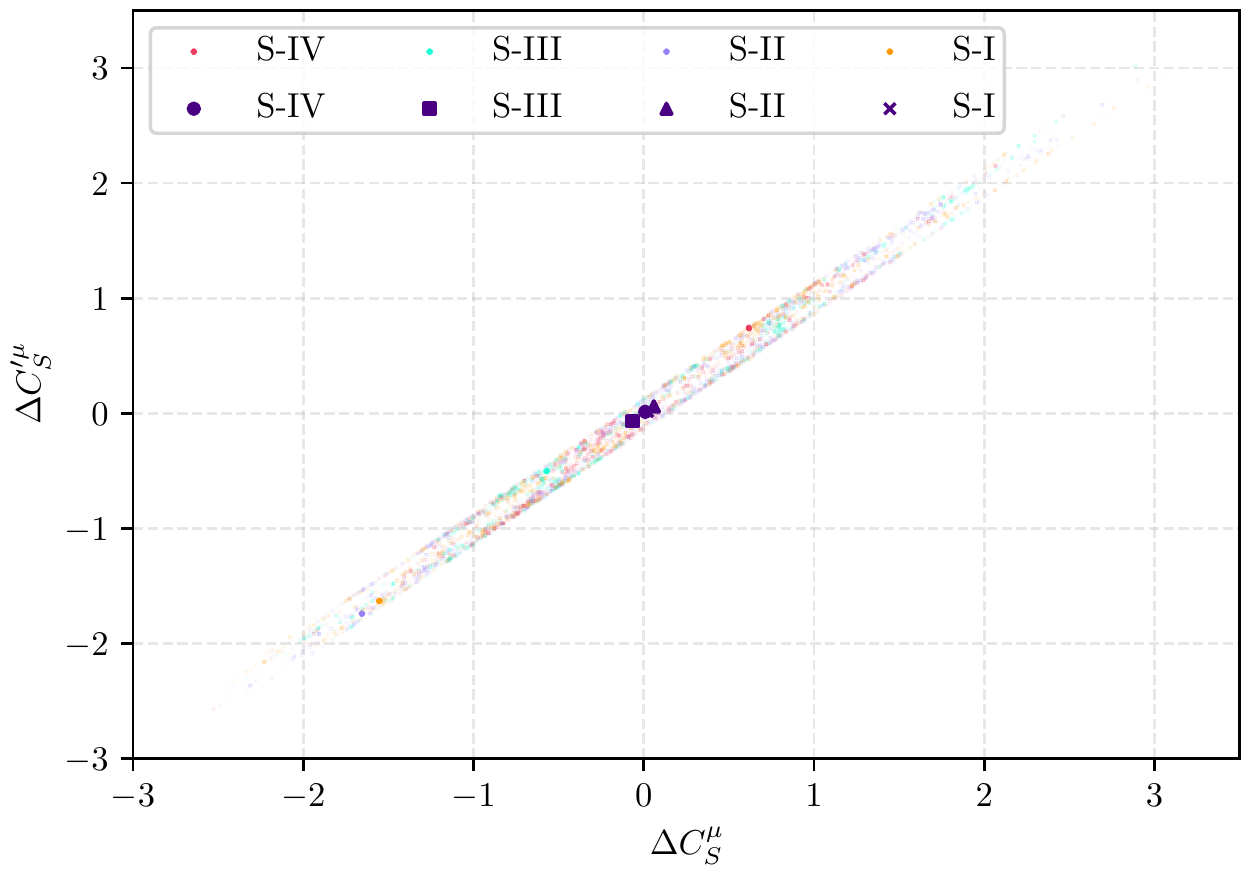}
		&	
		\includegraphics[width=0.42\linewidth]{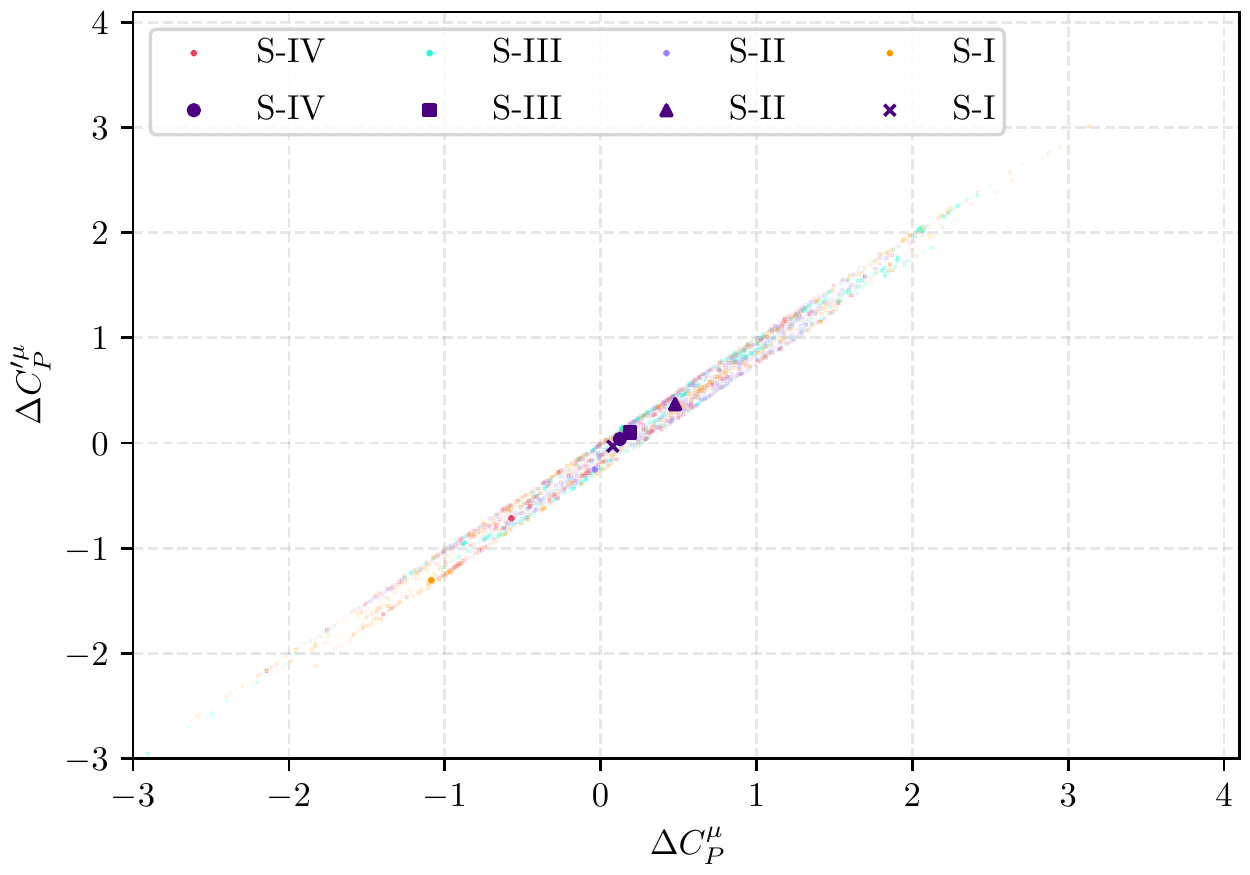}
		\\
		(e) & (f)
		\\
		\includegraphics[width=0.42\linewidth]{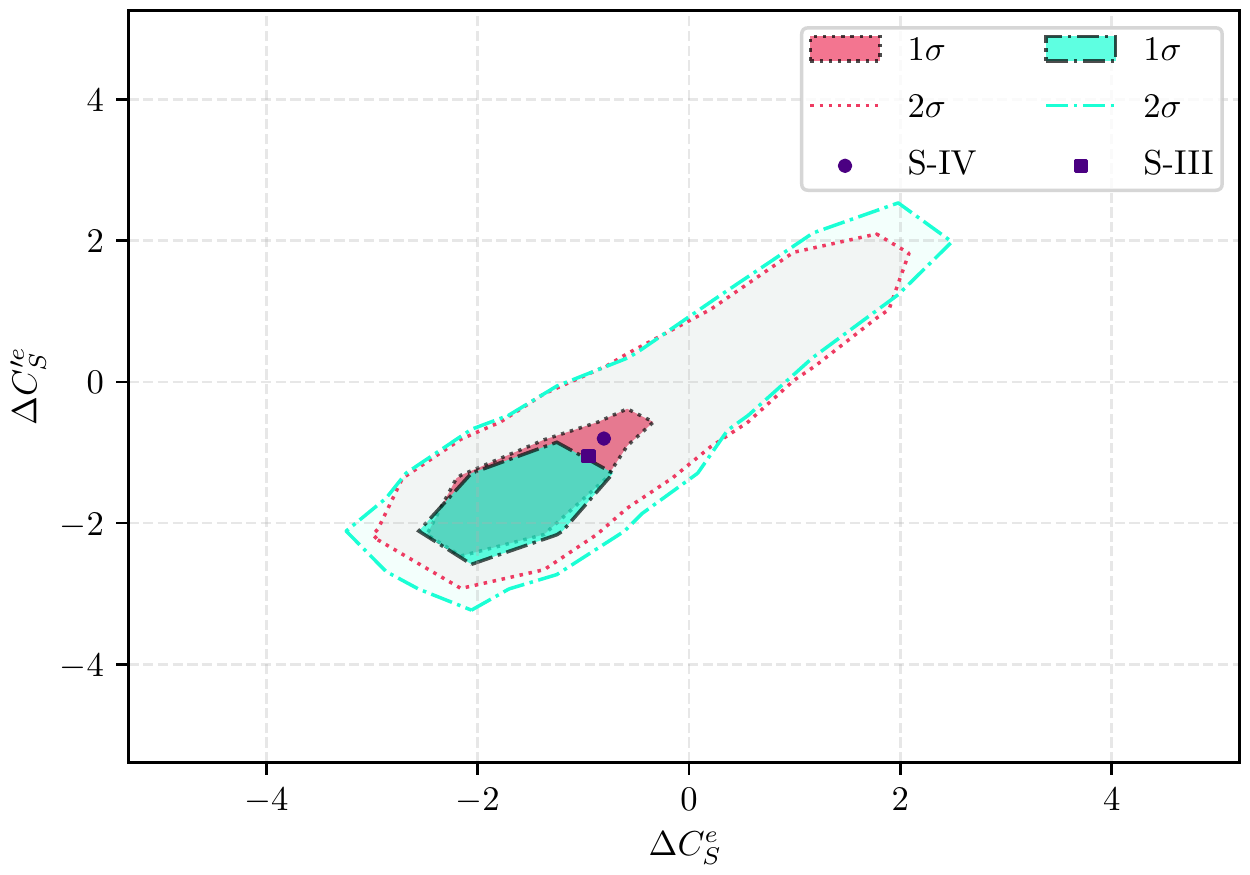}
		&	
		\includegraphics[width=0.42\linewidth]{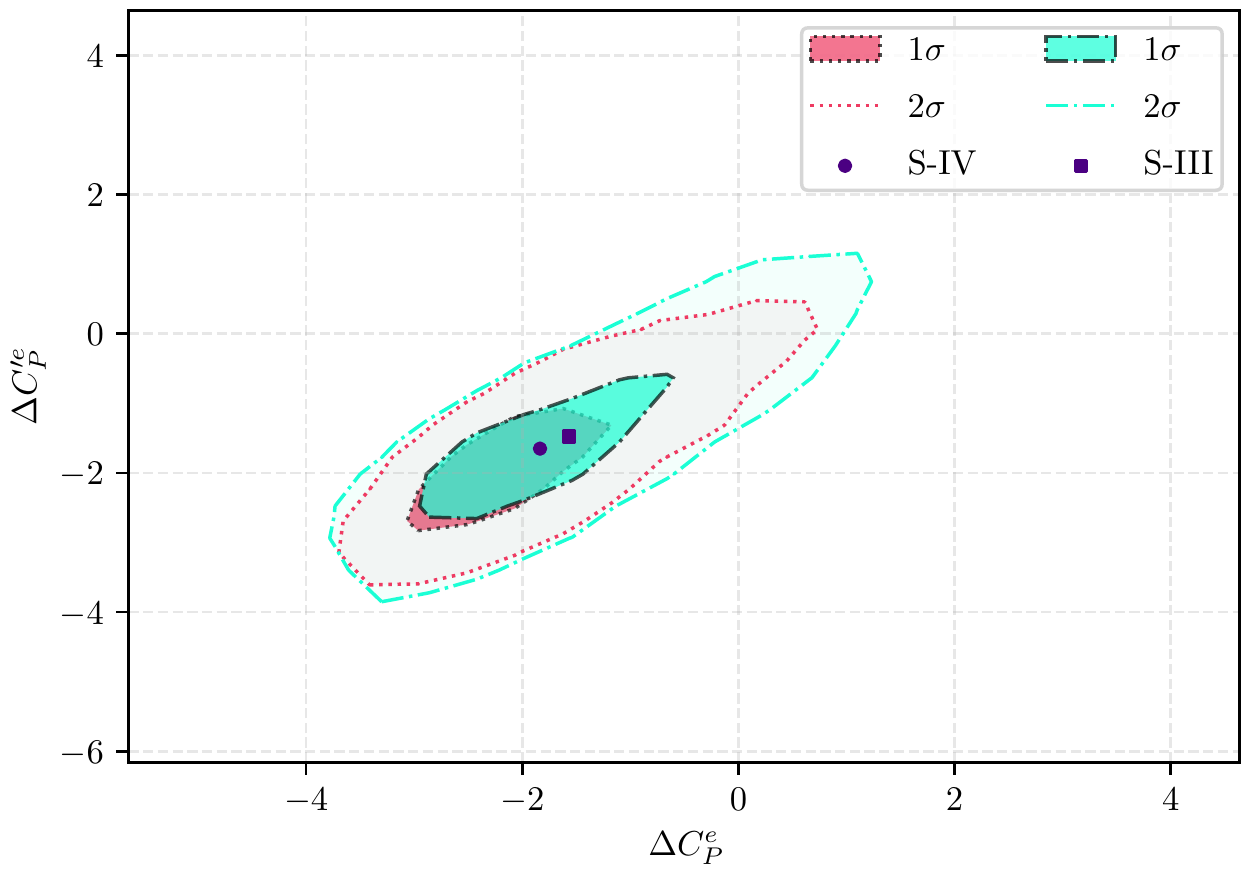}
	\end{array}
	$
	\caption{The correlations of corresponding WCs distinct from their chirality incorporating the updated $R_{K^{(*)}}$, 
		with a consideration of
		muon type operators ((a)-(d)) and typical electron type operators ((e)-(f)). The orange(solid), purple(dashed), green(dotdashed) and red(dotted) regions represent S-I, S-II, S-III and S-IV, respectively. The description of absence of S-I and S-II follows the Fig. \ref{Fig:flavor}.
	}
	\label{Fig:chiral}
\end{figure}
\noindent almost all of the two scenarios (S-III and S-IV), $\Delta C_9$ and $\Delta C_{S,P}$ deviate from this line in the $1\sigma$ region, while $\Delta C_{10}$ contains part of the "flavor identical line" in its $1\sigma$ region. But in $2\sigma$ regions, the flavor identical line can be contained by all $\Delta C_{9,10}$ and $\Delta C_{S,P}$. 
Therefore, at the $1\sigma$ level, the identical lepton flavor is only respected by $\Delta C_{10}$ but can be extended to all at the $2\sigma$ level.

\begin{figure}[t]
	\centering
	$
	\begin{array}{cc}
		(a)& (b)\\
		\includegraphics[width=0.42\linewidth]{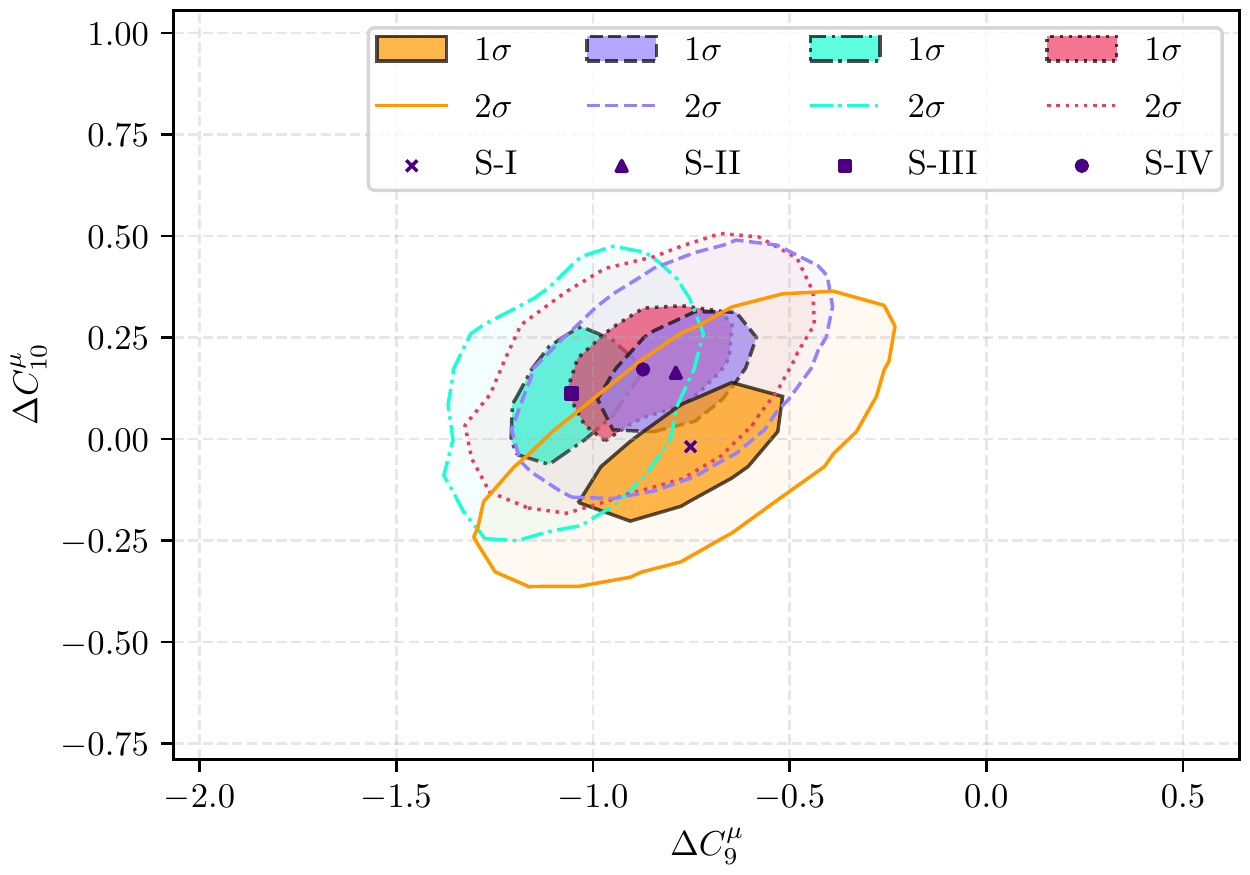}
		&\includegraphics[width=0.42\linewidth]{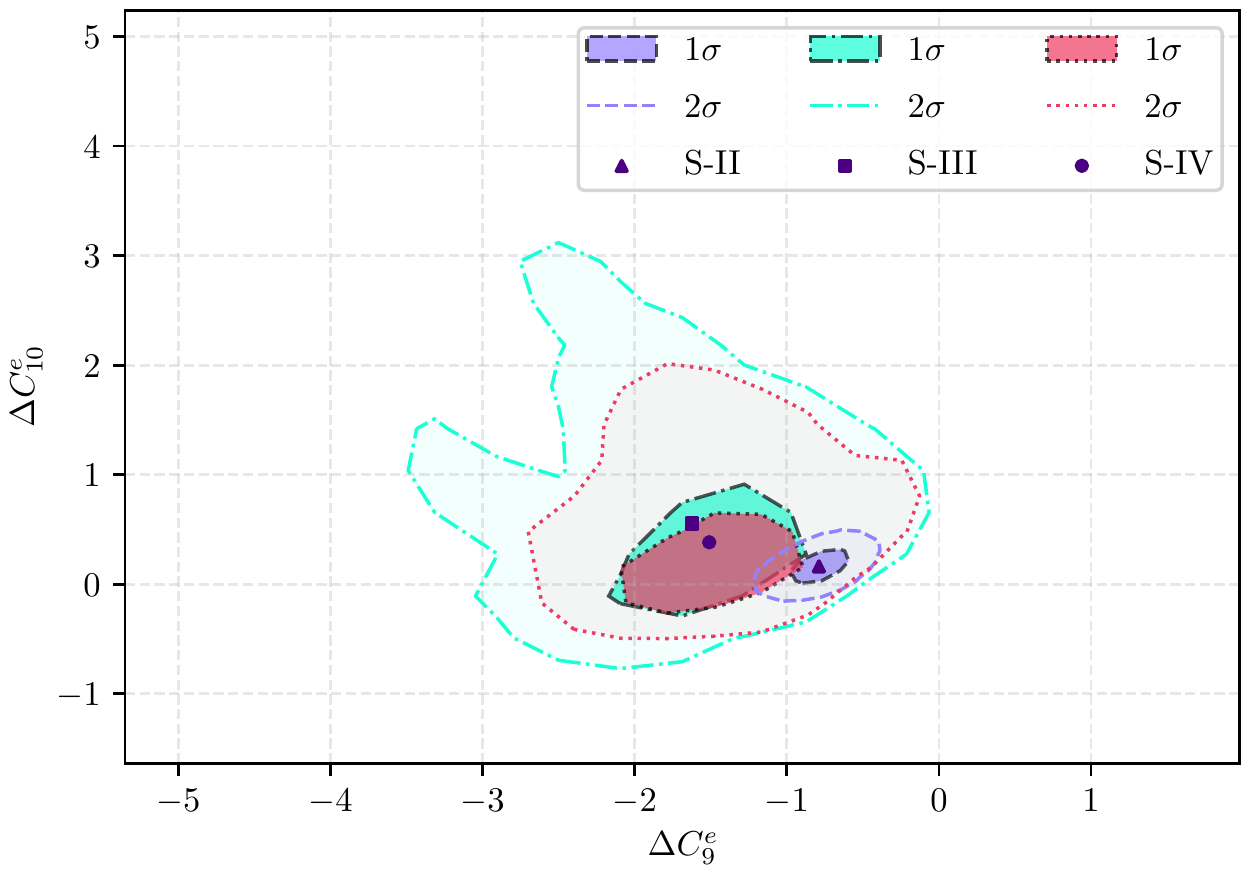}	
	\end{array}
	$
	\caption{The correlations of WCs involving more than $3\sigma$ deviations from SM predictions with the $R_{K^{(*)}}^{\text{2022}}$. Conventions of color, linetypes and markers remain the same as those in Fig. \ref{Fig:flavor} and \ref{Fig:chiral}. 
	} 
	\label{Fig:NP}
\end{figure}
The relations between operators with opposite chirality are also of interest. We show them in the form of 2-D correlations in Figure \ref{Fig:chiral}. Similar to the flavor situation, we take the line with slope of 1 and intercept of 0 as a criterion to judge the chirality dependence from data.
Apparently, the deviations of such a line, at the $2\sigma$ level, in all the four scenarios of $\Delta C_9^\mu$ indicate that $\Delta C_9^{\mu}$ and 
$\Delta C_9^{'\mu}$ are indeed two separated parameters. As for $\Delta C_{10}^\mu$, the identical chirality is not excluded at the $1\sigma$ region in S-I, 
and can be kept
at the $2\sigma$ regions in 
all the 4 
scenarios.
A strict respect of the criterion line can be found for the WCs of two scalar operators ($\Delta C_S$ and $\Delta C_P$) shown as (c) and (d) in Figure \ref{Fig:chiral}. This indicates that the chirality is indistinguishable for the two muon type scalar operators, although their fitted sizes are SM-like.
Due to the limited data about channels involving electrons, the surviving areas for $\Delta C_S^e-\Delta C_S^{'e}$ and $\Delta C_P^e-\Delta C_P^{'e}$ are much wider than the muon type operators, with the identical chirality line contained. 
Their sizes, which are consistent with SM at $2\sigma$ level, and the chirality relations are
anticipated to be improved more precisely when more data is accumulated.

\subsection{Discussions}	
\label{sub:disc}
Throughout the analyses in this work, the most important issue is whether the new physics possibility still exists 
by incorporating $R_{K^{(*)}}^{\text{2022}}$.
As shown in Fig. \ref{Fig:NP} explicitly,
although $\Delta C_{10}$ agrees with SM within $1\sigma$ in all the four scenarios for both muon and electron type,
 deviations from SM  in $\Delta C_{9}^\mu$ indeed exist for more than $4\sigma$ for most scenarios and
$\Delta C_{9}^e$ for around $3\sigma$ in Scenario II. 

The scalar operators are in general with small central values but large uncertainties as shown in Table \ref{tab:olddata} 
and \ref{tab:newdata}.  The 2-D correlations shown in Fig. \ref{Fig:chiral} apparently indicate the strong linear relations 
\begin{equation}
\Delta C_S^\mu = \Delta C_S^{'\mu},\quad
\Delta C_P^\mu = \Delta C_P^{'\mu}.
\end{equation}
On the other hand, if the emergence of new physics is described within the framework of SMEFT,  the relations
\begin{equation}
\Delta C_S^\mu= -\Delta C^\mu_P,\quad
\Delta C^{'\mu}_S = \Delta C^{'\mu}_P
\end{equation}
hold up to dimension-6 operators \cite{Jenkins:2017jig}, leading to
\begin{equation}
\Delta C_S^{(')\mu}=\Delta C_P^{(')\mu}=0.
\end{equation}
Note this feature for null scalar operators may violate in other non-SMEFT new physics\cite{Cata:2015lta}.

\section{Concluding Remarks}
\label{sec:con}
In this work, we perform global fits to two sets of datasets, one before and one after the release of $R_{K^{(*)}}^{\text{2022}}$ (denoted as
Dataset \textbf{A} and \textbf{B}), in four sets of working scenarios. In some of the working scenarios, we distinguish lepton flavor, which results in as many as 20 fitted parameters. Our numerical analysis based on Bayesian statistics  helps to interpret data and  the following points can be concluded
incorporating $R_{K^{(*)}}^{\text{2022}}$.
(i) The new physics possibility still exists.  As explicitly shown in Fig. \ref{Fig:NP}, current data still supports a deviation
 about or more than $4\sigma$ 
from SM for $\Delta C_9^\mu$. 
 (ii) The interpretation of flavor dependence in WCs is improved.  At $1\sigma$ level, the muon and electron flavor 
is distinguishable for $\Delta C_{9, S, P}$ but
indistinguishable  for $\Delta C_{10}$. 
And if allowing $2\sigma$ errors, lepton flavor is indistinguishable for all the four operators
and their corresponding WCs $\Delta C_{9,10,S,P}$.
(iii) The relation between operators and their chiral dual ones is specified. The WCs for scalar operators $\Delta C_{S}^\mu$ and  
$\Delta C_{P}^\mu$ are strictly chirality independent. At $1\sigma$ level, the data also indicates 
that to distinguish chirality is not necessary for $\Delta C_{S}^e$ and  
$\Delta C_{P}^e$. 
The WC $\Delta C_9^\mu$ differs from its dual one $\Delta C_9^{'\mu}$
at $2\sigma$ level in all the four scenarios, while $\Delta C_{10}^{\mu}$ distinguishes $\Delta C_{10}^{'\mu}$ in three of the four scenarios.
(iv) If the emergence of new physics is in terms of SMEFT, the scalar operators vanish.

Although the working scenario dependence indeed exists by comparing with detailed scenarios, it does not change the main features 
of global fits.  The obtained results from current model independent fits  provide useful inputs for new physics models  
or help to discriminate some of the models.

\acknowledgments
The authors benefited by discussions with Jibo He and Yue-Lin Sming Tsai.
This work was supported by NSFC  under Grant  No. U1932104.
\clearpage
\appendix
\section{Formulas for Involved Observables}
\label{app:Theo}
In this part, we provided a list about calculation of all observables we used in analysis, more detailed discussions are found in corresponding subsection.
\vspace{-1.0cm}
\subsection{\texorpdfstring{$B_{s,d}\to\ell\ell$}{Bsdll}}\label{app: Bsdll}
It is worth pointing out that our formula deduced previously called CP-averaged or 'experimental' branching ratio. The untagged or "theoretical" branching ratio is different from previous one through\cite{DeBruyn:2012wk} 
\begin{equation}
	\begin{aligned}
		\text{BR}^{\text{untag}}(B_s\to\mu^+\mu^-)=\left[\frac{1+\mathcal{A}_{\Delta \Gamma}y_s}{1-y_s^2}\right]\text{BR}(B_s\to\mu^+\mu^-),\\
		\mathcal{A}_{\Delta\Gamma}=\frac{|k_P|^2\cos(2\varphi_P)-|k_S|^2\cos(2\varphi_S)}{|k_P|^2+|k_S|^2},
	\end{aligned}
\end{equation}
with $ k_S, k_P, \varphi_S,\varphi_P $ which are given in\cite{DeBruyn:2012wk}.

\subsection{\texorpdfstring{$B \to V\ell^+\ell^-$}{BVll} formulae}
\label{app: BVll}
The full angular decay distribution of $ \bar{B}^0\to\bar{K}^{*0}\ell^+\ell^- $ which obtained from Buras\cite{Altmannshofer:2008dz} is shown as
\begin{equation}
	\begin{aligned}
		\frac{\dd^4\Gamma^\ell}{\dd q^2~\dd\cos\theta_\ell~\dd\cos\theta_{K^*}~\dd\phi}=\frac{9}{32\pi}J^\ell\left(q^2,\theta_\ell, \theta_{K^*},\phi\right),		
	\end{aligned}
\end{equation}
where 
\begin{equation}
	\begin{aligned}
		J^\ell\left(q^2,\theta_\ell,\theta_{K^*},\phi\right)&=J_{1;s}^\ell\sin^2\theta_{K^*}+J_{1;c}^\ell\cos^\theta_{K^*}+\left(J_{2;s}^\ell\sin^2\theta_{K^*}+J_{2;c}^\ell\cos^2\theta_{K^*}\right)\cos2\theta_\ell\\
		&+J_{3}^\ell\sin^2\theta_{K^*}\sin^2\theta_\ell\cos2\phi+J_{4}^\ell\sin2\theta_{K^*}\sin2\theta_\ell\cos\phi\\
		&+J_{5}^\ell\sin2\theta_{K^*}\sin\theta_\ell\cos\phi\\
		&+\left(J_{6;s}^\ell\sin^2\theta_{K^*}+J_{6;c}^\ell\cos^2\theta_{K^*}\right)\cos\theta_\ell+J_{7}^\ell\sin2\theta_{K^*}\sin\theta_\ell\sin\phi\\
		&+J_{8}^\ell\sin2\theta_{K^*}\sin2\theta_\ell\sin\phi+J_{9}^\ell\sin^2\theta_{K^*}\sin^2\theta_\ell\sin2\phi
	\end{aligned}\label{eq:angular_RK*}
\end{equation}
The corresponding expression for the CP-conjugated mode $ B^0\to K^{*0}\ell^+\ell^- $ is obtained form Eq.\eqref{eq:angular_RK*} by the replacements $ J^{(a)}_{1,2,3,4,7;\ell}\to \bar{J}^{(a)}_{1,2,3,4,7;\ell},~ J^{(a)}_{5,6,8,9;\ell}\to -\bar{J}^{(a)}_{5,6,8,9;\ell}. $ With the eight transverse amplitudes defined in the consecutive subsection, the angular coefficients $ J_i $ in \eqref{eq:angular_BVll} can be written as 
\begin{equation}
	\begin{aligned}
		J_{1;s}^\ell&=\frac{(2+\beta_\ell^2)}{4}\left[|A_{\perp;\ell}^\text{L}|^2+|A_{\parallel;\ell}^{\text{L}}|^2+\left(\text{L}\to\text{R}\right)\right]+\frac{4m_\ell^2}{q^2}\Re\left(A_{\perp;\ell}^{\text{L}}A_{\perp;\ell}^{\text{R$^\ast$}}+A_{\parallel;\ell}^{\text{L}}A_{\parallel;\ell}^{\text{R$^\ast$}}\right),\\
		J_{1;c}^\ell&=|A_{0;\ell}^\text{L}|^2+|A_{0;\ell}^\text{R}|^2+\frac{4m_\ell^2}{q^2}\left[|A_{t;\ell}|^2+2\Re\left(A_{0;\ell}^\text{L}A_{0;\ell}^\text{R$^\ast$}\right)\right]+\beta^2_{\ell}|A_{\text{S};\ell}|^2,\\
		J_{2;s}^\ell&=\frac{\beta_\ell^2}{4}\left[|A_{\perp;\ell}^\text{L}|^2+|A_{\parallel;\ell}^\text{L}|^2+\left(\text{L}\to\text{R}\right)\right],
		\quad J_{2;c}^\ell=-\beta_\ell^2\left[|A_{0;\ell}^\text{L}|^2+\left(\text{L}\to\text{R}\right)\right],\\
		J_{3}^\ell&=\frac{1}{2}\beta_\ell^2\left[|A_{\perp;\ell}^\text{L}|^2-|A_{\parallel;\ell}^{\text{L}}|^2+\left(\text{L}\to\text{R}\right)\right],
		\quad J_{4}^\ell=\frac{1}{\sqrt{2}}\beta_\ell^2\left[\Re\left(A_{0;\ell}^\text{L}A_{\parallel;\ell}^\text{L$^\ast$}\right)+\left(\text{L}\to\text{R}\right)\right],\\
		J_{5}^\ell&=\sqrt{2}\beta_\ell\left[\Re\left(A_{0;\ell}^\text{L}A_{\perp;\ell}^\text{L$^\ast$}\right)-\left(\text{L}\to\text{R}\right)-\frac{m_\ell}{\sqrt{q^2}}\Re\left(A_{\parallel;\ell}^\text{L}A_{\text{S};\ell}^\ast+A_{\parallel;\ell}^\text{R}A_{\text{S};\ell}^\ast\right)\right],\\
		J_{6;s}^\ell&=2\beta_\ell\left[\Re\left(A_{\parallel;\ell}^{\text{L}}A_{\perp;\ell}^{\text{L$^\ast$}}\right)-\left(\text{L}\to\text{R}\right)\right],
		\quad J_{6;c}^\ell=4\beta_\ell\frac{m_\ell}{\sqrt{q^2}}\left[\Re\left(A_{0;\ell}^{\text{L}}A_{\text{S};\ell}^{\text{$^\ast$}}\right)+\left(\text{L}\to\text{R}\right)\right],\\
		J_{7}^\ell&=\sqrt{2}\beta_\ell\left[\Im\left(A_{0;\ell}^\text{L}A_{\parallel;\ell}^{\text{L$^\ast$}}\right)-\left(\text{L}\to\text{R}\right)+\frac{m_\ell}{\sqrt{q^2}}\Im\left(A_{\perp;\ell}^\text{L}A_{\text{S};\ell}^\ast+A_{\perp;\ell}^\text{R}A_{\text{S};\ell}^\ast\right)\right],\\
		J_{8}^\ell&=\frac{1}{\sqrt{2}}\beta_\ell^2\left[\Im\left(A_{0;\ell}^\text{L}A_{\perp;\ell}^{\text{L$^\ast$}}\right)+\left(\text{L}\to\text{R}\right)\right],
		\quad J_{9}^\ell=\beta_\ell^2\left[\Im\left(A_{\parallel;\ell}^\text{L$^\ast$}A_{\perp;\ell}^\text{L}\right)+\left(\text{L}\to\text{R}\right)\right],
	\end{aligned}
\end{equation}
which further rely on various helicity amplitudes, giving
\begin{equation}
	\begin{aligned}
		A_{\perp;\ell}^{\text{L,R}}&=N_0^\ell\sqrt{2}\lambda M_B^2\left[\left[\left(C_9^{\ell}+C_9^{\prime\ell}\right)\mp\left(C_{10}^{\ell}+C_{10}^{\prime\ell}\right)\right]\frac{V(q^2)}{M_B+M_{K^\ast}}\right.\\
		&\hspace{4cm}\left.+\frac{2m_b}{q^2}\left(C_{7}^{\ell}+C_7^{\prime\ell}\right)T_1\left(q^2\right)\right]+\Delta A_{\perp}^{\text{L,R}},\\
		A_{\parallel;\ell}^\text{L,R}&=-N_0^\ell\sqrt{2}\left(M_B^2-M_{K^\ast}^2\right)\left[\left[\left(C_9^{\ell}-C_9^{\prime\ell}\right)\mp\left(C_{10}^{\ell}-C_{10}^{\prime\ell}\right)\right]\frac{A_1(q^2)}{M_B-M_{K^\ast}}\right.\\
		&\hspace{4cm}\left.\quad+\frac{2m_b}{q^2}\left(C_7^{\ell}-C_7^{\prime\ell}\right)T_2\left(q^2\right)\right]+\Delta A_{\parallel}^{\text{L,R}},\\
		A_{0;\ell}^{\text{L,R}}&=\frac{-N_0^\ell}{2M_{K^\ast}\sqrt{q^2}}\left\{\left[\left(C_9^{\ell\prime}-C_9^{\prime\ell}\right)\mp\left(C_{10}^{\ell}-C_{10}^{\prime\ell}\right)\right]\right.\\
		&\quad\left.\times\left[(M_B^2-M_{K^\ast}^2-q^2)(M_B+M_{K^\ast})A_1(q^2)-\lambda^2M_B^4\frac{A_2\left(q^2\right)}{M_B+M_{K^\ast}}\right]\right.\\
		&\quad\left.+2m_b\left(C^\ell_7-C_7^{\prime\ell}\right)\left[(M_B^2+3M_{K^\ast}^2-q^2)T_{2}\left(q^2\right)-\frac{\lambda^2M_B^4}{M_B^2-M_{K^\ast}^2}T_3\left(q^2\right)\right]\right\}+\Delta A_{0}^{\text{L,R}},\nonumber
	\end{aligned}
\end{equation}
\begin{equation}
	\begin{aligned}
		&A_{t;\ell}=\frac{N_0^\ell}{\sqrt{q^2}}\lambda M_B^2\left[2\left(C_{10}^{\ell}-C_{10}^{\prime\ell}\right)+\frac{q^2}{m_\ell}\left(C^\ell_P-C^{\prime\ell}_{P}\right)\right]A_{0}\left(q^2\right),\hspace{4cm}\\
		&A_{\text{S};\ell}=-2N_0^\ell\lambda M_B^2\left(C^\ell_\text{S}-C^{\prime\ell}_{\text{S}}\right)A_{0}\left(q^2\right),
	\end{aligned}
	\label{eq:helicity}
\end{equation}
associated with the constant 
\begin{equation}
	\begin{aligned}
		N^0_\ell=V_{tb}V_{ts}^\ast\left[\frac{G_F^2\alpha^2}{3\cdot2^{10}\pi^5M_B}q^2\lambda\beta_\ell\right]^{\frac12}
	\end{aligned}
\end{equation}
and parameters $ \lambda $ and $ \beta_\mu $  defined as
\begin{equation*}\label{eq:lambdabeta}
	\begin{aligned}
		\beta_\ell&=\sqrt{1-4\frac{m_\ell^2}{q^2}},~~
		\lambda\left(q^2,M_K^2\right)=\left[\left(1-\frac{q^2}{M_B^2}\right)^2 - 2\frac{m_K^2}{M_B^2}\left(1+\frac{q^2}{M_B^2}\right)+\frac{m_K^4}{M_B^4}\right]^{\frac12}.	
	\end{aligned}
\end{equation*}
The
corrections $ \Delta A $ in Eq. (\ref{eq:helicity}), originating from weak annihilation and spectator scattering \cite{Mahmoudi:2008tp}, are given as
\begin{equation}
	\begin{aligned}
		\Delta A_{\perp}^{L;R}&=\frac{2\sqrt{2}\lambda Nm_bM_B^2}{q^2}\left(\mathcal{T}_{\perp}^{(t),\text{nf+WA}}+\frac{\lambda_u}{\lambda_t}\mathcal{T}_{\perp}^{(u)}\right),\\
		\Delta A_{\parallel}^{L;R}&=\frac{-2\sqrt{2}E_{V}Nm_b(M_B^2-M_V^2)}{q^2M_B}\left(\mathcal{T}_{\perp}^{(t),\text{nf+WA}}+\frac{\lambda_u}{\lambda_t}\mathcal{T}_{\perp}^{(u)}\right),\\
		\Delta A_{0}^{L;R}&=-\frac{NM_B^2m_b}{\sqrt{q^2}M_V}\left\{\left[\left(M_B^2+3M_V^2-q^2\right)\frac{2E_V}{M_B^3}-\frac{\lambda^2M_B^2}{M_B^2-M_V^2}\right]\times\left(\mathcal{T}_{\perp}^{(t),\text{nf+WA}}+\frac{\lambda_u}{\lambda_t}\mathcal{T}_{\perp}^{(u)}\right)\right.\\
		&\quad\left.-\frac{\lambda^2M_B^2}{M_B^2-M_V^2}\times\left(\mathcal{T}_{\parallel}^{(t),\text{nf+WA}}+\frac{\lambda_u}{\lambda_t}\mathcal{T}_{\parallel}^{(u)}\right)\right\},
	\end{aligned}
	\label{eq:correction}
\end{equation}
with $ \lambda_i \equiv V_{ib}V_{is}^\ast$ and $ E_V=\frac{M_B^2+M_V^2-q^2}{2M_B} $. 
The amplitudes $\mathcal{T}_{\perp, \parallel}$ in Eq. (\ref{eq:correction}), containing contributions from both non-factorizable hard-spectator scattering 
and weak annihilation,  can be obtained by subtracting  factorizable contributions from the invariant amplitude calculated in
QCDF \cite{Beneke:2004dp,Beneke:2001at,Greub:2008cy}, giving
\begin{equation}
	\begin{aligned}
		\mathcal{T}_a=\xi_aC_a+\frac{\pi^2}{N_c}\frac{f_Bf_{V,a}}{MB}\Xi_a\sum_{\pm}\int_{0}^{\infty}\frac{d\omega}{\omega}\Phi_{B,\pm}(\omega)\int_{0}^{1}du\Phi_{K^\ast,a}(u)T_{a,\pm}(u,\omega)
	\end{aligned}
\end{equation}
with $ \Xi_\parallel=\frac{M_V}{E_V} $, $ \Xi_\perp =1$, $ N_c=3 $ and 
\begin{equation}
	\begin{aligned}
		C_{a}&=C_{a}^{(0)}+\frac{\alpha_s C_F}{4\pi}C_{a}^{(1)},\\
		T_{a,\pm}&=T_{a,\pm}^{(0)}+\frac{\alpha_sC_F}{4\pi}T_{a,\pm}^{(1)}.
	\end{aligned}
\end{equation}
In this work, we do not incorporate the  long-distance effect generated
from charm-loop, which has been considered in \cite{Khodjamirian:2010vf, Gubernari:2020eft}.
In large energy limit (LEL), the number of heavy-to-light transition form factors can be reduced from $7$ to $2$, corresponding
to transversal and longitudinal polarization of  $K^*$. The correspondence between the two sets of form factors \cite{Burdman:2000ku}  
is given as 
\begin{equation}
	\begin{aligned}
		V\left(q^2\right)&=\left(1+\frac{M_{\text{vec}}}{M}\right)\xi_{\perp},
		\quad A_{1}\left(q^2\right)=\frac{2E}{M+M_{\text{vec}}}\xi_{\perp},\\
		A_{2}\left(q^2\right)&=\left(1+\frac{M_{\text{vec}}}{M}\right)\left[\xi_{\perp}-\frac{M_{\text{vec}}}{E}\xi_{\parallel}\right],
		\quad A_{0}\left(q^2\right)=\left(1-\frac{M_{\text{vec}}^2}{ME}\right)\xi_{\parallel}+\frac{M_{\text{vec}}}{M}\xi_{\perp},\\
		T_1\left(q^2\right)&=\xi_{\perp},
		\quad T_2\left(q^2\right)=\left(1-\frac{q^2}{M^2-M_{\text{vec}}^2}\right)\xi_{\perp},
		\quad T_{3}\left(q^2\right)=\xi_{\perp}-\frac{M_{\text{vec}}}{E}\left(1-\frac{M_{\text{vec}}^2}{M^2}\right)\xi_{\parallel},\\
	\end{aligned}
\end{equation}
where $ M $ and $ M_\text{vec} $ stand for B mesons as well as final state vector mesons, respectively. 
To solve $ \xi_{\perp} $ and $ \xi_{\parallel} $,  we use form factors $ T_1 $ and $ A_0 $  \cite{Bharucha:2015bzk} 
in the form of simplified series expansion (SSE) in our practical analysis.
\subsection{\texorpdfstring{$ B\to P\ell^+\ell^- $}{BPll}}\label{app: BPll}
The angular distribution functions are described below\cite{Bobeth:2007dw}:
\begin{equation}
	\begin{aligned}
		I_a^\ell\left(q^2;C_{7,8,9,10,S,P}^{\prime}\right)&=\frac{q^2}{M_B^4}\left(\beta_\ell^3|F^\ell_S|^2+\beta_\ell|F^\ell_P|^2\right)+\frac{\lambda}{4}\left(\lambda\beta_\ell|C^\ell_{10}|^2+\lambda\beta_\ell\Bigg|C^\ell_9+\frac{2m_b\mathcal{T}_P^{(0)}}{M_B\xi_P}\Bigg|^2\right)\\
		&\hspace{1cm}+\frac{2m_\ell\beta_\ell\left(M_B^2-M_K^2+q^2\right)\Re\left(F^\ell_PF^{\ell\ast}_A\right)}{M_B^4}+\frac{4m_\ell^2\beta_\ell |F^\ell_A|^2}{M_B^2},\\		I_c^\ell\left(q^2;C_{7,8,9,10}^{\prime}\right)&=-\frac{\lambda}{4}\left(\lambda\beta_\ell^3|C^\ell_{10}|^2+\lambda\beta_\ell^3\Bigg|C^\ell_9+\frac{2m_b\mathcal{T}_P}{M_B\xi_P}\Bigg|^2\right),
	\end{aligned}
\end{equation}
where form factors are defined as
\begin{equation}
	\begin{aligned}
		&F^\ell_A=C^\ell_{10},\qquad F^\ell_T=0,\qquad F^\ell_{T5}=0,\\
		&F^\ell_P=\frac{M_B^2-M_K^2}{2m_b(m_b-m_s)}\frac{f_0}{f_+}(C_P^\ell+C_P^{\prime\ell})+m_lC^\ell_{10}\left[\frac{M_B^2-M_K^2}{q^2}\left(\frac{2E}{M_B}-1\right)-1\right],\qquad\\
		&F^\ell_S=\frac{M_B^2-M_K^2}{2m_b(m_b-m_s)}\frac{f_0\left(q^2\right)}{f_{+}\left(q^2\right)}\left(C^\ell_{S}+C^{\prime\ell}_{S}\right),\\
		& F^\ell_V=C^\ell_9+\frac{2m_b}{M_B}\frac{\mathcal{T}_P\left(q^2\right)}{\xi_P\left(q^2\right)},\quad\frac{f_0}{f_+}=\frac{2E}{M_B}\left[1+\mathcal{O}(\alpha_s)+\mathcal{O}\left(\frac{q^2}{M_B^2}\sqrt{\frac{\Lambda_{\text{QCD}}}{E}}\right)\right].
	\end{aligned}
\end{equation}
The invariant amplitude $ \mathcal{T}_P $ has the form similar with $\mathcal{T}_{\parallel}$ in $B\to V \ell^+ \ell^-$
 \cite{Beneke:2001at}. We preserve the leading order and next-leading order non-factorizable  contribution, giving
\begin{equation}
	\begin{aligned}
		\mathcal{T}_P(q^2)=&\xi_P\left[\left(C_7^{eff}+C_7^{\prime eff}\right)+\frac{M_B}{2m_b}Y(q^2)+\frac{\alpha_sC_F}{4\pi}C_\parallel^{nf}\right]\\
		&+\frac{\pi^2}{N_c}\frac{f_Bf_P}{M_B}\sum_{\pm}\int_{0}^{\infty}\frac{d\omega}{\omega}\Phi_{B,\pm}(\omega)\int_{0}^{1}du\Phi_{K,\parallel}(u)\left[T_{\parallel,\pm}^{(0)}+\frac{\alpha_sC_F}{4\pi}\left(T_{\parallel,\pm}^{(nf)}\right)\right](u,\omega).
	\end{aligned}
\end{equation}
The form factors based on SSE  \cite{Bobeth:2011nj,Khodjamirian:2010vf} can be parameterized as
\begin{equation}
	\begin{aligned}
		\xi_P&=f_+(q^2)		\equiv\frac{f_+(0)}{1-q^2/m^2_{\text{res}+}}\left\{
		1+a^1_+\left[z(q^2)-z(0)+\frac{1}{2}\left(z(q^2)^2-z(0)^2\right)\right]
		\right\},\\
		z(q^2)&=\frac{\sqrt{\tau_+-q^2}-\sqrt{\tau_+-\tau_0}}{\sqrt{\tau_+-q^2}+\sqrt{\tau_+-\tau_0}},~\tau_0=\sqrt{\tau_+}(\sqrt{\tau_+}-\sqrt{\tau_+-\tau_-}),~\tau_\pm=\left(M_B\pm M_P\right)^2,
	\end{aligned}
\end{equation} 
in which $ f_+(q^2=0) $ can be calculated by LCSR and listed in Table \ref{tab:FFs}.

\subsection{\texorpdfstring{$B\to X_s\ell^+\ell^-$}{BXsll}}
\label{app: BXsll}
Here we give the  definition of the important function $ \tilde{N} $  in  Eq. (\ref{eq:inclusive}),
\begin{equation}
	\begin{aligned}
		\tilde{N}&\equiv~\left(1+\frac{2\hat{m}_\ell^2}{\hat{s}}\right)\left[|C_{9}^{\text{new}}|^2\left(1+2\hat{s}\right)\left(1+\frac{\alpha_s}{\pi}\tau_{99}(\hat{s})\right)+4|C_{7}^{\text{new}}|^2\left(1+\frac{2}{\hat{s}}\right)\left(1+\frac{\alpha_s}{\pi}\tau_{77}(\hat{s})\right)\right.\\
		&+\left.12\Re(C_7^{\text{new}}C_9^{\text{new}\ast})\left(1+\frac{\alpha_s}{\pi}\tau_{79}\left(\hat{s}\right)\right)\right]+|C_{10}^{\text{new}}|^2\left[\left(1+2\hat{s}\right)+\frac{2\hat{m}_\ell^2}{\hat{s}}\left(1-4\hat{s}\right)\right]\left(1+\frac{\alpha_s}{\pi}\tau_{99}\left(\hat{s}\right)\right)\\
		&+\frac{3}{2}|m_bC_{S}|^2\left(\hat{s}-4\hat{m}_\ell^2\right)+\frac{3}{2}|m_bC_{P}|^2\hat{s}+6\Re\left(m_bC_{10}^{\text{new}}C_{P}^\ast\right)\hat{m}_\ell+\left(C_i\leftrightarrow C_i^\prime\right),
	\end{aligned}
\end{equation}
where $ \hat{m}_\ell $, $ \hat{s} $ and $ z $ has the form of 
\begin{equation}
	\begin{aligned}
		\hat{m}_\ell\equiv\frac{m_\ell}{m_{b;\text{pole}}},~\hat{s}\equiv\frac{s}{m_{b;\text{pole}}^2},~z\equiv\frac{m_{c;\text{pole}}^2}{m_{b;\text{pole}}^2}
	\end{aligned}
\end{equation}
and $ \tau_{77} $, $ \tau_{79} $ and $ \tau_{99} $ have been calculated  in \cite{Ghinculov:2003qd}. 
Especially, $ C_{7}^{\text{new}} $ and $ C_{9}^{\text{new}} $ are defined as
\begin{equation}
	\begin{aligned}
		&C_{7}^{\text{new}}(s) = \left(1+\frac{\alpha_s}{\pi}\sigma_7(s)\right)C_{7}^{\text{eff}}-\frac{\alpha_s}{4\pi}\left[C_1F_1^{(7)}(s) + C_2F_2^{(7)}(s) + C_8^{\text{eff}}F_8^{(7)}(s)\right],\\
		&C_{9}^{\text{new}}(s) = \left(1+\frac{\alpha_s}{\pi}\sigma_9(s)\right)C_{9}^{\text{eff}}-\frac{\alpha_s}{4\pi}\left[C_1F_1^{(9)}(s) + C_2F_2^{(9)}(s) + C_8^{\text{eff}}F_8^{(9)}(s)\right],\\
		&C_{9}^{\text{new}}(s) = \left(1+\frac{\alpha_s}{\pi}\sigma_9(s)\right)C_{10}^{\text{eff}},
	\end{aligned}
\end{equation}
with  $ \sigma_7,~\sigma_9 $ given in \cite{Ghinculov:2003qd} as well, 
 and $ F_{1;2;8}^{(7;9)} $ introduced in \cite{Beneke:2001at}. All the analytical formulas are well summarized in \cite{Mahmoudi:2008tp}.

\subsection{\texorpdfstring{$\Lambda_b\to\Lambda\ell^+\ell^- $}{LbL}}\label{app: LBL}
With naive factorization approximation, the 10 angular distribution function in bottomed baryon decays can be expressed as,
\begin{equation}
	\begin{aligned}
		J_{1ss}&=\frac14\left[|A_{\perp_1}^R|^2+|A_{\parallel_1}^R|^2+2|A_{\perp_0}^R|^2+2|A_{\parallel_0}^R|^2+(R\leftrightarrow L)\right],\\
		J_{2ss}&=\frac{\alpha_\Lambda}{2}\Re\left\{A_{\perp_1}^RA_{\parallel_1}^{\ast R}+2A_{\perp_0}^RA_{\parallel0}^{\ast R}+(R\leftrightarrow L)\right\},\\
		J_{1cc}&=\frac12\left[|A_{\perp_1}^R|^2+|A_{\parallel_1}^R|^2+\left(R\leftrightarrow L\right)\right],~~J_{1c}=-\Re\left\{A_{\perp_1}^RA_{\parallel_1}^{\ast R}-\left(R\leftrightarrow L\right)\right\},\\
		J_{2cc}&=+\alpha_\Lambda\Re\left\{A_{\perp_1}^RA_{\parallel_1}^{\ast R}+(R\leftrightarrow L)\right\},~~J_{2c}=-\frac{\alpha_\Lambda}{2}\left[|A_{\perp_1}^R|^2+|A_{\parallel_1}^R|^2-(R\leftrightarrow L)\right],\\	
		J_{3sc}&=+\frac{\alpha_\Lambda}{\sqrt{2}}\Im\left\{A_{\perp_1}^RA_{\perp_0}^{\ast R}-A_{\parallel_1}^RA_{\parallel_0}^{\ast R}+(R\leftrightarrow L)\right\},~~J_{3s}=\frac{\alpha_\Lambda}{\sqrt{2}}\Re\left\{A_{\perp_1}^RA_{\parallel_0}^{\ast R}-A_{\parallel_1}^RA_{\perp_0}^{\ast R}+(R\leftrightarrow L)\right\},\\
		J_{4sc}&=+\frac{\alpha_\Lambda}{\sqrt{2}}\Im\left\{A_{\perp_1}^RA_{\perp_0}^{\ast R}-A_{\parallel_1}^RA_{\parallel_0}^{\ast R}-(R\leftrightarrow L)\right\},~~J_{4s}=\frac{\alpha_\Lambda}{\sqrt{2}}\Re\left\{A_{\perp_1}^RA_{\parallel_0}^{\ast R}-A_{\parallel_1}^RA_{\perp_0}^{\ast R}-(R\leftrightarrow L)\right\},\nonumber	
	\end{aligned}
\end{equation}
where $ \alpha_\Lambda $ given in \cite{Detmold:2016pkz} and presented in Table \ref{tab:Input_para}. 
The definition of  amplitudes $A$ can be further defined based on helicity amplitudes, giving
\begin{equation}
	\begin{aligned}
		A_{\perp_0}^{L(R)}&=+\sqrt{2}N_1\left(C_{9,10,+}^{L(R)}H_0^V\left(+1/2,+1/2\right)-\frac{2m_b(C_7+C_7')}{q^2}H_0^{T}(+1/2,+1/2)\right),\\
		A_{\parallel_0}^{L(R)}&=-\sqrt{2}N_1\left(C_{9,10,+}^{L(R)}H_0^A\left(+1/2,+1/2\right)+\frac{2m_b(C_7-C_7')}{q^2}H_0^{T5}(+1/2,+1/2)\right),\\
		A_{\perp_1}^{L(R)}&=+\sqrt{2}N_1\left(C_{9,10,+}^{L(R)}H_+^V(-1/2,+1/2)-\frac{2m_b(C_7+C_7')}{q^2}H_+^T(-1/2,+1/2)\right),\\
		A_{\parallel_1}^{L(R)}&=-\sqrt{2}N_1\left(C_{9,10,-}^{L(R)}H_+^A\left(-1/2,+1/2\right)+\frac{2m_b(C_7-C_7')}{q^2}H_+^{T5}(-1/2,+1/2)\right),
	\end{aligned}
\end{equation}
with modified WCs and constant $N_1$
\begin{equation*}
\begin{aligned}
C_{9,10,+}^{L(R)}&=\left(C_9\mp C_{10}\right)+\left(C_9'\mp C_{10}'\right),\quad C_{9,10,-}^{L(R)}\left(C_9\mp C_{10}\right)-\left(C_9'\mp C_{10}'\right),\\	
N_1&=G_FV_{tb}V_{ts}^\ast\alpha_e\sqrt{\frac{q^2\lambda}{3\cdot2^{11}m_{\Lambda_b}\pi^5}},
\end{aligned}
\end{equation*}
and helicity amplitudes
\begin{equation}
	\begin{aligned}
		H_{t}^V\left(+1/2,+1/2\right)&=H_{t}^V\left(-1/2,-1/2\right)=f_t^V(q^2)\frac{m_{\Lambda_b}-m_\Lambda}{\sqrt{q^2}}\sqrt{s_+},\\
		H_{0}^V\left(+1/2,+1/2\right)&=H_{0}^V\left(-1/2,-1/2\right)=f_0^V(q^2)\frac{m_{\Lambda_b}+m_\Lambda}{\sqrt{q^2}}\sqrt{s_-},\\
		H_{+}^V\left(-1/2,+1/2\right)&=H_{-}^V\left(+1/2,-1/2\right)=-f_\perp^V(q^2)\sqrt{2s_-},\\
		H_{t}^A\left(+1/2,+1/2\right)&=-H_{t}^A\left(-1/2,-1/2\right)=f_t^A(q^2)\frac{m_{\Lambda_b}+m_\Lambda}{\sqrt{q^2}}\sqrt{s_-},\\
		H_{0}^A\left(+1/2,+1/2\right)&=-H_{0}^A\left(-1/2,-1/2\right)=f_0^A(q^2)\frac{m_{\Lambda_b}-m_\Lambda}{\sqrt{q^2}}\sqrt{s_+},\\
		H_{+}^A\left(-1/2,+1/2\right)&=-H_{-}^A\left(+1/2,-1/2\right)=-f_\perp^A(q^2)\sqrt{2s_-},\\
		H_{0}^T\left(+1/2,+1/2\right)&=-H_{0}^T\left(-1/2,-1/2\right)=-f_0^T\sqrt{q^2}\sqrt{s_-},\\
		H_{+}^T\left(-1/2,+1/2\right)&=-H_{-}^T\left(+1/2,-1/2\right)=f_\perp^T(q^2)\left(m_{\Lambda_b}+m_{\Lambda}\right)\sqrt{2s_-},\\	
		H_{0}^{T5}\left(+1/2,+1/2\right)&=-H_{0}^{T5}\left(-1/2,-1/2\right)=f_0^{T5}\sqrt{q^2}\sqrt{s_+},\\
		H_{+}^{T5}\left(-1/2,+1/2\right)&=-H_{-}^{T5}\left(+1/2,-1/2\right)=-f_\perp^{T5}(q^2)\left(m_{\Lambda_b}-m_{\Lambda}\right)\sqrt{2s_+}.
	\end{aligned}
\end{equation}
The form factors can be parameterized as
\begin{equation}
	\begin{aligned}
		f_i(q^2)=\frac{1}{1-q^2/(m_{\text{pole}}^{f_i})^2}\left[a_0^{f_i}+a_1^{f_i}z(q^2)\right].
	\end{aligned}
\end{equation}
and the detailed input parameters $a_{0,1}$ have been listed in Table \ref{tab:FFs}.

\newpage

\section{Experimental Data for Related Observables}
\label{app:exp_input}
Here  we summarize all the  experimental results  related to our analysis. The number of observables is 196 at total for new data, Dataset \textbf{B}, and 194  for Dataset \textbf{A}. 
The former one can be obtained via
replacing old $ R_{K^{(\ast)}} $ and the branching fraction of the corresponding electron mode  in the latter by the latest LHCb results \cite{LHCb:2022zom}. The detailed values have been presented in the following three tables (Table \ref{tab:old_data1}, \ref{tab:old_data2} and \ref{tab:new_data}) while the SM predictions in last column are calculated by our code supporting this analysis.

\begin{longtable}
	{p{0.21\linewidth} p{0.15\linewidth} p{0.28\linewidth} p{0.16\linewidth} p{0.16\linewidth}}
	\caption{The differential branching fractions part of Dataset \textbf{A}  in the unit of  $\text{GeV}^{-2} $.} 
	\label{tab:old_data1}
	\\
	\hline\hline
	Observable & $q^2$ (GeV$^{2}$) &Expt. value & This work &\textit{Flavio}\cite{Straub:2018kue}\\
	\endfirsthead\\
	
	\endhead
	\hline
	
	\multicolumn{5}{c}{LHCb $(B^+\to K^+\ell^+\ell^-)$\cite{LHCb:2021trn}}\\
	\hline
	$R_{K}$&$\left[1.1,6.0\right]$
	&$0.846^{+0.042+0.013}_{-0.039-0.012}$&$1.000\pm0.000$&$1.001\pm0.000$\\		
	$10^8d\mathcal{B}/dq^2|_{(K^+e^+e^-)}$&$\left[1.1,6.0\right]$
	&$2.86^{+0.15+0.13}_{-0.14-0.13}$&$3.647\pm1.147$&$3.484\pm0.647$\\
	\hline
	\multicolumn{5}{c}{LHCb $(B\to K^{(\ast)}\ell^+\ell^-)$\cite{LHCb:2021lvy}}\\
	\hline
	$R_{K^{\ast+}}$&$\left[0.045,6.0\right]$
	&$0.70^{+0.18+0.03}_{-0.13-0.04}$&$0.974\pm0.000$&$0.972\pm0.003$\\
	
	$R_{K_S}$&$\left[1.1,6.0\right]$
	&$0.66^{+0.20+0.02}_{-0.14-0.04}$&$1.000\pm0.000$&$1.001\pm0.000$\\
	
	$10^8d\mathcal{B}/dq^2|_{(K^0e^+e^-)}$
	&$\left[1.1,6.0\right]$&$2.6^{+0.6+0.1}_{-0.6-0.1}$&$3.383\pm1.045$&$3.230\pm0.531$\\
	
	$10^8d\mathcal{B}/dq^2|_{(K^{\ast+}e^+e^-)}$
	&$\left[0.045,6.0\right]$&$9.2^{+1.9+0.8}_{-1.8-0.6}$&$5.639\pm1.036$&$6.539\pm0.966$\\
	
	\hline
	\multicolumn{5}{c}{LHCb $(B^0\to K^{\ast0}\ell^+\ell^-)$\cite{LHCb:2017avl}}\\
	
	\hline
	\multirow{2}*{$R_{K^{\ast0}}$}&$\left[0.045,1.1\right]$
	&$ 0.66^{+0.11}_{-0.07}\pm0.03 $&$0.931\pm0.000$&$0.925\pm0.005$\\
	
	~&$\left[1.1,6.0\right]$
	&$0.69^{+0.11}_{-0.07}\pm0.05$&$0.996\pm0.000$&$0.996\pm0.001$\\
	\hline
	\multicolumn{5}{c}{Belle $(B \to K^{\ast}\ell^+\ell^-)$\cite{Belle:2019oag}}\\
	
	\hline
	\multirow{2}*{$R_{K^{\ast+}}$}
	&$[0.045,1.1]$&$0.62^{+0.60}_{-0.36}\pm0.09$&$0.932\pm0.000$&$0.925\pm0.005$\\
	
	~&$[1.1,6.0]$&$0.72^{+0.99}_{-0.44}\pm0.15$&$0.996\pm0.000$&$0.996\pm0.001$\\

	$10^7\mathcal{B}(K^{\ast+}e^+e^-)$
	&$[1.1,6.0]$&$1.7^{+1.0}_{-1.0}\pm0.2$&$2.227\pm0.464$&$2.546\pm0.396$\\

	$10^7\mathcal{B}(K^{\ast+}\mu^+\mu^-)$
	&$[1.1,6.0]$&$1.2^{+0.9}_{-0.7}\pm0.2$&$2.219\pm0.465$&$2.537\pm0.344$\\
	
	\hline
	
	\multirow{2}*{$R_{K^{\ast0}}$}
	&$[0.045,1.1]$&$0.46^{+0.55}_{-0.27}\pm0.13$&$0.931\pm0.000$&$0.925\pm0.004$\\
	~&$[1.1,6.0]$&$1.06^{+0.63}_{-0.38}\pm0.13$&$0.996\pm0.000$&$0.996\pm0.001$\\
	
	$10^7\mathcal{B}(K^{\ast0}e^+e^-)$
	&$[1.1,6.0]$&$1.8^{+0.6}_{-0.6}\pm0.2$&$2.035\pm0.430$&$2.331\pm0.338$\\
	
	$10^7\mathcal{B}(K^{\ast0}\mu^+\mu^-)$
	&$[1.1,6.0]$&$1.9^{+0.6}_{-0.5}\pm0.3$&$2.028\pm0.426$&$2.323\pm0.317$\\	
	
	\hline
	\multicolumn{5}{c}{Belle $(B\to K^\ast\gamma)$\cite{BelleII:2021tzi}}\\			
	\hline
	$10^5\mathcal{B}\left(K^{\ast0}\gamma\right)$
	&&$4.5\pm0.3\pm0.2$&$4.146\pm0.420$&$4.202\pm0.761$\\				
	
	$10^5\mathcal{B}\left(K^{\ast+}\gamma\right)$
	&&$5.2\pm0.4\pm0.3$&$4.474\pm0.454$&$4.271\pm0.792$\\			
	\hline
	\multicolumn{5}{c}{Belle $(B^+\to K^{+} \ell^+\ell^-)$\cite{BELLE:2019xld}}\\
	\hline
	$10^7\mathcal{B}(K^{+}\mu^+\mu^-)$&$[0.1,4.0]$
	&$1.76^{+0.41}_{-0.37}\pm0.04$&$1.444\pm0.434$&$1.370\pm0.222$\\
	
	$10^7\mathcal{B}(K^{0}_S\mu^+\mu^-)$&$[0.1,4.0]$
	&$0.62^{+0.30}_{-0.23}\pm0.02$&$0.670\pm0.202$&$0.635\pm0.111$\\
	
	$10^7\mathcal{B}(K^{+}e^+e^-)$&$[0.1,4.0]$
	&$1.80^{+0.33}_{-0.30}\pm0.05$&$1.446\pm0.442$&$1.371\pm0.232$\\
	
	$10^7\mathcal{B}(K^{0}_Se^+e^-)$&$[0.1,4.0]$
	&$0.38^{+0.25}_{-0.19}\pm0.01$&$0.671\pm0.201$&$0.636\pm0.115$\\
	
	$R_{K^+}$&$[0.1,4.0]$
	&$0.98^{+0.29}_{-0.26}\pm0.02$&$0.999\pm0.000$&$0.999\pm0.000$\\
	
	$R_{K^0_S}$&$[0.1,4.0]$
	&$1.62^{+1.31}_{-1.01}\pm0.02$&$0.999\pm0.000$&$0.999\pm0.000$\\
	\hline
	$10^7\mathcal{B}(K^{+}\mu^+\mu^-)$&$[1.0,6.0]$
	&$2.30^{+0.41}_{-0.38}\pm0.05$&$1.825\pm0.570$&$1.744\pm0.268$\\
	
	$10^7\mathcal{B}(K^{0}_S\mu^+\mu^-)$&$[1.0,6.0]$
	&$0.31^{+0.22}_{-0.16}\pm0.01$&$0.846\pm0.264$&$0.808\pm0.138$\\
	
	$10^7\mathcal{B}(K^{+}e^+e^-)$&$[1.0,6.0]$
	&$1.66^{+0.32}_{-0.29}\pm0.04$&$1.825\pm0.581$&$1.743\pm0.308$\\
	
	$10^7\mathcal{B}(K^{0}_Se^+e^-)$&$[1.0,6.0]$
	&$0.56^{+0.25}_{-0.20}\pm0.02$&$0.846\pm0.265$&$0.808\pm0.132$\\
	
	$R_{K^+}$&$[1.0,6.0]$
	&$1.39^{+0.36}_{-0.33}\pm0.02$&$1.000\pm0.000$&$1.001\pm0.000$\\
	
	$R_{K^0_S}$&$[1.0,6.0]$
	&$0.55^{+0.46}_{-0.34}\pm0.01$&$1.000\pm0.000$&$1.001\pm0.000$\\
	\hline
	
	\multicolumn{5}{c}{LHCb $(B^+\to K^{+} \mu^+\mu^-)$\cite{LHCb:2014cxe}}\\
	\hline
	\multirow{6}*{$10^9d\mathcal{B}/dq^2$}
	&$[1.1,2.0]$&$23.3\pm1.5\pm1.2$&$37.243\pm11.219$&$35.256\pm6.385$\\
	~&$[2.0,3.0]$&$28.2\pm1.6\pm1.4$&$36.911\pm11.308$&$35.095\pm6.056$\\
	~&$[3.0,4.0]$&$25.4\pm1.5\pm1.3$&$36.540\pm11.480$&$34.908\pm6.329$\\
	~&$[4.0,5.0]$&$22.1\pm1.4\pm1.1$&$36.128\pm11.715$&$34.689\pm5.610$\\
	~&$[5.0,6.0]$&$23.1\pm1.4\pm1.2$&$35.664\pm11.996$&$34.429\pm5.908$\\
	~&$[1.1,6.0]$&$24.2\pm0.7\pm1.2$&$36.482\pm11.472$&$34.868\pm5.777$\\
	\hline
	\multicolumn{5}{c}{LHCb $(B^0\to K^{0} \mu^+\mu^-)$\cite{LHCb:2014cxe}}  \\
	\hline
	\multirow{4}*{$10^9d\mathcal{B}/dq^2$}
	&$[0.1,2.0]$&$12.2^{+5.9}_{-5.2}\pm0.6$&$34.658\pm10.247$&$32.668\pm5.650$\\
	&$[2.0,4.0]$&$18.7^{+5.5}_{-4.9}\pm0.9$&$34.073\pm10.450$&$32.448\pm6.185$\\
	&$[4.0,6.0]$&$17.3^{+5.3}_{-4.8}\pm0.9$&$33.283\pm10.899$&$32.034\pm6.330$\\
	&$[1.1,6.0]$&$18.7^{+3.5}_{-3.2}\pm0.9$&$33.842\pm10.537$&$32.323\pm5.907$\\
	\hline
	\multicolumn{5}{c}{LHCb $(B^+\to K^{\ast+} \mu^+\mu^-)$\cite{LHCb:2014cxe}}  \\
	\hline
	\multirow{4}*{$10^9d\mathcal{B}/dq^2$}
	&$[0.1,2.0]$&$59.2^{+14.4}_{-13.0}\pm4.0$&$68.174\pm11.994$&$79.748\pm10.868$\\
	&$[2.0,4.0]$&$55.9^{+15.9}_{-14.4}\pm3.8$&$42.981\pm9.597$&$48.903\pm7.808$\\
	&$[4.0,6.0]$&$24.9^{+11.0}_{-9.6}\pm1.7$&$47.412\pm9.431$&$54.486\pm7.912$\\
	&$[1.1,6.0]$&$36.6^{+8.3}_{-7.6}\pm2.6$&$45.294\pm9.577$&$51.772\pm6.759$\\
	\hline
	\multicolumn{5}{c}{LHCb $(B^0\to K^{\ast0}\mu^+\mu^-)$\cite{LHCb:2016ykl}}\\
	\hline
	~&$\left[0.10,0.98\right]$&$1.016^{+0.067+0.029+0.069}_{-0.073-0.029-0.069}$&$0.881\pm0.144$&$1.063\pm0.146$\\
	~&$\left[1.1,2.5\right]$&$ 0.326^{+0.032+0.010+0.022}_{-0.031-0.010-0.022}$&$0.405\pm0.088$&$0.465\pm0.065$\\
	
	\multirow{2}*{$10^7d\mathcal{B}/dq^2$}&$\left[2.5,4.0\right]$
	&$ 0.334^{+0.031+0.009+0.023}_{-0.033-0.009-0.023} $&$0.393\pm0.085$&$0.448\pm0.062$\\
	
	~&$\left[4.0,6.0\right]$&$ 0.354^{+0.027+0.009+0.024}_{-0.026-0.009-0.024}$&$0.435\pm0.085$&$0.500\pm0.069$\\
	
	~&$\left[1.0,6.0\right]$&$ 0.342^{+0.017+0.009+0.023}_{-0.017-0.009-0.023}$&$0.414\pm0.086$&$0.474\pm0.073$\\
	\hline
	\multicolumn{5}{c}{CMS $(B^0\to K^{\ast0}\mu^+\mu^-)$ \cite{CMS:2015bcy}}  \\
	\hline
	\multirow{4}*{$10^8d\mathcal{B}/dq^2$}
	&$\left[1.0,2.0\right]$&$ 4.6^{+0.7}_{-0.7}\pm0.30 $&$4.216\pm0.090$&$4.855\pm0.666$\\
	
	&$\left[2.0,4.30\right]$&$ 3.3^{+0.5}_{-0.5}\pm0.2 $&$3.939\pm0.087$&$4.492\pm0.684$\\
	
	&$\left[4.30,6.00\right]$&$ 3.4^{+0.5}_{-0.5}\pm0.3 $&$4.398\pm0.086$&$5.056\pm0.774$\\
	
	&$\left[1.0,6.0\right]$&$ 3.6^{+0.3}_{-0.3}\pm0.2 $&$4.151\pm0.087$&$4.756\pm0.716$\\
	\hline
	\multicolumn{5}{c}{LHCb $(B^0_s\to \phi\mu^+\mu^-)$ \cite{LHCb:2021zwz}}\rule{0pt}{15pt}\\
	\hline
	\multirow{5}*{$10^{8}d\mathcal{B}/dq^2$}
	&$\left[0.1,0.98\right]$&$7.74\pm0.53\pm 0.12\pm 0.37$&$10.448\pm1.652$&$11.424\pm1.236$\\
	
	&$\left[1.1,2.5\right]$&$3.15\pm0.29\pm0.07\pm0.15$&$4.625\pm0.985$&$5.473\pm0.610$\\
	
	&$\left[2.5,4.0\right]$&$2.34\pm0.26\pm 0.05\pm0.11$&$4.405\pm0.942$&$5.166\pm0.662$\\
	
	&$\left[4.0,6.0\right]$&$3.11\pm0.24\pm0.06\pm0.15$&$4.820\pm0.922$&$5.529\pm0.788$\\
	
	&$\left[1.1,6.0\right]$&$2.88\pm0.15\pm0.05\pm0.14$&$4.637\pm0.944$&$5.402\pm0.559$\\
	\hline
	\multicolumn{5}{c}{LHCb $(\Lambda_b^0\to \Lambda\mu^+\mu^-)$\cite{LHCb:2015tgy}}  \\
	\hline
	\multirow{4}*{$10^7d\mathcal{B}/dq^2$}
	&$[1.1,6.0]$&$0.09^{+0.06+0.01}_{-0.05-0.01}\pm0.02$&$0.201\pm0.064$&$0.136\pm0.075$\\
	&$[0.1,2.0]$&$ 0.36^{+0.12+0.02}_{-0.11-0.02}\pm0.07 $&$0.192\pm0.062$&$0.088\pm0.053$\\
	&$[2.0,4.0]$&$ 0.11^{+0.12+0.01}_{-0.09-0.01}\pm0.02 $&$0.256\pm0.066$&$0.128\pm0.059$\\
	&$[4.0,6.0]$&$ 0.02^{+0.09+0.01}_{-0.00-0.01}\pm0.01 $&$0.213\pm0.063$&$0.103\pm0.051$\\
	\hline
	\multicolumn{5}{c}{BaBar $(B\to X_S\ell^+\ell^-)$\cite{BaBar:2013qry}}  \\
	\hline
	&$[1.0,6.0]$&$1.93^{+0.47+0.21}_{-0.45-0.16}\pm0.18$&$0.361\pm0.017$&$0.347\pm0.038$\\
	\multirow{3}*{$10^6d\mathcal{B}(X_s e^+e^-)/dq^2$}
	&$[0.1,2.0]$&$3.05^{+0.52+0.29}_{-0.49-0.21}\pm0.35$&$0.779\pm0.038$&$0.656\pm0.068$\\
	&$[2.0,4.3]$&$0.69^{+0.31+0.11}_{-0.28-0.07}\pm0.07$&$0.355\pm0.017$&$0.345\pm0.039$\\
	&$[4.3,6.8]$&$0.69^{+0.31+0.13}_{-0.29-0.10}\pm0.05$&$0.298\pm0.014$&$0.294\pm0.033$\\
	\hline
	\multirow{4}*{$10^6d\mathcal{B}(X_s \mu^+\mu^-)/dq^2$}
	&$[1.0,6.0]$&$0.66^{+0.82+0.30}_{-0.76-0.24}\pm0.07$&$0.361\pm0.017$&$0.334\pm0.032$\\
	&$[0.1,2.0]$&$1.83^{+0.90+0.30}_{-0.80-0.24}\pm0.20$&$0.782\pm0.038$&$0.622\pm0.066$\\
	&$[2.0,4.3]$&$-0.15^{+0.50+0.26}_{-0.43-0.14}\pm0.01$&$0.355\pm0.017$&$0.331\pm0.033$\\
	&$[4.3,6.8]$&$0.34^{+0.54+0.19}_{-0.50-0.15}\pm0.03$&$0.298\pm0.014$&$0.289\pm0.029$\\
	\hline
	\multicolumn{5}{c}{LHCb $(B^0\to \ell^+\ell^-)$ \cite{LHCb:2021awg}}\\
	\hline
	$10^9\mathcal{B}(B^0_s\to \mu^+\mu^-)$
	&&$3.09^{+0.46+0.15}_{-0.43-0.11}$&$3.681\pm0.020$&$3.672\pm0.152$\\
	$10^{10}\mathcal{B}(B^0_d\to \mu^+\mu^-)$
	&&$1.20^{+0.83+0.14}_{-0.74-0.14}$&$0.997\pm0.007$&$1.024\pm0.073$\\
	\hline
	\multicolumn{5}{c}{CMS $(B^0\to\ell^+\ell^-)$ \cite{CMS:2022mgd}}  \\
	\hline
	$10^9\mathcal{B}(B^0_s\to\mu^+\mu^-)$
	&&$3.83^{+0.38+0.19+0.14}_{-0.36-0.16-0.13}$&$3.681\pm0.020$&$3.672\pm0.152$\\
	$10^{10}\mathcal{B}(B^0_d\to\mu^+\mu^-)$
	&&$0.37^{+0.75+0.08}_{-0.67-0.09}$&$0.997\pm0.007$&$1.024\pm0.073$\\
	\hline
	\multicolumn{5}{c}{Belle $(B\to X_s\gamma)$\cite{Belle:2014nmp}}\\
	\hline
	Observable& $E_{\gamma}$ (GeV) & Expt. value&this work&\textit{Flavio}\cite{Straub:2018kue}\\
	\hline
	$10^4\mathcal{B}$
	&$>1.9$&$3.51\pm0.17\pm0.33$\\
	$10^6\text{(EXPLT. result)}$
	&$>1.6$&$375\pm18\pm35$&$296.1\pm38.0$&$330.8\pm22.9$\\
	\hline
	\multicolumn{5}{c}{Belle $(B\to \phi\gamma)$\cite{Belle:2014sac}}\\
	\hline
	Observable&  & Expt. value&this work&\textit{Flavio}\cite{Straub:2018kue}\\
	\hline
	$10^5\mathcal{B}$
	&&$3.6\pm0.5\pm0.3\pm0.6$&$3.348\pm0.526$&$4.072\pm0.510$\\
	\hline
\end{longtable}

\begin{longtable}
	{p{0.21\linewidth} p{0.15\linewidth} p{0.28\linewidth} p{0.16\linewidth} p{0.16\linewidth}}
	\caption{The part of angular distribution observables in Dataset \textbf{A}. } \label{tab:old_data2}\\
	\hline\hline
	Observable & $q^2$ (GeV$^{2}$) &Expt. value & this work &\textit{Flavio}\cite{Straub:2018kue}\\
	\endfirsthead\\
	
	\endhead
	
	\endfoot
	\hline
	\multicolumn{5}{c}{LHCb $(B^0_s\to \phi\mu^+\mu^-)$ \cite{LHCb:2021xxq}}\\
	\hline
	\multirow{4}*{$F_{\text{L}}$}
	&$\left[0.1,0.98\right]$&$0.254\pm0.045\pm 0.017$&$0.301\pm0.060$&$0.345\pm0.038$\\
	
	&$\left[1.1,4.0\right]$&$0.723\pm0.053\pm0.015$&$0.793\pm0.044$&$0.811\pm0.021$\\
	
	&$\left[4.0,6.0\right]$&$0.701\pm0.050\pm0.016$&$0.749\pm0.050$&$0.750\pm0.028$\\
	
	&$\left[1.1,6.0\right]$&$0.715\pm0.036\pm0.013$&$0.774\pm0.047$&$0.785\pm0.023$\\

	\hline
	\multicolumn{5}{c}{LHCb $(B^0\to K^{\ast0}\mu^+\mu^-)$\cite{LHCb:2020lmf}}\\
	\hline
	\multirow{4}*{$F_L$}
	&$[1.1,6.0]$&$ 0.700\pm0.025\pm0.013$ &$0.785\pm0.050$&$0.750\pm0.040$\\
	&$[1.1,2.5]$&$ 0.655\pm0.046\pm0.017 $&$0.776\pm0.051$&$0.760\pm0.039$\\
	&$[2.5,4.0]$&$ 0.756\pm0.047\pm0.023 $&$0.826\pm0.043$&$0.797\pm0.038$\\
	&$[4.0,6.0]$&$ 0.684\pm0.035\pm0.015 $&$0.762\pm0.054$&$0.712\pm0.047$\\
	\hline
	\multirow{4}*{$P_1$}
	&$[1.1,6.0]$&$ -0.079\pm0.159\pm0.021$&$-0.066\pm0.021$&$-0.113\pm0.036$\\
	&$[1.1,2.5]$&$ -0.617\pm0.296\pm0.023 $&$-0.001\pm0.001$&$0.024\pm0.053$\\
	&$[2.5,4.0]$&$ 0.168\pm0.371\pm0.043 $&$-0.064\pm0.021$&$-0.116\pm0.037$\\
	&$[4.0,6.0]$&$ 0.088\pm0.235\pm0.029 $&$-0.103\pm0.032$&$-0.178\pm0.055$\\
	\hline
	\multirow{4}*{$P_2$}
	&$[1.1,6.0]$&$ -0.162\pm0.050\pm0.012$&$-0.014\pm0.005$&$0.025\pm0.085$\\
	&$[1.1,2.5]$&$ -0.443\pm0.100\pm0.027 $&$-0.452\pm0.145$&$-0.451\pm0.013$\\
	&$[2.5,4.0]$&$ -0.191\pm0.116\pm0.043 $&$-0.114\pm0.033$&$-0.064\pm0.101$\\
	&$[4.0,6.0]$&$ 0.105\pm0.068\pm0.009 $&$0.268\pm0.086$&$0.293\pm0.074$\\
	\hline
	\multirow{4}*{$P_3$}
	&$[1.1,6.0]$&$ 0.085\pm0.090\pm0.005$&$0.001\pm0.0004$&$0.003\pm0.010$\\
	&$[1.1,2.5]$&$ 0.324\pm0.147\pm0.014 $&$0.001\pm0.001$&$0.004\pm0.021$\\
	&$[2.5,4.0]$&$ 0.049\pm0.195\pm0.014 $&$0.002\pm0.001$&$0.004\pm0.010$\\
	&$[4.0,6.0]$&$ -0.090\pm0.139\pm0.006 $&$0.001\pm0.0004$&$0.003\pm0.017$\\
	\hline
	\multirow{4}*{$P_4'$}
	&$[1.1,6.0]$&$ -0.298\pm0.087\pm0.016 $&$-0.338\pm0.090$&$-0.353\pm0.040$\\
	&$[1.1,2.5]$&$ -0.080\pm0.142\pm0.019 $&$-0.056\pm0.016$&$-0.061\pm0.044$\\
	&$[2.5,4.0]$&$ -0.435\pm0.169\pm0.035 $&$-0.374\pm0.101$&$-0.392\pm0.044$\\
	&$[4.0,6.0]$&$ -0.312\pm0.115\pm0.013 $&$-0.489\pm0.122$&$-0.503\pm0.029$\\
	\hline
	\multirow{4}*{$P_5'$}
	&$[1.1,6.0]$&$-0.114\pm0.068\pm0.026$&$-0.406\pm0.110$&$-0.447\pm0.096$\\
	&$[1.1,2.5]$&$ 0.365\pm0.122\pm0.013 $&$0.208\pm0.055$&$0.139\pm0.075$\\
	&$[2.5,4.0]$&$ -0.150\pm0.144\pm0.032 $&$-0.451\pm0.126$&$-0.501\pm0.102$\\
	&$[4.0,6.0]$&$ -0.439\pm0.111\pm0.036 $&$-0.752\pm0.191$&$-0.759\pm0.069$\\
	\hline
	&$[1.1,6.0]$&$-0.197\pm0.075\pm0.009$&$-0.045\pm0.011$&$-0.046\pm0.117$\\
	\multirow{2}*{$P_6'$}
	&$[1.1,2.5]$&$ -0.226\pm0.128\pm0.005 $&$-0.068\pm0.018$&$-0.069\pm0.083$\\
	&$[2.5,4.0]$&$ -0.155\pm0.148\pm0.024 $&$-0.051\pm0.013$&$-0.052\pm0.106$\\
	&$[4.0,6.0]$&$ -0.293\pm0.117\pm0.004 $&$-0.028\pm0.007$&$-0.030\pm0.121$\\
	\hline
	\multirow{4}*{$P_8'$}
	&$[1.1,6.0]$&$-0.020\pm0.089\pm0.009$&$-0.013\pm0.003$&$-0.015\pm0.031$\\
	&$[1.1,2.5]$&$ -0.366\pm0.158\pm0.005 $&$-0.015\pm0.004$&$-0.018\pm0.037$\\
	&$[2.5,4.0]$&$ 0.037\pm0.169\pm0.007 $&$-0.016\pm0.004$&$-0.017\pm0.038$\\
	&$[4.0,6.0]$&$ 0.166\pm0.127\pm0.004 $&$-0.010\pm0.002$&$-0.012\pm0.032$\\
	\hline
	\multicolumn{5}{c}{CMS $(B^0\to K^{\ast0}\mu^+\mu^-)$\cite{CMS:2017rzx}}\\
	\hline
	\multirow{3}*{$P_1$}
	&$\left[1.0,2.0\right]$&$ 0.12^{+0.46}_{-0.47}\pm0.10 $&$0.007\pm0.002$&$0.045\pm0.053$\\
	&$\left[2.0,4.30\right]$&$ -0.69^{+0.58}_{-0.27}\pm0.023 $&$-0.059\pm0.020$&$-0.105\pm0.037$\\
	&$\left[4.30,6.00\right]$&$ 0.53^{+0.24}_{-0.33}\pm0.19 $&$-0.104\pm0.033$&$-0.180\pm0.048$\\	
	\hline
	\multirow{3}*{$P_5'$}
	&$\left[1.0,2.0\right]$&$ 0.10^{+0.32}_{-0.31}\pm0.07 $&$0.352\pm0.101$&$0.289\pm0.061$\\
	&$\left[2.0,4.30\right]$&$ -0.57^{+0.34}_{-0.31}\pm0.18 $&$-0.398\pm0.108$&$-0.450\pm0.099$\\
	&$\left[4.30,6.00\right]$&$ -0.96^{+0.22}_{-0.21}\pm0.25 $&$-0.766\pm0.191$&$-0.771\pm0.077$\\
	\hline
	\multicolumn{5}{c}{CMS $(B^0\to K^{\ast0}\mu^+\mu^-)$ \cite{CMS:2015bcy}}  \\
	\hline
	\multirow{4}*{$F_L$}
	&$\left[1.0,2.0\right]$&$ 0.64^{+0.10}_{-0.09}\pm0.07 $&$0.739\pm0.057$&$0.724\pm0.052$\\
	&$\left[2.0,4.30\right]$&$ 0.80^{+0.08}_{-0.08}\pm0.06 $&$0.822\pm0.043$&$0.794\pm0.034$\\
	&$\left[4.30,6.00\right]$&$ 0.62^{+0.10}_{-0.09}\pm0.07 $&$0.756\pm0.054$&$0.704\pm0.055$\\
	&$\left[1.0,6.0\right]$&$ 0.73^{+0.05}_{-0.05}\pm0.04 $&$0.781\pm0.050$&$0.747\pm0.042$\\
	\hline
	\multirow{3}*{$A_{\text{FB}}$}
	&$\left[1.0,2.0\right]$&$ -0.27^{+0.17}_{-0.40}\pm0.07 $&$-0.143\pm0.038$&$-0.156\pm0.033$\\
	&$\left[2.0,4.30\right]$&$ -0.12^{+0.15}_{-0.17}\pm0.05 $&$-0.034\pm0.009$&$-0.026\pm0.029$\\
	&$\left[4.30,6.00\right]$&$ 0.01^{+0.15}_{-0.15}\pm0.03 $&$0.100\pm0.024$&$0.132\pm0.039$\\
	&$\left[1.0,6.0\right]$&$ -0.16^{+0.10}_{-0.09}\pm0.05 $&$-0.008\pm0.002$&$0.005\pm0.030$\\
	\hline
	\multicolumn{5}{c}{ATLAS $(B^0\to K^{\ast0}\mu^+\mu^-)$\cite{ATLAS:2018gqc}}\\
	\hline
	\multirow{3}*{$F_{\text{L}}$}
	&$\left[2.0,4.0\right]$&$ 0.64^{+0.11}_{-0.11}\pm0.05 $&$0.825\pm0.042$&$0.799\pm0.036$\\
	&$\left[4.0,6.0\right]$&$ 0.42^{+0.13}_{-0.13}\pm0.12 $&$0.762\pm0.053$&$0.712\pm0.048$\\
	&$\left[1.1,6.0\right]$&$ 0.56^{+0.07}_{-0.07}\pm0.06 $&$0.785\pm0.050$&$0.750\pm0.038$\\
	\hline
	\multirow{3}*{$P_1$}
	&$\left[2.0,4.0\right]$&$ -0.78^{+0.51}_{-0.51}\pm0.34 $&$-0.053\pm0.018$&$-0.095\pm0.039$\\
	&$\left[4.0,6.0\right]$&$ 0.14^{+0.43}_{-0.43}\pm0.26 $&$-0.103\pm0.033$&$-0.178\pm0.051$\\
	&$\left[1.1,6.0\right]$&$ -0.17^{+0.31}_{-0.31}\pm0.13 $&$-0.066\pm0.022$&$-0.113\pm0.033$\\
	\hline
	\multirow{3}*{$P_{4}'$}
	&$\left[2.0,4.0\right]$&$ -0.76^{+0.31}_{-0.31}\pm0.21 $&$-0.330\pm0.088$&$-0.347\pm0.044$\\
	&$\left[4.0,6.0\right]$&$ 0.64^{+0.33}_{-0.33}\pm0.18 $&$-0.489\pm0.125$&$-0.503\pm0.028$\\
	&$\left[1.1,6.0\right]$&$ 0.05^{+0.22}_{-0.22}\pm0.14 $&$-0.338\pm0.088$&$-0.353\pm0.034$\\
	\hline
	\multirow{3}*{$P_{5}'$}
	&$\left[2.0,4.0\right]$&$ -0.33^{+0.31}_{-0.31}\pm0.13 $&$-0.353\pm0.096$&$-0.410\pm0.107$\\
	&$\left[4.0,6.0\right]$&$ 0.26^{+0.35}_{-0.35}\pm0.18 $&$-0.752\pm0.196$&$-0.759\pm0.082$\\
	&$\left[1.1,6.0\right]$&$ 0.01^{+0.21}_{-0.21}\pm0.08 $&$-0.406\pm0.108$&$-0.447\pm0.092$\\
	\hline
	\multirow{3}*{$P_6'$}
	&$\left[2.0,4.0\right]$&$ 0.31^{+0.28}_{-0.28}\pm0.19 $&$-0.055\pm0.014$&$-0.056\pm0.099$\\
	&$\left[4.0,6.0\right]$&$ 0.06^{+0.27}_{-0.27}\pm0.13 $&$-0.028\pm0.006$&$-0.030\pm0.129$\\
	&$\left[1.1,6.0\right]$&$ 0.03^{+0.17}_{-0.17}\pm0.12 $&$-0.045\pm0.011$&$-0.046\pm0.088$\\
	\hline
	\multirow{3}*{$P_8'$}
	&$\left[2.0,4.0\right]$&$ 1.07^{+0.41}_{-0.41}\pm0.39 $&$-0.016\pm0.004$&$-0.018\pm0.037$\\
	&$\left[4.0,6.0\right]$&$ -0.24^{+0.42}_{-0.42}\pm0.09 $&$-0.010\pm0.002$&$-0.012\pm0.032$\\
	&$\left[1.1,6.0\right]$&$ 0.23^{+0.28}_{-0.28}\pm0.20 $&$-0.013\pm0.003$&$-0.015\pm0.037$\\
	\hline
	\multicolumn{5}{c}{Belle $(B^0\to K^{\ast0}e^+e^-)$\cite{Belle:2016fev}}\\
	\hline
	\multirow{3}*{$P_4^{\mu'}$}
	&$\left[1.0,6.0\right]$&$ -0.22^{+0.35}_{-0.34}\pm0.15 $&$-0.326\pm0.086$&$-0.341\pm0.045$\\
	&$\left[0.10,4.00\right]$&$ -0.38^{+0.50}_{-0.48}\pm0.12 $&$-0.026\pm0.008$&$-0.028\pm0.030$\\
	&$\left[4.00,8.00\right]$&$ -0.07^{+0.32}_{-0.31}\pm0.07 $&$-0.503\pm0.124$&$-0.518\pm0.026$\\
	\hline
	\multirow{3}*{$P_4^{e'}$}
	&$\left[1.0,6.0\right]$&$ -0.72^{+0.40}_{-0.39}\pm0.06 $&$-0.323\pm0.093$&$-0.338\pm0.037$\\
	&$\left[0.10,4.00\right]$&$ 0.34^{+0.41}_{-0.45}\pm0.11 $&$-0.004\pm0.002$&$-0.004\pm0.030$\\
	&$\left[4.00,8.00\right]$&$ -0.52^{+0.24}_{-0.22}\pm0.03 $&$-0.503\pm0.124$&$-0.518\pm0.021$\\
	\hline
	\multirow{3}*{$P_5^{\mu'}$}
	&$\left[1.0,6.0\right]$&$ 0.43^{+0.26}_{-0.28}\pm0.10 $&$-0.382\pm0.104$&$-0.423\pm0.082$\\
	&$\left[0.10,4.00\right]$&$ 0.42^{+0.39}_{-0.39}\pm0.14 $&$0.205\pm0.061$&$0.156\pm0.064$\\
	&$\left[4.00,8.00\right]$&$ -0.03^{+0.31}_{-0.30}\pm0.09 $&$-0.802\pm0.198$&$-0.795\pm0.070$\\
	\hline
	\multirow{3}*{$P_5^{e'}$}
	&$\left[1.0,6.0\right]$&$ -0.22^{+0.39}_{-0.41}\pm0.03 $&$-0.375\pm0.107$&$-0.416\pm0.092$\\
	&$\left[0.10,4.00\right]$&$ 0.51^{+0.39}_{-0.46}\pm0.09 $&$0.219\pm0.063$&$0.174\pm0.063$\\
	&$\left[4.00,8.00\right]$&$ -0.52^{+0.28}_{-0.26}\pm0.03 $&$-0.799\pm0.197$&$-0.792\pm0.057$\\
	\hline
	\multirow{3}*{$Q_4$}
	&$\left[1.0,6.0\right]$&$ 0.498^{+0.527}_{-0.527}\pm0.166 $&$-0.003\pm0.127$&$-0.003\pm0.0002$\\
	&$\left[0.10,4.00\right]$&$ -0.723^{+0.676}_{-0.676}\pm0.163 $&$-0.022\pm0.008$&$-0.024\pm0.003$\\
	&$\left[4.00,8.00\right]$&$ -0.448^{+0.392}_{-0.392}\pm0.076 $&$-0.000\pm0.175$&$-0.000\pm0.000$\\
	\hline
	\multirow{3}*{$Q_5$}
	&$\left[1.0,6.0\right]$&$ 0.656^{+0.485}_{-0.485}\pm0.103 $&$-0.007\pm0.149$&$-0.007\pm0.001$\\
	&$\left[0.10,4.00\right]$&$ -0.097^{+0.601}_{-0.601}\pm0.164 $&$-0.014\pm0.088$&$-0.018\pm0.006$\\
	&$\left[4.00,8.00\right]$&$ 0.498^{+0.410}_{-0.410}\pm0.095 $&$-0.003\pm0.279$&$-0.003\pm0.000$\\
	\hline
	
	\multicolumn{5}{c}{LHCb $(B^+\to K^{\ast+}\mu^+\mu^-)$\cite{LHCb:2020gog}}\\  
	\hline
	\multirow{4}*{$F_L$}
	&$[1.1,2.5]$&$0.54^{+0.18}_{-0.19}\pm0.03$&$0.784\pm0.049$&$0.768\pm0.041$\\
	&$[2.5,4.0]$&$0.17^{+0.24}_{-0.14}\pm0.04$&$0.829\pm0.042$&$0.800\pm0.034$\\
	&$[4.0,6.0]$&$0.67^{+0.11}_{-0.14}\pm0.03$&$0.764\pm0.053$&$0.714\pm0.051$\\
	&$[1.1,6.0]$&$0.59^{+0.10}_{-0.10}\pm0.03$&$0.788\pm0.049$&$0.754\pm0.038$\\
	\hline
	\multirow{4}*{$P_1$}
	&$[1.1,2.5]$&$1.60^{+4.92}_{-1.75}\pm0.32$&$-0.001\pm0.0004$&$0.022\pm0.049$\\
	&$[2.5,4.0]$&$-0.29^{+1.43}_{-1.04}\pm0.22$&$-0.064\pm0.021$&$-0.118\pm0.036$\\
	&$[4.0,6.0]$&$-1.24^{+0.99}_{-1.17}\pm0.29$&$-0.102\pm0.032$&$-0.178\pm0.049$\\
	&$[1.1,6.0]$&$-0.51^{+0.56}_{-0.54}\pm0.08$&$-0.066\pm0.022$&$-0.115\pm0.032$\\
	\hline
	\multirow{4}*{$P_2$}
	&$[1.1,2.5]$&$-0.28^{+0.24}_{-0.42}\pm0.15$&$-0.453\pm0.154$&$-0.451\pm0.016$\\
	&$[2.5,4.0]$&$0.03^{+0.26}_{-0.25}\pm0.11$&$-0.109\pm0.033$&$-0.055\pm0.107$\\
	&$[4.0,6.0]$&$-0.15^{+0.19}_{-0.20}\pm0.06$&$0.268\pm0.087$&$0.295\pm0.064$\\
	&$[1.1,6.0]$&$-0.13^{+0.13}_{-0.13}\pm0.05$&$-0.011\pm0.002$&$0.032\pm0.080$\\
	\hline
	&$[1.1,2.5]$&$-0.09^{+0.70}_{-0.99}\pm0.18$&$0.001\pm0.0004$&$0.004\pm0.021$\\
	&$[2.5,4.0]$&$-0.45^{+0.50}_{-0.62}\pm0.20$&$0.002\pm0.0005$&$0.004\pm0.011$\\
	\multirow{2}*{$P_3$}
	&$[4.0,6.0]$&$0.52^{+0.82}_{-0.62}\pm0.15$&$0.001\pm0.0004$&$0.003\pm0.014$\\
	&$[1.1,6.0]$&$0.12^{+0.27}_{-0.28}\pm0.04$&$0.001\pm0.0004$&$0.003\pm0.010$\\
	\hline
	\multirow{4}*{$P_4'$}
	&$[1.1,2.5]$&$-0.58^{+0.62}_{-0.56}\pm0.11$&$-0.052\pm0.015$&$-0.063\pm0.043$\\
	&$[2.5,4.0]$&$-0.81^{+1.09}_{-0.84}\pm0.14$&$-0.371\pm0.098$&$-0.391\pm0.044$\\
	&$[4.0,6.0]$&$-0.79^{+0.47}_{-0.28}\pm0.09$&$-0.487\pm0.120$&$-0.502\pm0.027$\\
	&$[1.1,6.0]$&$-0.41^{+0.28}_{-0.28}\pm0.07$&$-0.335\pm0.096$&$-0.353\pm0.042$\\
	\hline
	\multirow{4}*{$P_5'$}
	&$[1.1,2.5]$&$0.88^{+0.70}_{-0.71}\pm0.10$&$0.180\pm0.050$&$0.113\pm0.113$\\
	&$[2.5,4.0]$&$-0.87^{+1.00}_{-1.68}\pm0.09$&$-0.467\pm0.125$&$-0.517\pm0.098$\\
	&$[4.0,6.0]$&$-0.25^{+0.32}_{-0.40}\pm0.09$&$-0.756\pm0.187$&$-0.764\pm0.083$\\
	&$[1.1,6.0]$&$-0.07^{+0.25}_{-0.25}\pm0.04$&$-0.421\pm0.123$&$-0.461\pm0.086$\\
	\hline
	&$[1.1,2.5]$&$0.25^{+1.22}_{-1.32}\pm0.08$&$-0.059\pm0.017$&$-0.054\pm0.083$\\
	&$[2.5,4.0]$&$-0.37^{+1.59}_{-3.91}\pm0.05$&$-0.049\pm0.012$&$-0.044\pm0.102$\\
	\multirow{2}*{$P_6'$}
	&$[4.0,6.0]$&$-0.09^{+0.40}_{-0.41}\pm0.05$&$-0.029\pm0.007$&$-0.028\pm0.113$\\
	&$[1.1,6.0]$&$-0.21^{+0.23}_{-0.23}\pm0.04$&$-0.043\pm0.012$&$-0.039\pm0.092$\\
	\hline
	\multirow{4}*{$P_8'$}
	&$[1.1,2.5]$&$0.12^{+0.75}_{-0.76}\pm0.05$&$-0.021\pm0.005$&$-0.027\pm0.040$\\
	&$[2.5,4.0]$&$0.12^{+7.89}_{-4.95}\pm0.07$&$-0.016\pm0.004$&$-0.018\pm0.034$\\
	&$[4.0,6.0]$&$-0.15^{+0.44}_{-0.48}\pm0.05$&$-0.011\pm0.002$&$-0.011\pm0.033$\\
	&$[1.1,6.0]$&$0.03^{+0.26}_{-0.28}\pm0.06$&$-0.015\pm0.004$&$-0.017\pm0.036$\\
	\hline
	\multicolumn{5}{c}{LHCb $(B^0\to K^{\ast0}e^+e^-)$\cite{LHCb:2020dof}}\\  
	\hline
	$F_L$&$[0.0008,0.257]$&$0.54^{+0.18}_{-0.19}\pm0.03$&$0.077\pm0.026$&$0.050\pm0.013$\\
	$A_T^{\Re}$&$[0.0008,0.257]$&$0.17^{+0.24}_{-0.14}\pm0.04$&$-0.032\pm0.012$&$-0.024\pm0.001$\\
	$A_T^2$&$[0.0008,0.257]$&$0.67^{+0.11}_{-0.14}\pm0.03$&$-0.000\pm0.000$&$-0.002\pm0.021$\\
	$A_T^{\Im}$&$[0.0008,0.257]$&$0.59^{+0.10}_{-0.10}\pm0.03$&$0.001\pm0.000$&$0.032\pm0.020$\\
	\hline
	\multicolumn{5}{c}{LHCb $(B^0_s\to \phi\gamma)$\cite{LHCb:2019vks} }\\  
	\hline
	$S_{\phi\gamma}$&&$0.43\pm0.30\pm0.11$&$0.001\pm0.000$&$-0.000\pm0.0002$\\
	$A_{\text{CP}}$&&$0.11\pm0.29\pm0.11$&$0.000\pm0.000$&$0.004\pm0.002$\\
	$A_{\Delta\Gamma}$&&$-0.67^{+0.37}_{-0.41}\pm0.17$&$0.029\pm0.000$&$0.031\pm0.020$\\
	\hline
\end{longtable}
\begin{table}[h]
	\caption{The updated components related to LHCb new results in Dataset \textbf{B}, where the 
	branching fractions are in the unit of  $\text{GeV} ^{-2} $.} \label{tab:new_data}
	\resizebox{0.95 \textwidth}{!}
	{\begin{tabular}{p{0.25\linewidth} p{0.15\linewidth} p{0.20\linewidth} p{0.16\linewidth} p{0.16\linewidth}}
		\hline
		\multicolumn{5}{c}{LHCb $(B\to K^{(\ast)}\ell^+\ell^-)$\cite{LHCb:2020gog}}\\ 
		\hline
		Observable & $q^2$ (GeV$^{2}$) &Expt. value&this work&\textit{Flavio}\cite{Straub:2018kue} \\
		\hline
		$ R_{K} $
		& $\left[0.1,1.1\right]$ & $ 0.994^{+0.090+0.029}_{-0.082-0.027} $ &$0.994\pm0.000$&$0.993\pm0.000$\\
		$ R_{K} $
		& $\left[1.1,6.0\right]$ & $ 0.949^{+0.042+0.022}_{-0.041-0.022} $ &$1.000\pm0.000$&$1.001\pm0.000$\\
		$ R_{K^{\ast}} $
		& $\left[0.1,1.1\right]$ & $ 0.927^{+0.093+0.036}_{-0.087-0.035} $ &$0.983\pm0.000$&$0.983\pm0.001$\\
		$ R_{K^{\ast}} $
		& $\left[1.1,6.0\right]$ & $ 1.027^{+0.072+0.027}_{-0.068-0.026}$ &$0.996\pm0.000$&$0.996\pm0.001$\\
		$10^9d\mathcal{B}(K^+ e^+e^-)/dq^2$
		& $\left[1.1,6.0\right]$ & $ 25.5^{+1.3}_{-1.2}\pm1.1 $  &$36.474\pm11.5$&$34.841\pm6.065$\\
		$10^9d\mathcal{B}(K^{\ast0} e^+e^-)/dq^2$
		& $\left[1.1,6.0\right]$ & $ 33.3^{+2.7}_{-2.6}\pm2.2 $  &$41.541\pm8.73$&$47.580\pm7.031$\\
		\hline
	\end{tabular}}
\end{table}

\normalem
\clearpage
\bibliography{reference_V2}
\end{document}